\documentclass[superscriptaddress,twocolumn,showpacs,a4paper,
amssymb,amsmath,nobibnotes,aps,prd,
showkeys,
nofootinbib,notitlepage]{revtex4-1}
\usepackage{verbatim}
\usepackage[T1]{fontenc}
\usepackage[utf8]{inputenc}
\usepackage[american]{babel}
\usepackage{epsfig}
\usepackage{graphicx}
\usepackage{booktabs}
\usepackage{multirow}
\usepackage{dcolumn}
\usepackage{amsmath}
\usepackage{mathtools}
\usepackage{amsfonts}
\usepackage{amssymb}
\usepackage{ulem}
\usepackage{epstopdf}
\usepackage{bm}
\usepackage{siunitx}
\usepackage{braket}
\usepackage{enumitem}
\usepackage{soul}
\usepackage{fixmath}
\usepackage[table]{xcolor}
\usepackage{color}
\usepackage{transparent}
\usepackage{pifont}
\usepackage{enumitem}

\definecolor{navyblue}{rgb}{0.0, 0.0, 0.5}
\definecolor{royalblue}{rgb}{0.25, 0.41, 0.88}
\definecolor{cadmiumgreen}{rgb}{0.0, 0.42, 0.24}
\definecolor{blue-violet}{rgb}{0.54, 0.17, 0.89}
\definecolor{darkviolet}{rgb}{0.58, 0.0, 0.83}
\definecolor{orange(colorwheel)}{rgb}{1.0, 0.5, 0.0}

\usepackage{hyperref}
\hypersetup{
    colorlinks=true, 
    linkcolor=royalblue, 
    citecolor=magenta}

\usepackage{booktabs}
\usepackage{multirow}
\usepackage{dcolumn}
\usepackage{colortbl}

\begin{document}

\title{Dynamical dark energy  confronted with multiple CMB missions}

\author{Mahdi Najafi}
\email{mahdinajafi12676@yahoo.com}
\affiliation{Department of Physics, K.N. Toosi University of Technology, P.O. Box 15875-4416, Tehran, Iran}
\affiliation{PDAT Laboratory, Department of Physics, K.N. Toosi University of Technology, P.O. Box 15875-4416, Tehran, Iran}

\author{Supriya Pan}
\email{supriya.maths@presiuniv.ac.in}
\affiliation{Department of Mathematics, Presidency University, 86/1 College Street,  Kolkata 700073, India}
\affiliation{Institute of Systems Science, Durban University of Technology, PO Box 1334, Durban 4000, Republic of South Africa}

\author{Eleonora Di Valentino}
\email{e.divalentino@sheffield.ac.uk}
\affiliation{School of Mathematics and Statistics, University of Sheffield, Hounsfield Road, Sheffield S3 7RH, United Kingdom}

\author{Javad T. Firouzjaee}
\email{firouzjaee@kntu.ac.ir}
\affiliation{Department of Physics, K.N. Toosi University of Technology, P.O. Box 15875-4416, Tehran, Iran}
\affiliation{PDAT Laboratory, Department of Physics, K.N. Toosi University of Technology, P.O. Box 15875-4416, Tehran, Iran}
\affiliation{School of Physics, Institute for Research in Fundamental Sciences (IPM), P.O. Box 19395-5531, Tehran, Iran}

\begin{abstract}
The measurements of the cosmic microwave background (CMB) have played a significant role in understanding the nature of dark energy. In this article, we investigate the dynamics of the dark energy equation of state, utilizing high-precision CMB data from multiple experiments. We begin by examining the Chevallier-Polarski-Linder (CPL) parametrization, a commonly used and recognized framework for describing the dark energy equation of state. We then explore the general Exponential parametrization, which incorporates CPL as its first-order approximation, and extensions of this parametrization that incorporate nonlinear terms. We constrain these scenarios using CMB data from various missions, including the Planck 2018 legacy release, the Wilkinson Microwave Anisotropy Probe (WMAP), the Atacama Cosmology Telescope (ACT), and the South Pole Telescope (SPT), as well as combinations with low-redshift cosmological probes such as Baryon Acoustic Oscillations (BAO) and the Pantheon sample. While the $\Lambda$CDM cosmology remains consistent within the 68\% confidence level, we observe that the extensions of the CPL parametrization are indistinguishable for Planck data. However, for ACT and SPT data, the inclusion of additional terms begins to reveal a peak in $w_{\rm a, DE}$ that was previously unconstrained.

\end{abstract}
\maketitle

\section{Introduction}

One of the noteworthy discoveries of the twentieth century is the detection of the late-time accelerating expansion of the universe, which opened a new chapter in modern cosmology~\cite{SupernovaSearchTeam:1998fmf,SupernovaCosmologyProject:1998vns}. This phenomenon requires the existence of some 'exotic' kind of fluid(s) with large negative pressure. Despite the lack of a clear understanding of the origin of such exotic fluid(s) based on current observational evidence, there are several possible explanations for this accelerating expansion of the universe. The simplest approach is to modify the matter sector of the universe by introducing some exotic fluid(s), referred to as 'dark energy' (DE), while assuming that the gravitational sector of our universe is perfectly described by Einstein's General Relativity (GR).
In this framework, the evolution of DE density could either remain constant or vary with cosmic time. The cosmological constant represents the simplest DE candidate, with a density that remains constant throughout the evolution of the universe. However, there are numerous examples of DE models that exhibit time-dependent behavior~\cite{Copeland:2006wr,Bamba:2012cp}. Conversely, modifications of GR or new gravitational theories beyond GR can also give rise to effective exotic types of fluids with negative pressure (often termed as geometrical DE), which consequently serve as a source for the late-time cosmic acceleration. Pursuing this avenue, various geometrical DE models have been proposed~\cite{Nojiri:2006ri,Sotiriou:2008rp,DeFelice:2010aj,Clifton:2011jh,Capozziello:2011et,Cai:2015emx,Nojiri:2017ncd}.
In a spatially flat, homogeneous, and isotropic Friedmann-Lema\^{i}tre-Robertson-Walker (FLRW) universe, if we consider Einstein's GR in the background and assume a DE fluid alongside normal matter (NM) in the matter sector, the Friedmann equations take the form:
$\rho_{\rm NM} + \rho_{\rm DE} = \frac{3 H^2}{\kappa^2}$,
and
$p_{\rm NM} + p_{\rm DE} = -\frac{2\dot{H} + 3 H^2}{\kappa^2}$,
where $H$ is the Hubble rate of the FLRW universe, an overhead dot represents the cosmic time derivative, $\kappa^2$ is Einstein's gravitational constant, and $\rho_i, p_i$ are respectively the energy density and pressure of the $i$-th fluid ($i = {\rm NM}, {\rm DE}$). 
Now, if the gravitational sector is described by a modified version of GR or by a completely different gravity theory, then the Friedmann equations in the FLRW universe can be expressed similarly to those in the case of GR~\cite{Nojiri:2017ncd}:

\begin{eqnarray}\label{mg-efe}
 \rho_{\rm NM} + \rho_{\rm eff} = \frac{3 H^2}{\kappa^2},\quad p_{\rm NM} + p_{\rm eff}  = - \frac{2\dot{H} + 3 H^2}{\kappa^2},
\end{eqnarray}
where $\rho_{\rm eff}$ and $p_{\rm eff}$ are respectively the energy density and pressure of the effective fluid that mimics the effects of DE in GR. For example, if one considers a modified version of GR described by $F(R) = R + f(R)$ (where $R$ is the Ricci scalar), then $\rho_{\rm eff}$ and $p_{\rm eff}$ will contain the terms responsible for $f(R)$.
In reality, one can propose any DE (in terms of a specific energy-momentum tensor) or geometrical DE model (assuming any modified version of GR or a completely new gravitational theory) and investigate the underlying model with the observational data. Among these existing cosmological models, the $\Lambda$-Cold Dark Matter ($\Lambda$CDM) cosmological model within the framework of GR stands out. In this model, $\Lambda > 0$ serves as the source of the DE (the energy density of DE, i.e., $\Lambda/\kappa^2$, is constant over time), and DM is pressure-less. $\Lambda$CDM represents the simplest cosmological scenario that agrees with several, but not all, observational datasets. However, $\Lambda$CDM has faced challenges for various reasons~\cite{Abdalla:2022yfr}, prompting suggestions for a revision of $\Lambda$CDM cosmology based on existing discrepancies between early and late cosmological probes.

Along this line, a popular approach to revise the $\Lambda$CDM model is to parametrize the equation of state of the dark component (either DE or geometrical DE), which is defined as the ratio of its pressure to the energy density. Parametrizing the equation of state of DE or geometrical DE can offer a potential route to understand the inherent nature of the dark sector. By examining the gravitational equations (\ref{mg-efe}), one can reconstruct the underlying gravitational theory through $w_{\text{eff}} = p_{\text{eff}} / \rho_{\text{eff}}$. Over the last several years, various forms of $w_{\text{DE}}$ ($w_{\text{eff}}$) have been explored with observational data (see, for example,~\cite{Efstathiou:1999tm,Chevallier:2000qy,Astier:2000as,Weller:2001gf,Linder:2002et,Rubano:2003er,Guo:2004fq,Feng:2004ff,Nesseris:2005ur,Jassal:2005qc,Gong:2005de,Nojiri:2006ww,Kurek:2007bu,Barboza:2008rh,Kurek:2008qt,Barboza:2009ks,Li:2011dr,Li:2012vn,Pace:2011kb,Feng:2011zzo,Feng:2012gf,Akarsu:2015yea,Chavez:2016epc,Pantazis:2016nky,Yang:2017amu,Yang:2017alx,Pan:2017zoh,Rezaei:2017yyj,Mehrabi:2018oke,Yang:2018qmz,Yang:2018prh,Panotopoulos:2018sso,Pan:2019hac,Chudaykin:2020ghx,Yang:2021flj,Yang:2021eud,Yang:2020ope,Benaoum:2020qsi,Wang:2022xdw,Yang:2022kho,Rezaei:2023xkj,Escamilla:2023oce,Adil:2023exv,Rezaei:2024vtg,Escamilla:2024olw,Moshafi:2024guo,Reyhani:2024cnr} and the references therein).
In this article, we consider an Exponential model for the DE equation of state~\cite{Dimakis:2016mip,Pan:2019brc}, which, up to its first order of expansion, returns the well-known Chevallier-Polarski-Linder parametrization~\cite{Chevallier:2000qy,Linder:2002et}. The second and third terms of the expansion add corrections to the CPL parametrization. Our aim is to constrain the generalized Exponential form and its expansion up to the third order using cosmic microwave background (CMB) observations from multiple missions. With increasing sensitivity in astronomical data and the availability of various datasets, it is crucially important to examine whether the latest cosmological probes are offering any new information in this direction.

The article is organized as follows. In Section~\ref{sec-2}, we describe the gravitational equations of a general DE model, considering its evolution at both the background and perturbation levels. We then introduce the DE parametrizations that we wish to study. In Section~\ref{sec-dataset}, we describe the cosmological probes and the numerical methodology adopted to constrain the present DE parametrizations.

\section{Parametrized Dark Energy}
\label{sec-2}

As mentioned in the introduction, we adopt the spatially flat FLRW line element to describe the geometrical configuration of our universe, with its gravitational sector assumed to be governed by Einstein's General Relativity. The fluids in the matter sector, including relativistic radiation, pressureless matter, and a DE component with an evolving equation of state $w_{\text{DE}}(z)$, do not interact with each other. The complete set of equations describing this underlying scenario is given by:

\begin{eqnarray}
    H^2 = \frac{\kappa^2}{3} (\rho_r + \rho_m + \rho_{\rm DE}),\label{efe-1}\\
    2\dot{H} + 3 H^2 = - \kappa^2 (p_r + p_m + p_{\rm DE}),\label{efe-2}
\end{eqnarray}

where $\rho_i$ and $p_i$ represent, respectively, the energy density and pressure of the $i$-th fluid ($i$ denotes radiation ($r$), pressureless matter ($m$), and dark energy ($\text{DE}$)).
Note that from Bianchi's identity, one can derive the conservation equation for the total fluid $\sum_{i} \dot{\rho}_{i} + 3 H  \sum_{i} (p_i + \rho_i) = 0$ . Since the fluids do not interact with each other, one can write down the conservation equation for each individual fluid as follows:
\begin{eqnarray}
\dot{\rho}_{i} + 3 H  (p_i + \rho_i) = 0. \label{balance}      
\end{eqnarray}
Now, for radiation, $p_r = \rho_r/3$; for pressure-less matter, $p_m = 0$; and for DE with a dynamical equation of state $w_{\text{DE}}(z)$, one can use the balance equation (\ref{balance}) to find:

\begin{eqnarray}
    \rho_r = \rho_{r,0} (a/a_0)^{-4}, \quad \rho_m  = \rho_{m,0} (a/a_0)^{-3},\\
    \rho_{\rm DE} = \rho_{\rm DE,0}\; \exp \left[ 3 \; \int_{a}^{a_0} \frac{1+w_{\rm DE}(a^\prime)}{a'} da^\prime \right],
\end{eqnarray}
where $\rho_{i,0}$ refers to the present value of $\rho_{i}$ and $a_0$ is the scale factor at the current time (hereafter we set $a_0 = 1$). 
Thus, given a specific form of $w_{\text{DE}}(z)$, one can, in principle, determine the evolution of the energy density of the DE sector either analytically or numerically. Finally, using the Friedmann equation (\ref{efe-1}), one can obtain the expansion scale factor. This implies that the evolution of the cosmological scenario driven by the DE equation of state can be derived at the background level. However, along with the evolution at the background level, it is essential to understand how the model affects the structure formation of the universe. In other words, it is crucial to investigate whether the prescribed DE equation of state can lead to a stable cosmological scenario at the level of perturbations as well.
The next paragraph provides the key cosmological equations in this context. To incorporate the evolution of the underlying model at the level of perturbations, we consider the following metric:
\begin{eqnarray}
\label{perturbed-metric}
ds^2 = a^2(\tau) \left [-d\tau^2 + (\delta_{ij}+h_{ij}) dx^idx^j  \right], 
\end{eqnarray}
where $\tau$ is the conformal time, $\delta_{ij}$ and $h_{ij}$ are respectively the unperturbed and perturbed metric tensors. 
For the perturbed metric (\ref{perturbed-metric}), utilizing the conservation equation for the $i$-th fluid, i.e., $T^{\mu \nu}_{; \nu} = 0$, the Einstein's equations in the synchronous gauge of Fourier space $\kappa$ can be expressed as~\cite{Mukhanov:1990me,Ma:1995ey,Malik:2008im}:
\begin{eqnarray}
\delta'_{i}  = - (1+ w_{i})\, \left(\theta_{i}+ \frac{h'}{2}\right) - 
3\mathcal{H}\left(\frac{\delta p_i}{\delta \rho_i} - w_{i} \right)\delta_i \nonumber \\ -  9 
\mathcal{H}^2\left(\frac{\delta p_i}{\delta \rho_i} - c^2_{a,i} \right) (1+w_i) 
\frac{\theta_i}
{{\kappa}^2}, \label{per1} \\
\theta'_{i}  = - \mathcal{H} \left(1- 3 \frac{\delta p_i}{\delta
\rho_i}\right)\theta_{i} 
+ \frac{\delta p_i/\delta \rho_i}{1+w_{i}}\, {\kappa}^2\, \delta_{i} 
-{\kappa}^2\sigma_i,\label{per2}
\end{eqnarray}
where a prime attached to any quantity denotes the derivative with respect to the conformal time $\tau$; $\mathcal{H}= a^{\prime}/a$ denotes the conformal Hubble factor; $\delta_i = \delta \rho_i/\rho_i$ denotes the density perturbation; $h = h^{j}{j}$ is the trace of the metric perturbations $h{ij}$; $\theta_{i}\equiv i \kappa^{j} v_{j}$ refers to the divergence of the $i$-th fluid velocity; $\sigma_i$ stands for the anisotropic stress of the $i$-th fluid, which we have not considered further.
Additionally, $c_{a,i}^2 = \dot{p}_i/\dot{\rho}_i$ represents the adiabatic speed of sound of the $i$-th fluid. For an imperfect fluid, $c^2_{s} = \delta p_i / \delta \rho_i$ stands for the sound speed of the $i$-th fluid. Hence, one can establish the following relation 
$ c^2_{a,i} =  w_i - \frac{w_i^{\prime}}{3\mathcal{H}(1+w_i)}. $
For a constant DE equation of state, $p_{\rm DE} = w_{\rm DE} \rho_{\rm DE}$, $c_s^2 = w_{\rm DE} < 0$. Consequently, the sound speed in this medium becomes imaginary, leading to an unphysical situation. In order to avoid this, we fix $c_s^2 =1$ similar to Ref.~\cite{Valiviita:2008iv}.

With the complete cosmological equations at both the background and perturbation levels, one can proceed to investigate the underlying cosmological scenario for any given DE equation of state. In the era of precise cosmology, it has become easier to constrain cosmological parameters with higher accuracy through observations. However, as suggested by~\cite{Dimakis:2016mip,Pan:2019brc}, one can consider the DE equation of state to have the following dynamical {\bf Exponential} form:
\begin{align}\label{Exponential}
    w_{\rm DE}  = w_{\rm 0, DE} - w_{\rm a, DE}+w_{\rm a, DE} \exp\left(\frac{z}{1+z} \right),
\end{align}
where $w_{\rm 0, DE}$ is the present value of the DE equation of state, i.e., $w_{\rm DE}(z=0)=w_{\rm 0, DE}$; $w_{\rm a, DE}$ is another free parameter; and $1+z = a_0/a$. 
This parametrization has two benefits: first, at its first order of approximation, it will return the well-known alternative dynamical DE model, Chevallier-Polarski-Linder ({\bf CPL}) parametrization~\cite{Chevallier:2000qy, Linder:2002et}:
\begin{align}\label{cpl}
    w_{\rm DE}  = w_{\rm 0, DE} + w_{\rm a, DE}\; \frac{z}{1+z}.
\end{align}
Second, it will enable us to see the effects of higher orders of correction on cosmological parameters. To this end, we expand (\ref{Exponential}) up to second and third orders:
\begin{eqnarray}
w_{\rm DE} (z) &=& w_{\rm 0, DE} + w_{\rm a, DE} \left[\frac{z}{1+z} + \frac{1}{2!} \left(\frac{z}{1+z} \right)^2\right], \label{cpl-1A}\\
w_{\rm DE} (z) &=& w_{\rm 0, DE} + w_{\rm a, DE} \Bigg[\frac{z}{1+z} + \frac{1}{2!} \left(\frac{z}{1+z} \right)^2 + \nonumber \\ && + \frac{1}{3!} \left(\frac{z}{1+z} \right)^3 \Bigg].\label{cpl-1B}
\end{eqnarray}
We label the equation of state (\ref{cpl-1A}) as {\bf CPL-A} and the equation of state (\ref{cpl-1B}) as {\bf CPL-B}. This approach enables the investigation of both the influence of diverse cosmological probes on parameters and the assessment of higher-order corrections to the CPL model. In summary, expressing the DE equation of state in exponential form facilitates the study of parameter constraints using different observational datasets and allows observation of the effects of higher-order corrections to the CPL model through the expansion of Equation (\ref{Exponential}).

\section{Observational Datasets and Statistical methodology}
\label{sec-dataset}

In this section, we describe the cosmological probes obtained from various observational surveys and the numerical methodology considered to constrain the proposed DE scenarios. Below, we outline the cosmological probes.

\begin{itemize}

    \item {\bf Cosmic microwave background (CMB) data}: One of the main components of this article involves comparing CMB data from multiple independent experiments, as detailed below.

    \begin{itemize}

\item {\bf Planck 2018 legacy release:} The baseline likelihoods include~\cite{Aghanim:2019ame,Aghanim:2018oex}: \texttt{\textbf{Commander}} likelihood, providing low multipoles TT data in the range of $2\leq \ell \leq 29$; \texttt{\textbf{SimAll}} likelihood, offering low multipoles EE data in the range of $2\leq \ell \leq 29$; \texttt{{\tt Plik}} TT,TE,EE likelihood, providing high multipoles TT, TE, and EE data in the range of $30\leq \ell \leq 2508$ for TT and $30\leq \ell \leq 1996$ for TE and EE; and \textbf{lensing} reconstruction likelihood, which can provide complementary information to the Planck CMB power spectra. 

\item {\bf Atacama Cosmology Telescope (ACT):} We consider the temperature and polarization anisotropy DR4 likelihood from CMB measurements made by the Atacama Cosmology Telescope from 2017 to 2021~\cite{ACT:2020gnv,ACT:2020frw}.

\item {\bf South Pole Telescope (SPT):} 
 We analyse the temperature and polarization (TT TE EE) likelihood from CMB measurements made from the ~South Pole Telescope\cite{SPT-3G:2022hvq}. 

\item  {\bf Wilkinson Microwave Anisotropy
Probe (WMAP):} For a complete analysis of Planck-independent CMB measurements~\cite{WMAP:2012fli}, we utilized the temperature and polarization WMAP nine-year data at low-$\ell$, excluding TE data due to the possibility of dust contamination~\cite{Planck:2013win}. Consequently, we set the minimum multipole in TE at $\ell=24$, enabling us to combine the WMAP data with ACT and SPT likelihoods, considering a Gaussian prior $\tau = 0.065\pm 0.015$, as conducted in the ACT DR4 analysis~\cite{ACT:2020gnv}.
\end{itemize}

We label the Planck dataset simply as \textbf{Planck}, while we refer the combination of ACT+$\tau$ prior+WMAP and SPT+$\tau$ prior+WMAP as {\bf ATW} and {\bf STW} respectively.

\item {\bf Baryon acoustic oscillations (BAO) data:} We have used both isotropic and anisotropic distance and expansion rate measurements of the SDSS-IV eBOSS survey~\cite{Alam:2020sor}.

\item {\bf Pantheon sample of Type Ia Supernovae (SNeIa):}  We have used the Pantheon catalog, consisting 1048 B-band observation of relative magnitudes of Type Ia Supernovae distributed in the redshift interval $z = [0.01, 2.3]$~\cite{Scolnic:2017caz}.

\end{itemize}

To constrain the underlying scenarios, we utilized a modified version of the publicly available cosmological code \texttt{CAMB}~\cite{Lewis:1999bs,Howlett:2012mh}. For exploring the posterior distributions of a given parameter space, we employed the publicly available Monte Carlo Markov Chain package \texttt{Cobaya}~\cite{Lewis:2002ah, Lewis:2013hha}, generating parameter spaces with the speed hierarchy implementation "fast dragging"~\cite{Neal:2005}, which includes a convergence diagnostic following the Gelman and Rubin statistic~\cite{Gelman:1992zz}. In Table~\ref{tab-priors}, we present the flat priors on the following free parameters: \textbf{i)} $\Omega_\mathrm{b} h^2$ representing the physical baryon density, \textbf{ii)} $\Omega_\mathrm{c} h^2$ representing the physical cold dark matter density, \textbf{iii)} $A_\mathrm{s}$ representing the amplitude of the primordial scalar perturbations, \textbf{iv)} $n_s$ representing the scalar spectral index, \textbf{v)} $\tau$ representing the optical depth, \textbf{vi)} $\theta_{MC}$ representing the ratio of the sound horizon to the angular diameter distance, \textbf{vii)} $w_{\rm 0, DE}$ representing the present value of $w_{\rm DE}$, and \textbf{viii)} $w_{\rm a, DE}$ describing the evolution with redshift of $w_{\rm DE}$. 

For model comparison, in addition to the relative best-fit ${\Delta\chi^2}$, we calculate the relative log-Bayesian evidence $\ln \mathcal{B}_{ij}$ using the \texttt{MCEvidence} package\footnote{\href{https://github.com/yabebalFantaye/MCEvidence}{github.com/yabebalFantaye/MCEvidence}}~\cite{Heavens:2017hkr,Heavens:2017afc}. A negative value indicates that $\Lambda$CDM is preferred over the tested model. To interpret the results, we refer to the revised Jeffreys' scale by Trotta~\cite{Kass:1995loi,Trotta:2008qt}: the evidence is inconclusive if $0 \leq | \ln \mathcal{B}_{ij}|  < 1$, weak if $1 \leq | \ln \mathcal{B}_{ij}|  < 2.5$, moderate if $2.5 \leq | \ln \mathcal{B}_{ij}|  < 5$, strong if $5 \leq | \ln \mathcal{B}_{ij}|  < 10$, and very strong if $| \ln \mathcal{B}_{ij} | \geq 10$.

\begin{table}
\begin{tabular} { l  c}
\hline 
 Parameter &  Prior\\
\hline

$\Omega_\mathrm{b} h^2$  & $[0.005 , 0.1]  $ \\

$\Omega_\mathrm{c} h^2$  & $[0.01 , 0.99]  $ \\

$\log(10^{10} A_\mathrm{s})$ & $[1.61 , 3.91]$\\

$n_\mathrm{s}   $ & $[0.8 , 1.2]  $ \\

$\tau$            &  $[0.01,0.8]$         \\

$100\theta_\mathrm{MC}$  & $[0.5 , 10]  $ \\

$w_{\rm 0, DE}$  & $[-3 , 1]  $ \\

$w_{\rm a, DE}$  & $[-3 , 2]  $ \\
\hline
\end{tabular}
\caption{Flat priors on the free parameters considered in the analyses. For alternative CMB likelihoods (ACT and SPT), Gaussian prior $\tau = 0.065\pm0.015$ is considered. }
\label{tab-priors}
\end{table}

\section{Results}
\label{sec-results}

\begin{figure} 
\begin{center}
\includegraphics[width=0.44\textwidth]{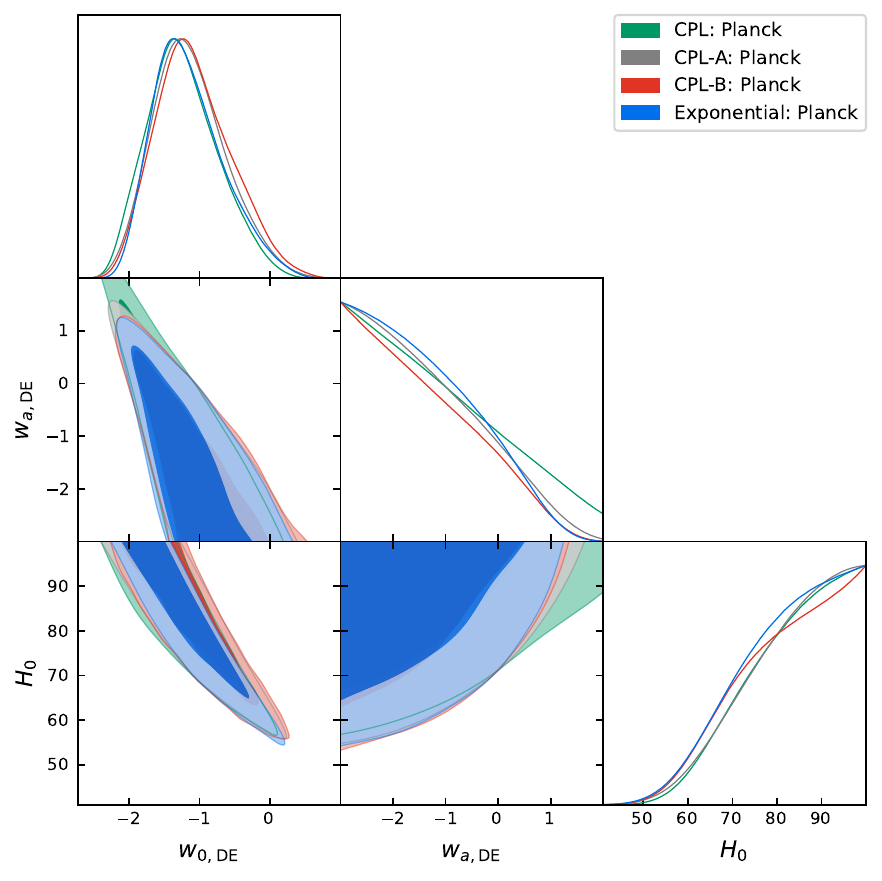}
\includegraphics[width=0.44\textwidth]{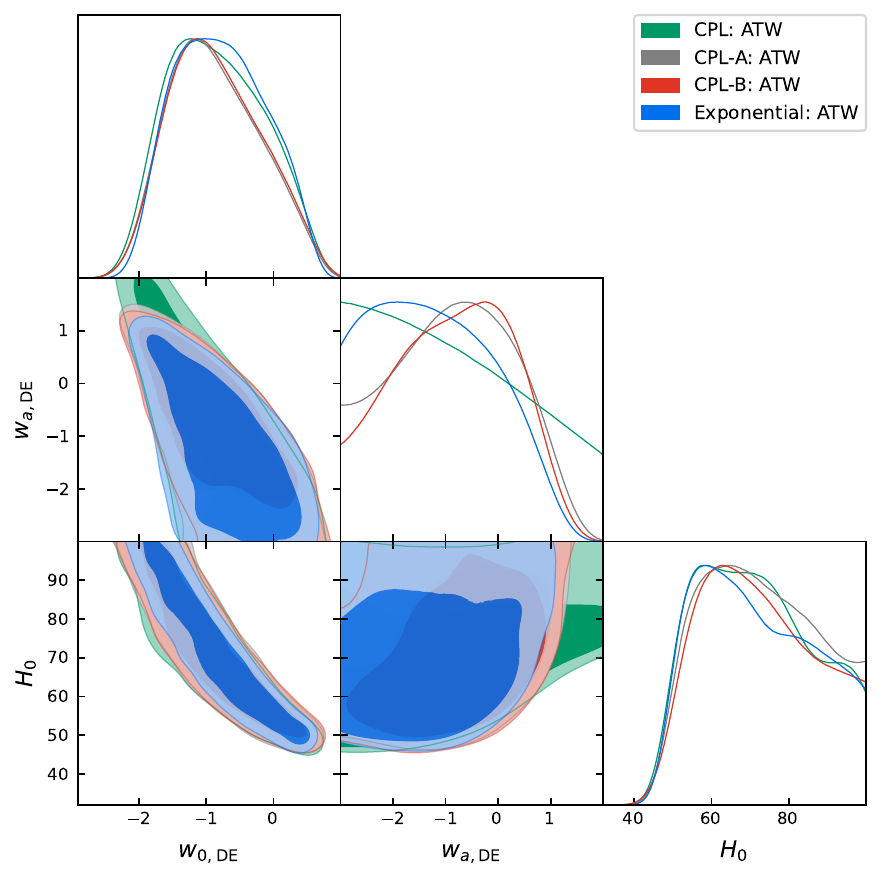}
\includegraphics[width=0.44\textwidth]{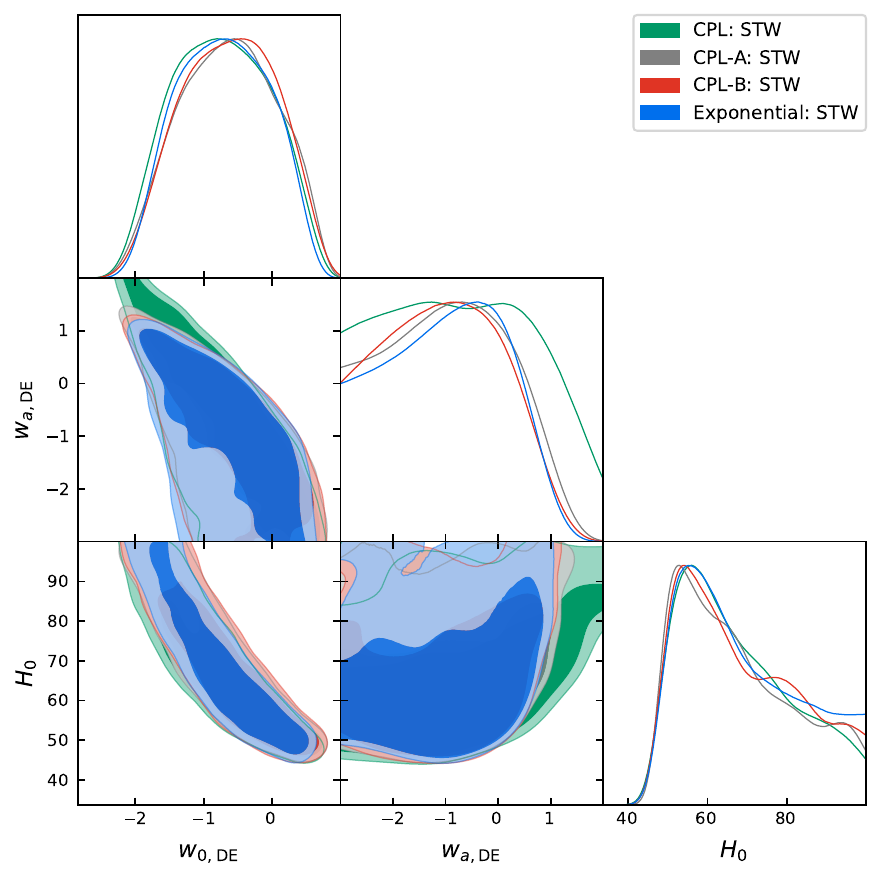}
\end{center}
\caption{In this figure we show how different parametrizations respond to  CMB alone datasets, i.e. Planck alone, ATW, STW,  in terms of the parameters $w_{\rm 0, DE}, w_{\rm a, DE}$ and $H_0$.  This figure also shows a comparison between these parametrizations.  }
\label{Fig:CMB}
\end{figure}

\begin{figure*} 
\begin{center}
\includegraphics[width=0.47\textwidth]{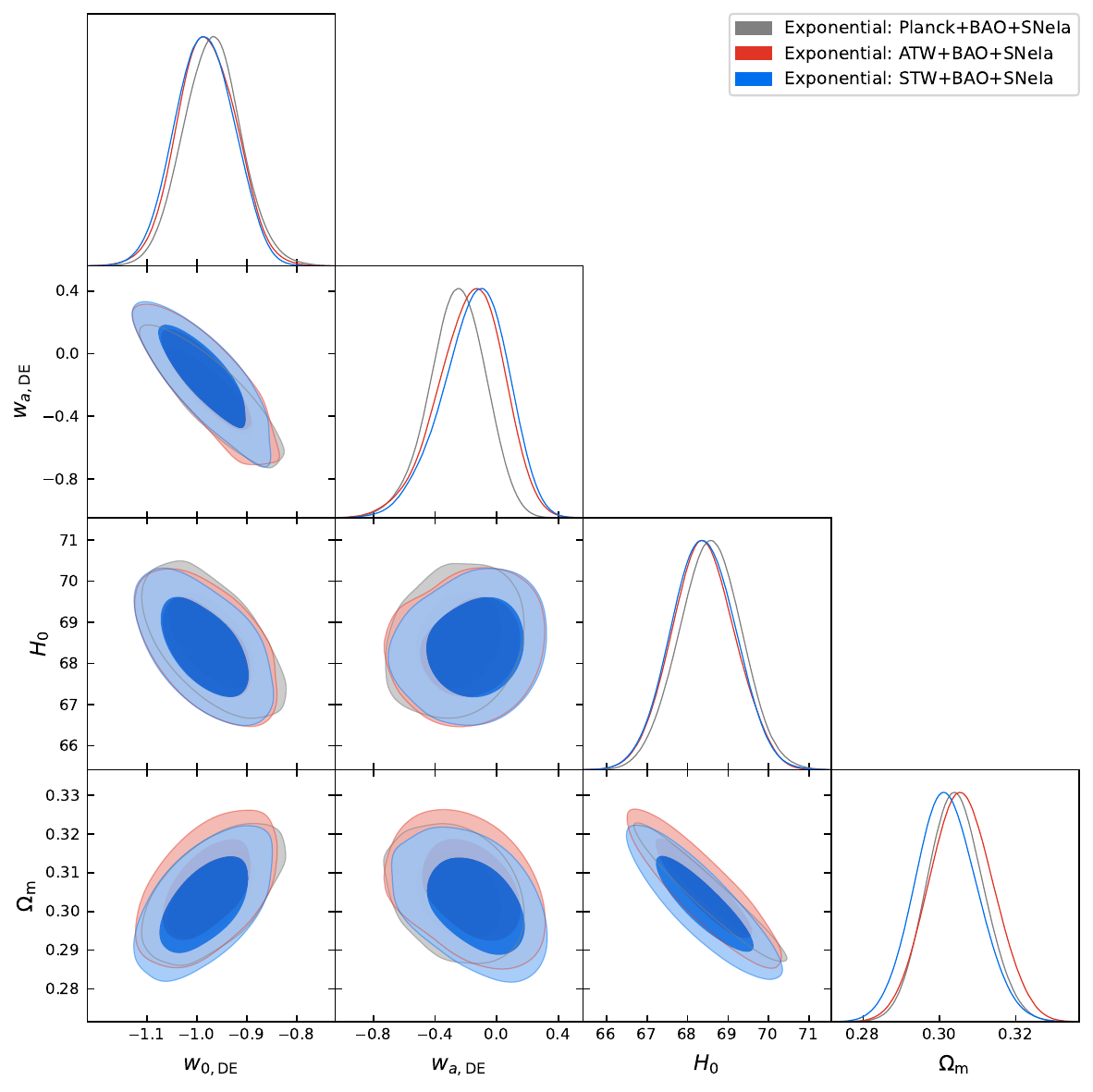}
\includegraphics[width=0.47\textwidth]{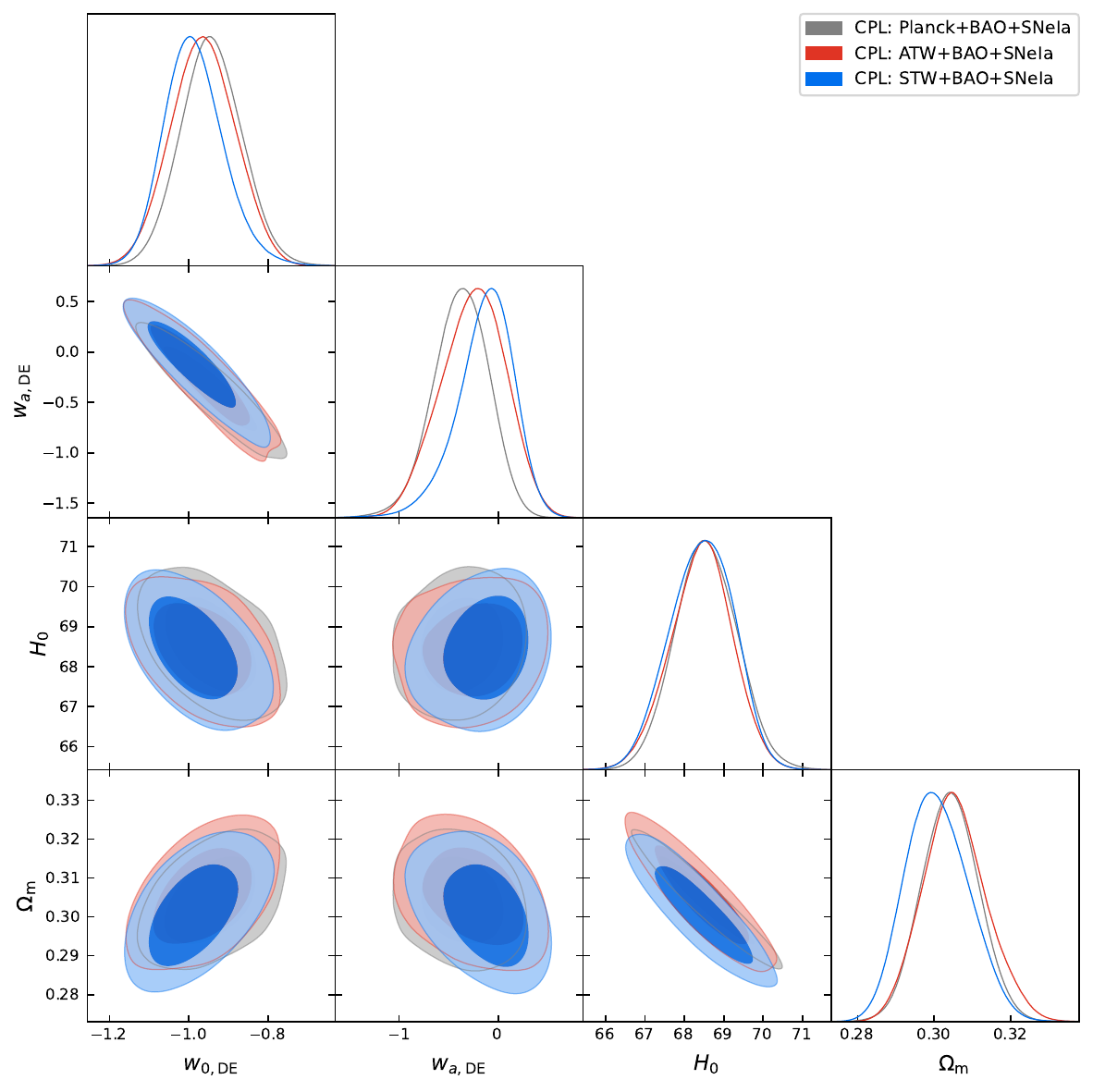}
\includegraphics[width=0.47\textwidth]{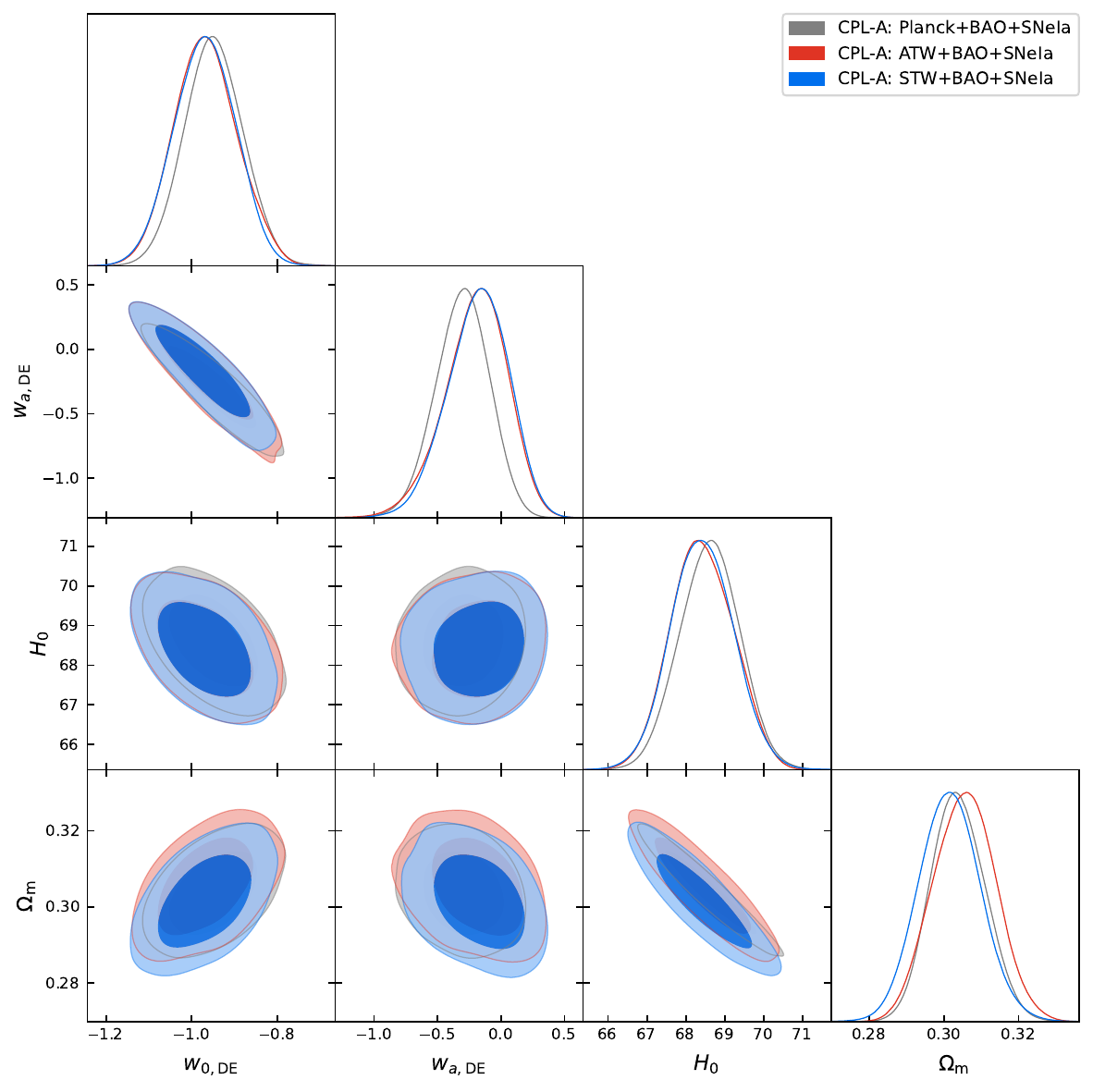}
\includegraphics[width=0.47\textwidth]{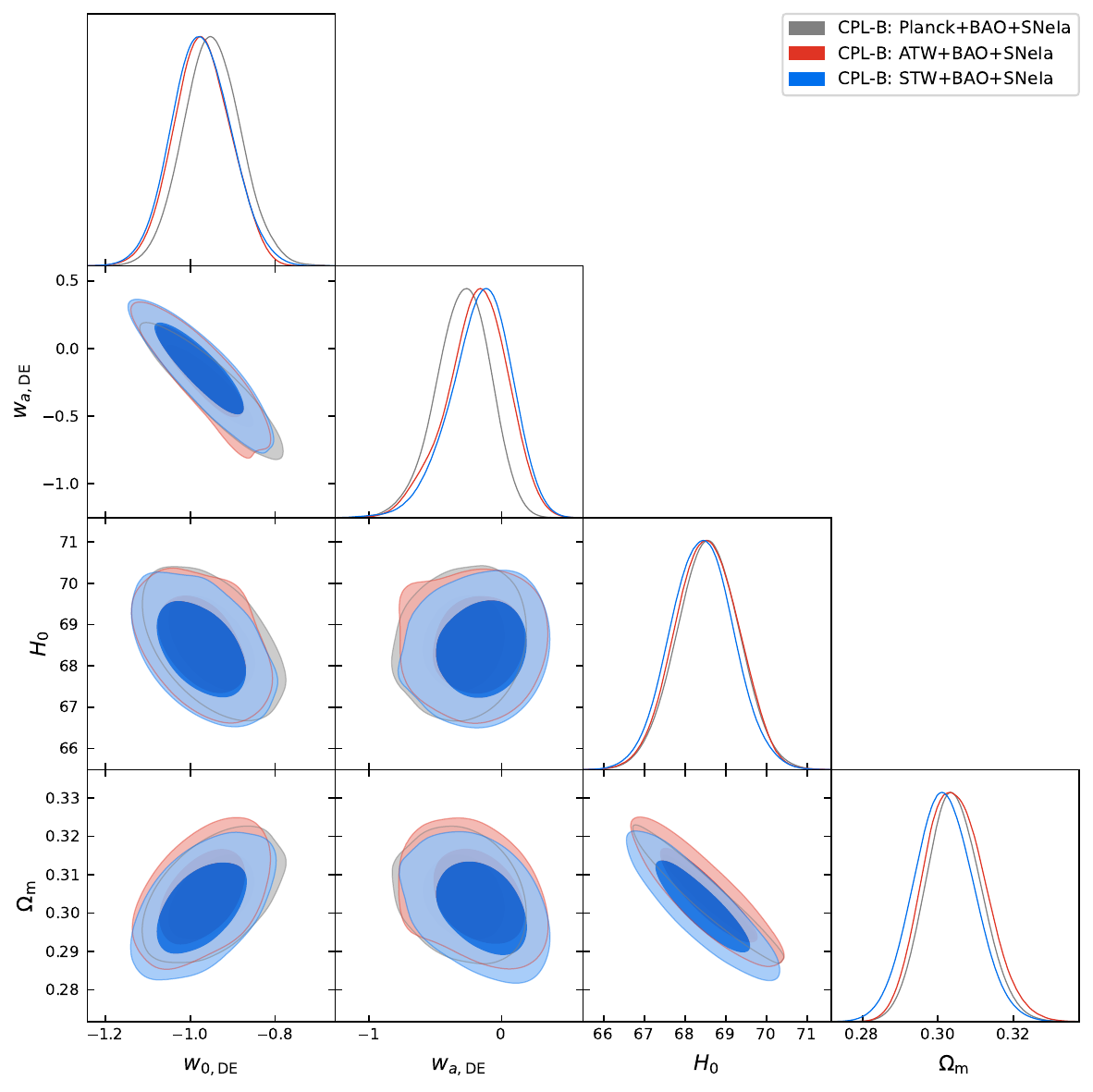}
\end{center}
\caption{The plots display the behaviour of the parameters $w_{\rm 0, DE}, w_{\rm a, DE}$, $H_0$ and $\Omega_{\rm m}$ within the proposed DE parametrizations for Planck+BAO+SNeIa, ATW+BAO+SNeIa and STW+BAO+SNeIa. }
\label{Fig:CMB+BAO+SN}
\end{figure*}

\begin{figure*} 
\begin{center}
\includegraphics[width=\textwidth]{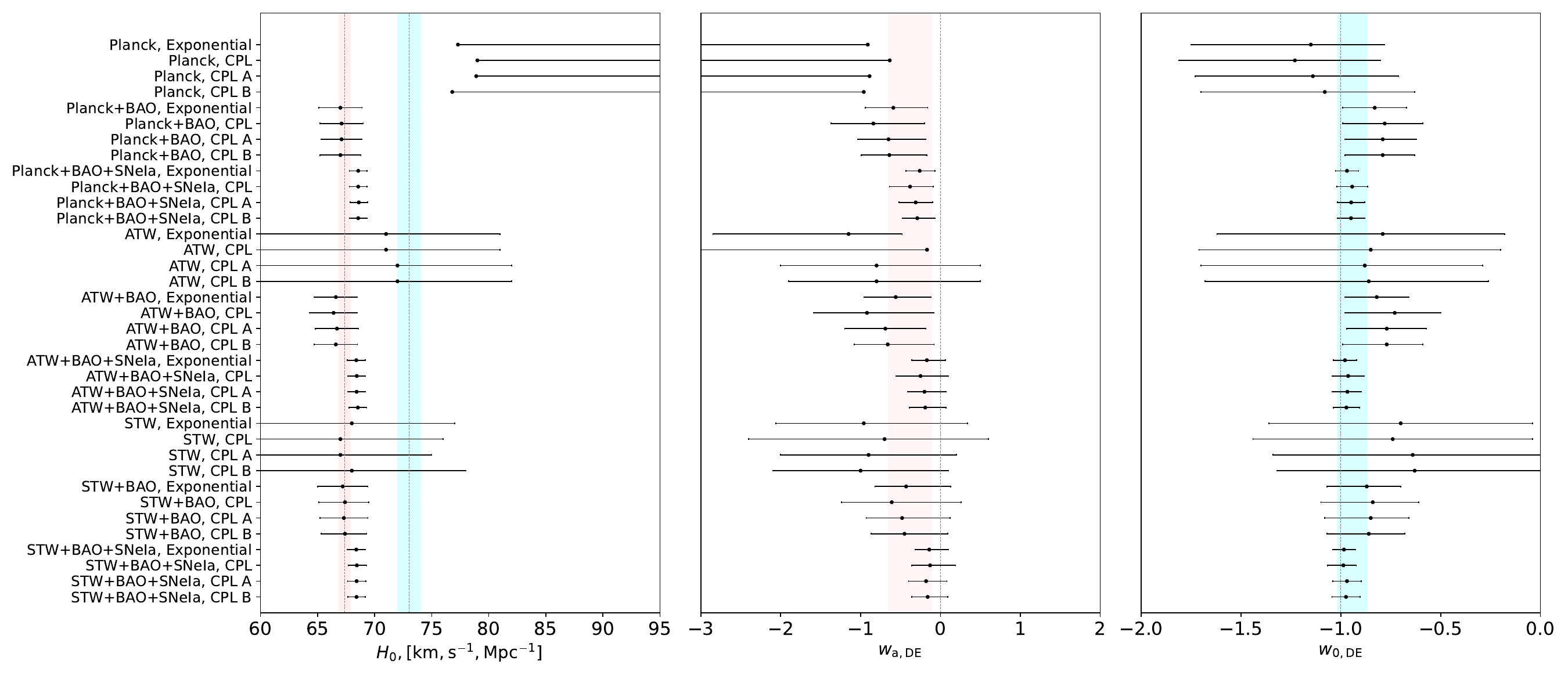}
\end{center}
\caption{Whisker plot showing the 68\% CL constraints or upper limits on $H_0$ (left panel), $w_{\rm a, DE}$ (middle panel), and $w_{\rm 0, DE}$ (right panel) from Exponential, CPL, CPL-A, and CPL-B parametrizations considering various observational datasets. In the left panel, the cyan vertical band corresponds to the $H_0$ value from the SH0ES collaboration ($H_0 = 73.04 \pm 1.04$ km$s^{-1}$Mpc$^{-1}$ at 68\% CL)~\cite{Riess:2021jrx}, and the reddish vertical band corresponds to the $H_0$ value (within the $\Lambda$CDM paradigm) from the Planck collaboration ($H_0 = 67.36 \pm 0.54 $ km$s^{-1}$Mpc$^{-1}$ at 68\% CL)~\cite{Planck:2018vyg}. For the middle panel, the dotted vertical line corresponds to $w_{\rm a, DE} = 0$, and the reddish vertical band corresponds to the 68\% CL constraints for Planck+BAO+SNeIa dataset combination assuming CPL. Finally, for the right panel, the dotted vertical line corresponds to $w_{\rm 0, DE} = -1$, and the cyan vertical band corresponds to the 68\% CL constraints for Planck+BAO+SNeIa dataset combination assuming CPL. }
\label{fig:whisker}
\end{figure*}

\begin{figure} 
\includegraphics[width=0.44\textwidth]{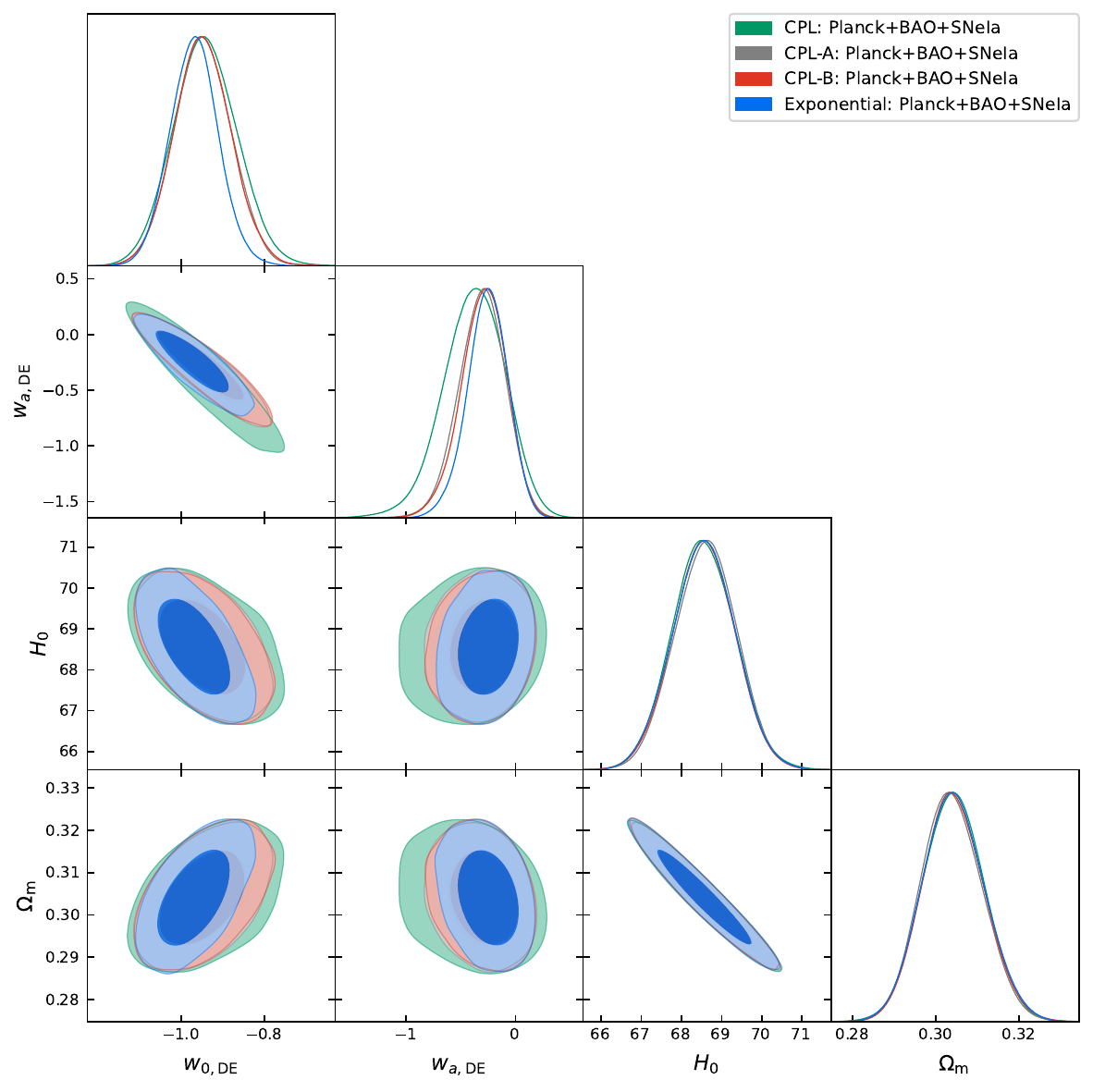}
\includegraphics[width=0.44\textwidth]{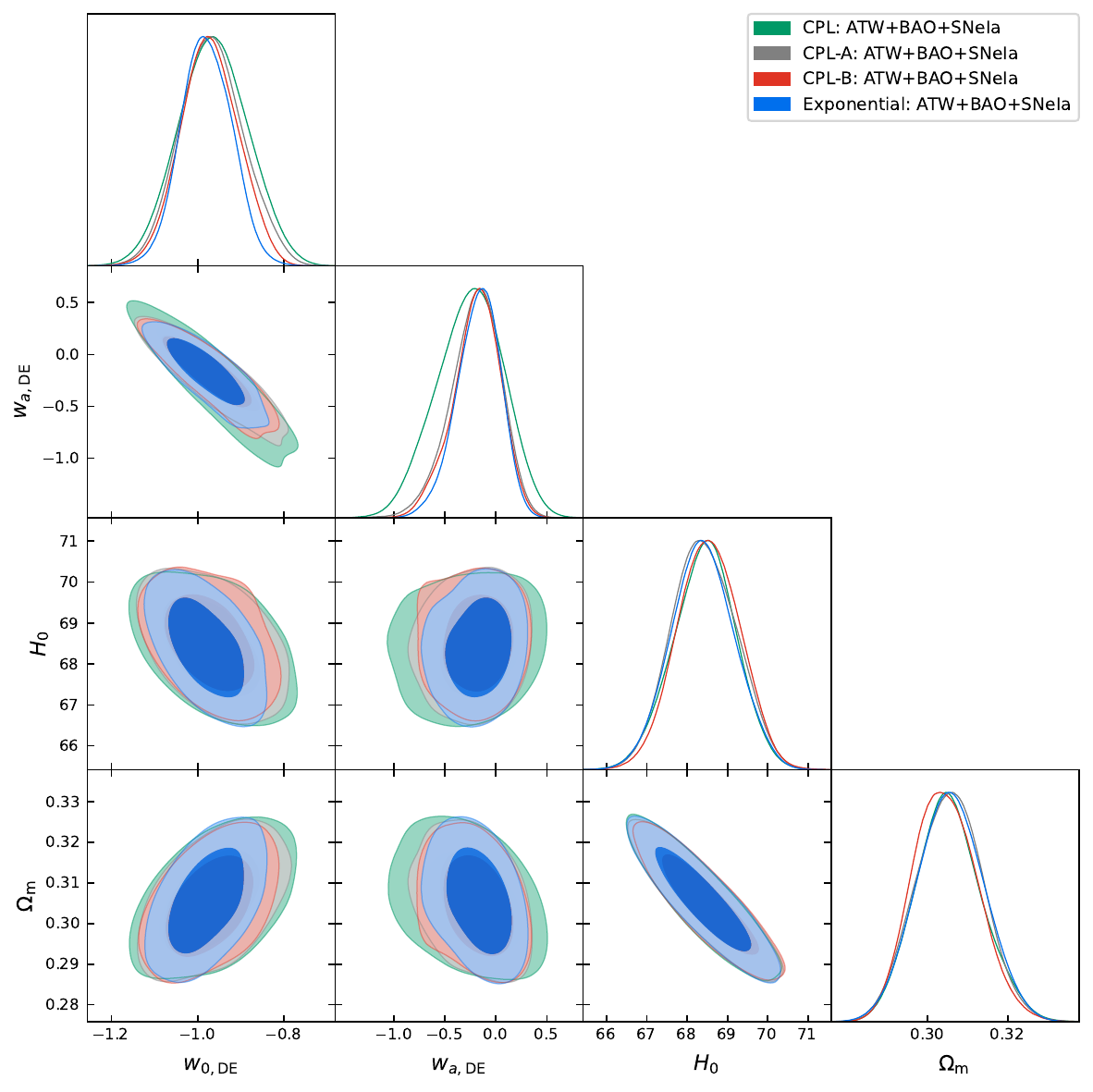}
\includegraphics[width=0.44\textwidth]{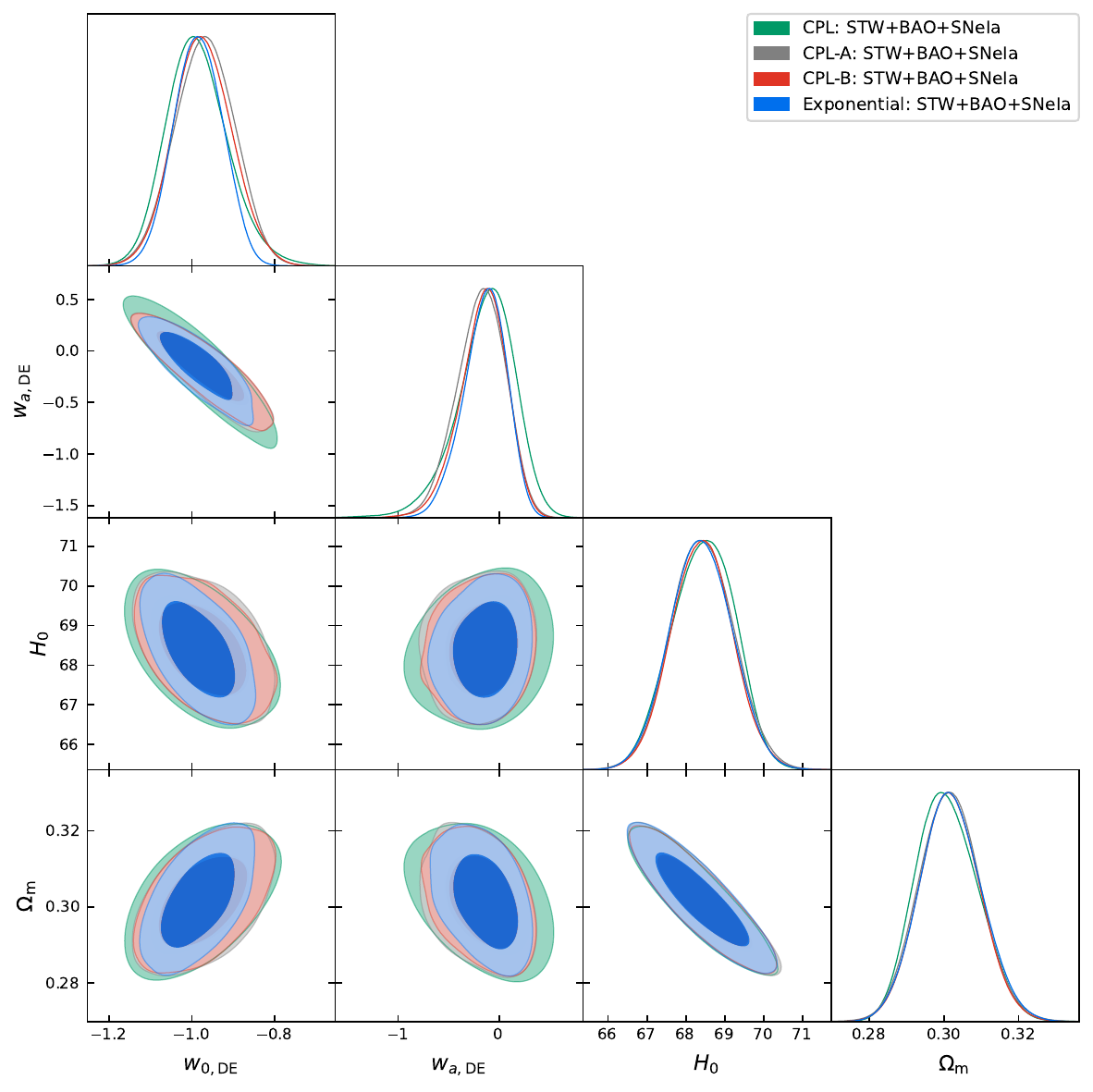}
\caption{The plots offer a comparison between the proposed DE parametrizations considering  the combined datasets, namely, Planck+BAO+SNeIa (upper plot), ATW+BAO+SNeIa (middle plot) and STW+BAO+SNeIa (lower plot). }
\label{Fig:CMB+BAO+SN2}
\end{figure}

In this section, we summarize the constraints on all the dark energy (DE) parametrizations proposed in this work. To compare the key parameters of all the parametrizations, we have utilized some common datasets, namely: i) Planck, ATW, STW; ii) Planck+BAO, ATW+BAO, STW+BAO; iii) Planck+BAO+SNeIa, ATW+BAO+SNeIa, STW+BAO+SNeIa. As the article is focused on the Exponential model and its various extensions, including the well-known CPL parametrization, in light of multiple CMB probes and their combinations with other external cosmological probes, our primary focus is to first understand how different CMB probes affect the DE parameters $w_{\rm 0, DE}$ and $w_{\rm a, DE}$, and then to understand how these parameters affect other cosmological parameters, e.g., $H_0$, $S_8$, $\Omega_\mathrm{m}$.
Tables~\ref{table-exp} (for the Exponential model),~\ref{table-CPL} (for CPL),~\ref{table-CPLA} (for CPL-A), and~\ref{table-CPLB} (for CPL-B) display the constraints on the free and derived parameters of these DE parametrizations. Regarding the graphical presentations, we have summarized them into four combined figures:

\begin{enumerate}
    \item Fig.~\ref{Fig:CMB} indicates the behavior of the dark energy parameters ($w_{\rm 0, DE}$, $w_{\rm a, DE}$) and $H_0$ for all the parametrizations considering only the CMB datasets (i.e., Planck, ATW, STW). This figure also provides a comparison among all these parametrizations for the CMB alone dataset.
    
    \item Fig.~\ref{Fig:CMB+BAO+SN} shows the behavior of $w_{\rm 0, DE}$, $w_{\rm a, DE}$, $H_0$, $\Omega_{\rm m}$ for each parametrization considering only the combined dataset CMB (Planck/ATW/STW)+BAO+SNIa.
    
    \item Fig.~\ref{fig:whisker} presents the whisker graph at 68\% CL of $w_{\rm 0, DE}$, $w_{\rm a, DE}$, and $H_0$, offering a quick summary of all the parametrizations in terms of these parameters.
    
    \item Fig.~\ref{Fig:CMB+BAO+SN2} compares all the parametrizations considering three combined datasets: Planck+BAO+SNIa, ATW+BAO+SNIa, STW+BAO+SNIa.
\end{enumerate}

In what follows, we describe the results extracted from each parametrization.

\begin{table*}[!t]
\begin{center}
\renewcommand{\arraystretch}{1.4}
\caption{68\% CL constraints, upper, and lower limits on the free and derived parameters of the cosmological scenario in which DE assumes the Exponential parametrization using all the dataset combinations. Note that $\Delta\chi^2 = $ $\chi^2$ (Model) $-$ $\chi^2$ ($\Lambda$CDM), and hence, $\Delta\chi^2 <0$ means that the model is preferred over the $\Lambda$CDM. 
} 

\resizebox{\textwidth}{!}{                                              
\begin{tabular}{cccccccccccc}   
\hline 
 Parameter &  Planck & ATW & STW &  Planck+BAO & ATW+BAO & STW+BAO &  Planck+BAO+SNeIa & ATW+BAO+SNeIa & STW+BAO+SNeIa \\
\hline
$\Omega_\mathrm{c} h^2$ & $0.1191\pm 0.0012$ & $0.1200\pm 0.0027$ &$0.1170\pm 0.0028$& $0.1201\pm 0.0011$ & $0.1206\pm 0.0019$ &$0.1187\pm 0.0021$& $0.1199\pm 0.0010$ & $0.1199\pm 0.0018$ &$0.1181\pm 0.0019$\\

$\Omega_\mathrm{b} h^2$ & $0.02244\pm 0.00015$ & $0.02237\pm 0.00020$ & $0.02238\pm 0.00021$& $0.02238\pm 0.00014$ & $0.02234\pm 0.00018$ & $0.02235\pm 0.00020 $& $0.02239\pm 0.00014$ & $0.02238\pm 0.00019$ & $0.02235\pm 0.00020 $\\

$100\theta_\mathrm{MC}$ & $1.04100\pm 0.00031$ &  $1.04169\pm 0.00065$ & $1.04010\pm 0.00069$& $1.04090\pm 0.00029$ &  $1.04156\pm 0.00062$ & $1.04001\pm 0.00065$& $1.04091\pm 0.00030$ &  $1.04165^{+0.00058}_{-0.00065}$ & $1.04004\pm 0.00066$\\

$\tau$ & $0.0518\pm 0.0076$ & $0.062\pm 0.014$ & $0.058\pm 0.014$ & $0.0530\pm 0.0074$ & $0.063\pm 0.013$ & $0.059\pm 0.014$& $0.0537\pm 0.0074$ & $0.062\pm 0.014$ & $0.059\pm 0.014$\\

$n_\mathrm{s}   $ & $0.9672\pm 0.0041$ & $0.9729^{+0.0058}_{-0.0066}$ & $0.9678\pm 0.0068$& $0.9649^{+0.0041}_{-0.0037}$ & $0.9717\pm 0.0052$ & $0.9651\pm 0.0059$& $0.9651\pm 0.0039$ & $0.9729\pm 0.0052$ & $0.9660\pm 0.0058$\\

$\log(10^{10} A_\mathrm{s})$ & $3.037\pm 0.015$ & $3.066\pm 0.027$ & $3.043\pm 0.028$& $3.042\pm 0.014$ & $3.071^{+0.027}_{-0.024}$ & $3.050\pm 0.027$& $3.043\pm 0.014$ & $3.067\pm 0.027$ & $3.049\pm 0.027$\\

$w_{\rm 0, DE}$ & $-1.15^{+0.37}_{-0.60}$ & $-0.79^{+0.61}_{-0.83}$ & $-0.70\pm 0.66$& $-0.83\pm 0.16$ & $-0.82\pm 0.16$ & $-0.87^{+0.17}_{-0.20}$& $-0.969\pm 0.058$ & $-0.979\pm 0.058$ & $-0.984\pm 0.058$\\

$w_{\rm a, DE}$ & $< -0.910$ & $-1.15^{+0.67}_{-1.7}$ & $-0.96^{+1.3}_{-1.1}$& $-0.59^{+0.43}_{-0.35}$ & $-0.56^{+0.44}_{-0.40}$ & $-0.43^{+0.56}_{-0.39}$& $-0.26^{+0.19}_{-0.17}$ & $-0.17^{+0.23}_{-0.19}$ & $-0.14^{+0.24}_{-0.18}$\\

$\Omega_\mathrm{m}         $ & $0.223^{+0.018}_{-0.082}$ & $0.321^{+0.067}_{-0.17}$ & $0.343^{+0.084}_{-0.20}$& $0.320^{+0.017}_{-0.019}$ & $0.325^{+0.019}_{-0.021}$ & $0.315^{+0.020}_{-0.025}$& $0.3043\pm 0.0074$ & $0.3057\pm 0.0083$ & $0.3018\pm 0.0082$\\

$\sigma_8                  $ & $0.935^{+0.12}_{-0.052}$ & $0.85^{+0.13}_{-0.16}$ & $0.796^{+0.096}_{-0.17}$& $0.812\pm 0.016$ & $0.825\pm 0.017$ & $0.809\pm 0.018$& $0.8239\pm 0.0096$ & $0.832\pm 0.016$ & $0.814\pm 0.017$\\

$S_8                       $ & $0.785^{+0.026}_{-0.050}$ & $0.837\pm 0.053$ & $0.806^{+0.059}_{-0.052}$& $0.838\pm 0.014$ & $0.858\pm 0.026$ & $0.828\pm 0.029$& $0.830\pm 0.011$ & $0.840\pm 0.021$ & $0.817\pm 0.021$\\

$H_0$    & $> 77.3$ & $71^{+10}_{-20}$ & $68^{+9}_{-20}$& $67.0\pm 1.9$ & $66.6\pm 1.9$ & $67.2\pm 2.2$& $68.56\pm 0.78$ & $68.39\pm 0.78$ & $68.39\pm 0.78$\\

$r_\mathrm{drag}$ & $147.26\pm 0.27$ & $147.10\pm 0.63$ & $147.88\pm 0.70$& $147.08\pm 0.25$ & $146.97\pm 0.46$ & $147.48\pm 0.55$& $147.10\pm 0.24$ & $147.12\pm 0.47$ & $147.63\pm 0.52$\\

\hline
$\chi^2                       $ & $2768.3$ & $5837.2$ & $7417.7$& $2793.7$ & $5857.4$ & $7435.5$& $3828.5$ & $6892.8$ & $8472.7$\\
$\Delta\chi^2                      $ & $-4.4$ & $0.05$ & $1.2$& $-3.0$ & $-1.0$ & $-3.0$& $-3.5$ & $-0.3$ & $-0.9$\\
\hline
\end{tabular}
 }      
\label{table-exp} 
\end{center}
\end{table*}

\subsection{Exponential Parametrization}

This parametrization is the generalized version of all the parametrizations explored in this work, and hence, it is of special importance in this article. We also note that for the first time constraints on this parametrization using multiple CMB probes are given in this work.
Table~\ref{table-exp} summarizes the constraints on this parametric DE model considering various CMB datasets and their combinations with other astronomical probes, described above. Figures~\ref{Fig:CMB},~\ref{Fig:CMB+BAO+SN},~\ref{fig:whisker}, and~\ref{Fig:CMB+BAO+SN2} summarize all the findings graphically.

If we focus solely on the CMB datasets, we observe a moderate shift among them, around $1-2\sigma$, on most of the parameters reported in Table~\ref{table-exp}, such as, $100\theta_\mathrm{MC}$, $\log(10^{10} A_\mathrm{s})$, or $S_8$. Due to the correlations among these parameters, Planck alone yields a very low value for the matter density compared to the ground-based telescopes explored here, resulting in a lower limit for $H_0$. This contrasts with the findings from ATW and STW, where a higher value for $\Omega_\mathrm{m}$ is observed, compensated by a lower value for $H_0$, which is now constrained.

Regarding the DE parameters, from Planck alone, a phantom behavior of $w_{\rm 0, DE}$ ($ = -1.15^{+0.37}_{-0.60}$ at 68\% CL) for the present epoch is found, which is not supported by any of the ground-based experiments. The mean value from ground-based experiments lies in the quintessence regime, although all experiments are consistent with a cosmological constant (i.e., $w_{\rm 0, DE} =-1$). As for $w_{\rm a, DE}$, only an upper limit is obtained from Planck alone, but this becomes constrained at 68\% CL for ATW and STW. For ATW, we get evidence of $w_{\rm a, DE} \neq 0$ at more than 68\% CL, meaning that evidence of a dynamical DE exists; however, for STW, $w_{\rm a, DE} = 0$ is allowed within 68\% CL. 
Figure~\ref{Fig:CMB} gives a comprehensive understanding of the behavior of these DE parameters and their dependence on $H_0$ for different CMB datasets explored here.

When the BAO data are included, while we can still see a shift for the same parameters that move in the CMB only case, we find that the DE parameters, as well as the matter density $\Omega_\mathrm{m}$ and the Hubble constant $H_0$, are very well constrained and in complete agreement between Planck+BAO, ATW+BAO, and STW+BAO. Focusing on the DE parameters, we notice that the mean values of $w_{\rm 0, DE}$ are quintessential for all the cases; however, specifically for Planck+BAO and ATW+BAO, $w_{\rm 0, DE}$ remains strictly in the quintessence regime at 68\% CL, while STW+BAO allows $w_{\rm 0, DE} < -1$ at 68\% CL. 
Looking at the constraints on $w_{\rm a, DE}$ for all three cases, we see that $w_{\rm a, DE} \neq 0$ (in fact, $w_{\rm a, DE} < 0$) at more than 68\% CL for both Planck+BAO and ATW+BAO. However, for STW+BAO, $w_{\rm a, DE} = 0$ is allowed within 68\% CL. Hence, a dynamical phantom nature of $w_{\rm DE}$ at slightly more than $1\sigma$ is indicated in the past for Planck+BAO and ATW+BAO.
Finally, when SNeIa data are added, all the parameters shift back in agreement with the $\Lambda$CDM case (except for the case with Planck+BAO+SNeIa which recommends a mild non-zero value of $w_{\rm a, DE}$ ($w_{\rm a, DE} = -0.26^{+0.19}_{-0.17}$ at 68\% CL for Planck+BAO+SNeIa) and hence a dynamical DE is endorsed), and they are very much consistent among the different CMB datasets, as we can see in Fig.~\ref{Fig:CMB+BAO+SN}.

In order to have a smooth and quick understanding of the nature of the DE parameters $w_{\rm 0, DE}$, $w_{\rm a, DE}$, as well as the Hubble constant $H_0$ within this parametrized framework, in Fig.~\ref{fig:whisker} we show the whisker plot at 68\% CL on these parameters. As seen from Fig.~\ref{fig:whisker}, when BAO and BAO+SNeIa datasets are added to CMB (Planck/ATW/STW), the resulting scenario becomes close and consistent with the $\Lambda$CDM model.

As expected, since this model has two additional degrees of freedom, it is statistically disfavored compared to the $\Lambda$CDM scenario, as shown in Table~\ref{tab:bayesian1}, if we compute the Bayesian evidence. However, we have a modest improvement in the $\chi^2$ as shown in Table~\ref{table-exp}.

\begin{table*}[!t]
\begin{center}
\renewcommand{\arraystretch}{1.4}
\caption{Like Table~\ref{table-exp} but for the CPL parametrization.}
\resizebox{\textwidth}{!}{                                              
\begin{tabular}{cccccccccccc}   
\hline 
 Parameter &  Planck & ATW & STW &  Planck+BAO & ATW+BAO & STW+BAO &  Planck+BAO+SNeIa & ATW+BAO+SNeIa & STW+BAO+SNeIa \\
\hline
$\Omega_\mathrm{c} h^2$ & $0.1192\pm 0.0012$ & $0.1201\pm 0.0027$ &$0.1173\pm 0.0029$ & $0.1200\pm 0.0010$ & $0.1208\pm 0.0020$ &$0.1186\pm 0.0021$ & $0.1199\pm 0.0010$ &  $0.1199^{+0.0020}_{-0.0017}$ & $0.1179\pm 0.0019$ \\

$\Omega_\mathrm{b} h^2$ & $0.02244\pm 0.00015$ & $0.02236\pm 0.00021$ & $0.02236\pm 0.00021$ & $0.02237\pm 0.00014$ & $0.02235\pm 0.00018$ & $0.02234\pm 0.00020$ &  $0.02238\pm 0.00014$ &  $0.02237\pm 0.00019$ & $0.02238\pm 0.00020$ \\

$100\theta_\mathrm{MC}$ & $1.04099\pm 0.00031$ &  $1.04168\pm 0.00065$ & $1.04010\pm 0.00070$ & $1.04090\pm 0.00030$ &  $1.04163\pm 0.00061$ & $1.03998\pm 0.00066$ & $1.04090\pm 0.00030$ & $1.04167\pm 0.00064$ & $1.04015\pm 0.00068$  \\

$\tau$ & $0.0523\pm 0.0073$ & $0.061\pm 0.014$ & $0.058\pm 0.014$ & $0.0530\pm 0.0076$ & $0.064\pm 0.014$ & $0.059\pm 0.014$ & $0.0535\pm 0.0074$ & $0.063\pm 0.014$ & $0.060\pm 0.013$ \\

$n_\mathrm{s}   $ & $0.9668\pm 0.0042$ & $0.9728\pm 0.0063$ & $0.9674\pm 0.0070$ & $0.9650\pm 0.0038$ & $0.9713^{+0.0047}_{-0.0054}$ & $0.9651\pm 0.0062$ & $0.9653\pm 0.0038$ & $0.9731\pm 0.0051$ & $0.9664\pm 0.0058$\\

$\log(10^{10} A_\mathrm{s})$ & $3.038\pm 0.014$ & $3.065\pm 0.027$ & $3.045\pm 0.028$ & $3.041\pm 0.015$ & $3.072\pm 0.026$ & $3.050\pm 0.027$ & $3.042\pm 0.015$ & $3.069\pm 0.026$ & $3.051\pm 0.026$\\

$w_{\rm 0, DE}$ & $-1.23^{+0.43}_{-0.58}$ & $-0.85^{+0.65}_{-0.86}$ & $-0.74\pm 0.70$ & $-0.78^{+0.19}_{-0.21}$ & $-0.73^{+0.23}_{-0.25}$ & $-0.84^{+0.23}_{-0.26}$ & $-0.943\pm 0.077$ & $-0.963\pm 0.080$ & $-0.987^{+0.065}_{-0.080}$\\

$w_{\rm a, DE}$ & $< -0.635$ & $< -0.168$ & $-0.7^{+1.3}_{-1.7}$ & $-0.84^{+0.64}_{-0.53}$& $-0.92^{+0.84}_{-0.67}$ & $-0.61^{+0.87}_{-0.63}$ & $-0.38^{+0.29}_{-0.26}$ & $-0.25^{+0.35}_{-0.31}$ & $-0.13^{+0.32}_{-0.23}$\\

$\Omega_\mathrm{m}         $ & $0.215^{+0.016}_{-0.073}$ & $0.322^{+0.062}_{-0.18}$ & $0.351^{+0.081}_{-0.20}$ & $0.318^{+0.018}_{-0.020}$ & $0.327\pm 0.023$ & $0.313\pm0.023$ & $0.3042\pm 0.0073$ &$0.3054^{+0.0076}_{-0.0088}$ & $0.3009^{+0.0076}_{-0.0090}$\\

$\sigma_8                  $ & $0.943^{+0.11}_{-0.047}$ & $0.85^{+0.12}_{-0.16}$ & $0.785^{+0.091}_{-0.16}$ & $0.813\pm 0.017$ & $0.823\pm 0.018$ & $0.809\pm 0.018$ & $0.823\pm 0.010$ & $0.834\pm 0.016$ & $0.813\pm 0.017$\\

$S_8                       $ & $0.780^{+0.023}_{-0.046}$ & $0.833\pm 0.051$ & $0.807^{+0.056}_{-0.049}$ & $0.836\pm 0.014$ & $0.859\pm 0.027$ & $0.826\pm 0.028$  & $0.829\pm 0.011$ &  $0.841^{+0.021}_{-0.019}$ & $0.814^{+0.020}_{-0.023}$\\

$H_0$    & $> 79.0$ & $71^{+10}_{-20}$ & $67^{+9}_{-20}$ & $67.1\pm 1.9$ & $66.4\pm 2.1$ & $67.4^{+2.1}_{-2.3}$  & $68.56\pm 0.78$ &  $68.43\pm 0.78$ & $68.44^{+0.86}_{-0.76}$\\

$r_\mathrm{drag}$ & $147.23\pm 0.27$ & $147.08\pm 0.66$ & $147.83\pm 0.73$ & $147.09\pm 0.24$ & $146.92\pm 0.51$ & $147.52\pm 0.55$ & $147.12\pm 0.24$ & $147.13^{+0.42}_{-0.48}$ & $147.67\pm 0.49$\\

\hline
$\chi^2                       $ & $2768.9$ & $5837.4$ & $7417.8$ & $2793.8$ & $5857.5$ & $7438.6$ & $3828.2$ & $6892.6$ & $8473.6$\\
$\Delta\chi^2                      $ & $-3.8$ & $0.2$ & $1.3$ & $-2.9$ & $-0.9$ & $0.1$ & $-3.7$ & $-0.5$ & $0.041$\\
\hline
\end{tabular}
 }      
\label{table-CPL} 
\end{center}
\end{table*}


\begin{table*}[!t]
\begin{center}
\renewcommand{\arraystretch}{1.4}
\caption{Like Table~\ref{table-exp} but for the CPL-A parametrization.} 
\resizebox{\textwidth}{!}{                                              
\begin{tabular}{cccccccccccc}   
\hline 
 Parameter &  Planck & ATW & STW &  Planck+BAO & ATW+BAO & STW+BAO &  Planck+BAO+SNeIa & ATW+BAO+SNeIa & STW+BAO+SNeIa \\
\hline
$\Omega_\mathrm{c} h^2$ & $0.1192\pm 0.0012$ & $0.1201\pm 0.0027$ &$0.1172\pm 0.0028$ & $0.1200\pm 0.0011$ & $0.1207\pm 0.0020$ &$0.1187^{+0.0021}_{-0.0020}$ & $0.1199\pm 0.0010$ &  $0.1199\pm 0.0018$ & $0.1181\pm 0.0019$ \\

$\Omega_\mathrm{b} h^2$ & $0.02244\pm 0.00015$ & $0.02238\pm 0.00020$ & $0.02237\pm 0.00021$ & $0.02238\pm 0.00014$ & $0.02235\pm 0.00019$ & $0.02234\pm 0.00021$ &  $.02238\pm 0.00014$ &  $0.02238\pm 0.00018$ & $0.02236\pm 0.00020$ \\

$100\theta_\mathrm{MC}$ & $1.04100\pm 0.00030$ &  $1.04169\pm 0.00065$ & $1.04009\pm 0.00069$ & $1.04091\pm 0.00030$ &  $1.04154\pm 0.00064$ & $1.03997\pm 0.00065$ & $1.04091\pm 0.00028$ & $1.04169\pm 0.00062$ & $1.04001\pm 0.00065$  \\

$\tau$ & $0.0518\pm 0.0074$ & $0.062\pm 0.014$ & $0.058\pm 0.014$ & $0.0530\pm 0.0073$ & $0.063\pm 0.014$ & $0.059\pm 0.014$ & $0.0533\pm 0.0073$ & $00.063\pm 0.014$ & $0.059\pm 0.014$ \\

$n_\mathrm{s}   $ & $0.9669\pm 0.0042$ & $0.9729\pm 0.0063$ & $0.9675\pm 0.0069$ & $0.9651\pm 0.0038$ & $0.9716\pm 0.0053$ & $0.9652\pm 0.0060$ & $0.9652\pm 0.0038$ & $0.9730^{+0.0050}_{-0.0055}$ & $0.9662\pm 0.0058$\\

$\log(10^{10} A_\mathrm{s})$ & $3.037\pm 0.015$ & $3.068\pm 0.027$ & $3.043\pm 0.027$ & $3.041\pm 0.014$ & $3.071\pm 0.028$ & $3.050\pm 0.027$ & $3.042\pm 0.014$ & $3.069\pm 0.027$ & $3.050\pm 0.027$\\

$w_{\rm 0, DE}$ & $-1.14^{+0.43}_{-0.59}$ & $-0.88^{+0.59}_{-0.82}$ & $-0.64\pm 0.70$ & $-0.79^{+0.17}_{-0.19}$ & $-0.77\pm 0.20$ & $-0.85^{+0.19}_{-0.23}$ & $-0.948\pm 0.068$ & $-0.967^{+0.069}_{-0.077}$ & $-0.969\pm 0.071$\\

$w_{\rm a, DE}$ & $< -0.888$ & $-0.8^{+1.3}_{-1.2}$ & $-0.9\pm 1.1$ & $-0.65^{+0.47}_{-0.39}$& $-0.69\pm 0.51$ & $-0.48^{+0.60}_{-0.45}$ & $-0.31\pm 0.21$ & $-0.20^{+0.27}_{-0.21}$ & $-0.18^{+0.26}_{-0.22}$\\

$\Omega_\mathrm{m}         $ & $0.217^{+0.016}_{-0.075}$ & $0.311^{+0.056}_{-0.17}$ & $0.353^{+0.083}_{-0.21}$ & $0.318^{+0.016}_{-0.019}$ & $0.324^{+0.019}_{-0.022}$ & $0.314^{+0.019}_{-0.023}$ & $0.3038^{+0.0069}_{-0.0077}$ &$0.3055\pm 0.0082$ & $0.3016\pm 0.0082$\\

$\sigma_8                  $ & $0.941^{+0.11}_{-0.049}$ & $0.86\pm 0.13$ & $0.786^{+0.085}_{-0.17}$ & $0.812\pm 0.016$ & $0.825\pm 0.019$ & $0.810\pm 0.018$ & $0.8240\pm 0.0097$ & $0.833\pm 0.016$ & $0.815\pm 0.016$\\

$S_8                       $ & $0.781^{+0.023}_{-0.046}$ & $0.831\pm 0.052$ & $0.808^{+0.060}_{-0.048}$ & $0.836\pm 0.014$ & $0.857\pm 0.026$ & $0.827\pm 0.027$ & $0.829\pm 0.011$ &  $0.841\pm 0.020$ & $0.817\pm 0.020$\\

$H_0                       $ & $> 78.9$ & $72^{+10}_{-20}$ & $67^{+8}_{-20}$ & $67.1\pm 1.8$ & $66.7\pm 1.9$ & $67.3\pm 2.1$ & $68.61\pm 0.76$ &  $68.42\pm 0.79$ & $68.42\pm 0.80$\\

$r_\mathrm{drag}$ & $147.24\pm 0.26$ & $147.08\pm 0.63$ & $147.86\pm 0.71$ & $147.09\pm 0.25$ & $146.94\pm 0.49$ & $147.48^{+0.50}_{-0.56}$ & $147.11\pm 0.24$ & $147.11\pm 0.46$ & $147.62\pm 0.50$\\

\hline
$\chi^2                       $ & $2770.5$ & $5836.4$ & $7417.4$ & $2794.3$ & $5857.2$ & $7436.8$ & $3828.2$ & $6892.8$ & $8473.8$\\
$\Delta\chi^2                      $ & $-2.2$ & $-0.7$ & $0.9$ & $-2.4$ & $-1.2$ & $-1.7$ & $-3.8$ & $-0.3$ & $0.2$\\
\hline
\end{tabular}
 }      
\label{table-CPLA} 
\end{center}
\end{table*}

\begin{table*}[!t]
\begin{center}
\renewcommand{\arraystretch}{1.4}
\caption{Like Table~\ref{table-exp} but for the CPL-B parametrization.
 } 
\resizebox{\textwidth}{!}{                                              
\begin{tabular}{cccccccccccc}   
\hline 
 Parameter &  Planck & ATW & STW &  Planck+BAO & ATW+BAO & STW+BAO &  Planck+BAO+SNeIa & ATW+BAO+SNeIa & STW+BAO+SNeIa \\
\hline
$\Omega_\mathrm{c} h^2$ & $0.1191\pm 0.0012$ & $0.1201\pm 0.0028$ &$0.1172\pm 0.0029$ & $0.1201\pm 0.0010$ & $0.1207\pm 0.0019$ &$0.1186\pm 0.0020$ & $0.1199\pm 0.0010$ &  $0.1199\pm 0.0018$ & $0.1181\pm 0.0019$ \\

$\Omega_\mathrm{b} h^2$ & $0.02245\pm 0.00015$ & $0.02238\pm 0.00020$ & $0.02237\pm 0.00022$ & $0.02238\pm 0.00014$ & $0.02234\pm 0.00018$ & $0.02234\pm 0.00020$ &  $0.02238\pm 0.00014$ &  $0.02239\pm 0.00018$ & $0.02236\pm 0.00020$ \\

$100\theta_\mathrm{MC}$ & $1.04099\pm 0.00031$ &  $1.04172^{+0.00068}_{-0.00061}$ & $1.04008\pm 0.00068$ & $1.04089\pm 0.00029$ &  $1.04159\pm 0.00060$ & $1.03996\pm 0.00065$ & $1.04091\pm 0.00030$ & $1.04169\pm 0.00060$ & $1.04004\pm 0.00065$  \\

$\tau$ & $0.0523\pm 0.0075$ & $0.061\pm 0.014$ & $0.057\pm 0.015$  & $0.0529\pm 0.0073$ & $0.063\pm 0.014$ & $0.059\pm 0.014$ & $0.0535\pm 0.0074$ & $0.061\pm 0.014$ & $0.060\pm 0.014$ \\

$n_\mathrm{s}   $ & $0.9672\pm 0.0041$ & $0.9729\pm 0.0064$ & $0.9672\pm 0.0068$ & $0.9648\pm 0.0038$ & $0.9716\pm 0.0052$ & $0.9653\pm 0.0061$ & $0.9652\pm 0.0038$ & $0.9731\pm 0.0052$ & $0.9661\pm 0.0058$\\

$\log(10^{10} A_\mathrm{s})$ & $3.038\pm 0.014$ & $3.066\pm 0.026$ & $3.043\pm 0.028$ & $3.041\pm 0.014$ & $3.072\pm 0.027$ & $3.049\pm 0.026$ & $3.042\pm 0.014$ & $3.065\pm 0.027$ & $3.050\pm 0.026$\\

$w_{\rm 0, DE}$ & $-1.08^{+0.45}_{-0.62}$ & $-0.86^{+0.60}_{-0.82}$ & $-0.63\pm 0.69$ & $-0.79^{+0.16}_{-0.19}$ & $-0.77^{+0.18}_{-0.22}$ & $-0.86^{+0.18}_{-0.21}$ & $-0.949\pm 0.068$ & $-0.972\pm 0.066$ & $-0.974\pm 0.070$\\

$w_{\rm a, DE}$ & $< -0.960$ & $-0.8^{+1.3}_{-1.1}$ & $-1.0\pm 1.1$ & $-0.64^{+0.47}_{-0.35}$& $-0.66^{+0.58}_{-0.42}$ & $-0.45^{+0.54}_{-0.42}$ & $-0.29^{+0.22}_{-0.19}$ & $-0.19^{+0.26}_{-0.20}$ & $-0.16^{+0.25}_{-0.20}$\\

$\Omega_\mathrm{m}         $ & $0.224^{+0.020}_{-0.083}$ & $0.314^{+0.060}_{-0.17}$ & $0.348^{+0.087}_{-0.20}$ & $0.319^{+0.016}_{-0.019}$ & $0.325\pm 0.021$ & $0.313\pm 0.021$ & $0.3043\pm 0.0073$ &$0.3045^{+0.0075}_{-0.0086}$ & $0.3016\pm 0.0080$\\

$\sigma_8                  $ & $0.933^{+0.12}_{-0.057}$ & $0.86^{+0.13}_{-0.15}$ & $0.792^{+0.098}_{-0.17}$ & $0.812\pm 0.016$ & $0.825\pm 0.018$ & $0.809\pm 0.018$ & $0.8236\pm 0.0098$ & $0.832\pm 0.015$ & $0.814\pm 0.016$\\

$S_8                       $ & $0.784^{+0.025}_{-0.050}$ & $0.833\pm 0.049$ & $0.808^{+0.059}_{-0.054}$ & $0.837\pm 0.014$ & $0.858\pm 0.026$ & $0.825\pm 0.026$ & $0.829\pm 0.011$ &  $0.838\pm 0.020$ & $0.816\pm 0.020$\\

$H_0                       $ & $> 76.8$ & $72^{+10}_{-20}$ & $68^{+10}_{-20}$ & $67.0\pm 1.8$ & $66.6\pm 1.9$ & $67.4^{+1.9}_{-2.1}$ & $68.56\pm 0.77$ &  $68.53\pm 0.77$ & $68.41\pm 0.77$\\

$r_\mathrm{drag}$ & $147.26\pm 0.27$ & $147.05\pm 0.69$ & $147.83\pm 0.73$ & $147.07\pm 0.25$ & $146.95^{+0.44}_{-0.49}$ & $147.52\pm 0.52$ & $147.11\pm 0.24$ & $147.11\pm 0.47$ & $147.63\pm 0.49$\\

\hline
$\chi^2                       $ & $2770.9$ & $5837.2$ & $7417.8$ & $2792.0$ & $5857.3$ & $7438.9$ & $3828.3$ & $6892.9$ & $8473.9$\\
$\Delta\chi^2                      $ & $-1.8$ & $0.0053$ & $1.3$ & $-4.7$ & $-1.0$ & $0.4$ & $-3.6$ & $-0.2$ & $0.3$\\
\hline
\end{tabular}
 }      
\label{table-CPLB} 
\end{center}
\end{table*}

\subsection{CPL}

In Table~\ref{table-CPL}, we present the results of this parametric DE model considering different datasets as mentioned above, and Figs.~\ref{Fig:CMB},~\ref{Fig:CMB+BAO+SN},~\ref{fig:whisker},~\ref{Fig:CMB+BAO+SN2} summarize all the findings graphically.

We start with the constraints from CMB alone (i.e., Planck, ATW, and STW).
Looking at the constraints on $w_{\rm 0, DE}$ from all the three datasets, we notice that for Planck alone, the mean value of $w_{\rm 0, DE}$ indicates a highly phantom nature of the DE, while within 68\% CL, the quintessence scenario is also allowed ($w_{\rm 0, DE} = -1.23^{+0.43}_{-0.58}$ at 68\% CL for Planck alone). Because of this phantom nature of $w_{\rm 0, DE}$, the expansion rate of the universe at present time quantified by $H_0$ takes a very high value, and consequently, due to the existing geometric degeneracy between $H_0$ and $\Omega_\mathrm{m}$, the model leads to a lower value of the matter density parameter ($\Omega_\mathrm{m}  = 0.215^{+0.016}_{-0.073}$).
In contrast, for both ATW and STW, DE exhibits its quintessential character ($w_{\rm 0, DE} = -0.85^{+0.65}_{-0.86}$ for ATW, and $w_{\rm 0, DE} = -0.74\pm 0.70$ for STW) although $w_{\rm 0, DE} < -1$ is allowed in both cases.
For the other parameter, $w_{\rm a, DE}$, we notice that for both Planck alone and ATW, it has an upper limit but STW constrains it at 68\% CL ($w_{\rm a, DE} = -0.7^{+1.3}_{-1.7}$). It is interesting to note that while neither Planck nor ATW can constrain $w_{\rm a, DE}$, STW can. In addition, our impression of $H_0$ remains similar to what we observed in the Exponential case for the same datasets.

When the BAO data are added to all variants of the CMB data (i.e., Planck, ATW, STW), referring to Table~\ref{table-CPL}, we find that the mean values of $w_{\rm 0, DE}$ remain in the quintessence regime for all cases. However, $w_{\rm 0, DE} > -1$ strictly at 68\% CL for Planck+BAO and ATW+BAO, while STW+BAO allows $w_{\rm 0, DE} < -1$ at 68\% CL.
Looking at the constraints on $w_{\rm a, DE}$ for all three cases, we see that $w_{\rm a, DE} \neq 0$ (in fact, $w_{\rm a, DE} < 0$) at more than 68\% CL for both Planck+BAO and ATW+BAO. However, for STW+BAO, $w_{\rm a, DE} = 0$ is allowed within 68\% CL. Hence, a phantom nature of $w_{\rm DE}$ is indicated in the past for Planck+BAO and ATW+BAO, while STW+BAO is consistent with a cosmological constant.

Finally, when the SNeIa dataset is added (see Table~\ref{table-CPL}), we do not find any strong evidence beyond the $\Lambda$CDM model, because $w_{\rm 0, DE}$ for Planck+BAO+SNeIa, ATW+BAO+SNeIa, and STW+BAO+SNeIa is consistent with $-1$, and $w_{\rm a, DE} = 0$ is allowed for ATW+BAO+SNeIa and STW+BAO+SNeIa datasets. However, for Planck+BAO+SNeIa, we notice $w_{\rm a, DE} = -0.38^{+0.29}_{-0.26}$ at 68\% CL, and therefore the slight dynamical nature of DE is not strictly ruled out for this combined dataset. In fact, since $w_{\rm a, DE} < 0$ at slightly more than 68\% CL, a slight phantom dynamical DE is suggested in this case.

As expected, this parametrization is also statistically disfavored compared to the $\Lambda$CDM scenario, as shown in Table~\ref{tab:bayesian1}, from a Bayesian evidence point of view, but slightly favored by a $\chi^2$ analysis, as shown in Table~\ref{table-CPL}. However, this parametrization is indistinguishable from the Exponential one, or the other CPL extensions, from a Bayesian point of view.

\subsection{CPL-A}

In Table~\ref{table-CPLA}, we show the constraints on this parametrization for various datasets, and Figs.~\ref{Fig:CMB},~\ref{Fig:CMB+BAO+SN},~\ref{fig:whisker},~\ref{Fig:CMB+BAO+SN2} display the results graphically.

Focusing on the CMB constraints (Table~\ref{table-CPLA}), we find that the results on $w_{\rm 0, DE}$ in this parametrization are similar to what we have observed with the \textbf{CPL} parametrization (see Table~\ref{table-CPL}). That means, for Planck alone, the tendency of $w_{\rm 0, DE}$ being in the phantom regime is observed ($w_{\rm 0, DE} = -1.14^{+0.43}_{-0.59}$ at 68\% CL), while within 68\% CL, the cosmological constant is equally allowed. However, ATW and STW agree well with the quintessential DE. Concerning the $w_{\rm a, DE}$ parameter, Planck alone gives its upper limit, while for the remaining CMB experiments, it is well constrained within 68\% CL. Our observation on $H_0$ remains the same (see Table~\ref{table-CPL}). We also refer to Fig.~\ref{Fig:CMB} for a quick visualization of these parameters for these CMB datasets.

When adding BAO measurements to the CMB, referring to Table~\ref{table-CPLA}, we find that the mean values of $w_{\rm 0, DE}$ lie in the quintessence regime ($w_{\rm 0, DE} > -1$) for all three datasets. Furthermore, $w_{\rm 0, DE}$ remains strictly in the quintessence regime for Planck+BAO and ATW+BAO at the 68\% CL constraints, while for STW+BAO, $w_{\rm 0, DE} < -1$ is allowed.
Focusing on $w_{\rm a, DE}$, we see that $w_{\rm a, DE} \neq 0$ (in fact, $w_{\rm a, DE} < 0$) at more than 68\% CL for both Planck+BAO and ATW+BAO, while for STW+BAO, $w_{\rm a, DE} = 0$ is allowed within 68\% CL. Hence, a mild phantom nature of $w_{\rm DE}$ is indicated in the past for both Planck+BAO and ATW+BAO at slightly more than 68\% CL. Additionally, we notice that $H_0$ takes similar values as in the $\Lambda$CDM model but with much relaxed constraints.

The inclusion of SNeIa to the former dataset combinations, referring to Table~\ref{table-CPLA} and Fig.~\ref{Fig:CMB+BAO+SN}, does not add anything new compared to our observations with the CPL case, except for slightly stronger error bars. However, similar to the earlier two parametrizations, namely, Exponential and CPL, we notice here that Planck+BAO+SNeIa suggests the possibility of a dynamical DE at slightly more than 68\% CL, while ATW+BAO+SNeIa and STW+BAO+SNeIa are close to and consistent with the $\Lambda$CDM model expectations.
We refer to Fig.~\ref{fig:whisker} for a quick view of the DE parameters and $H_0$ for all the datasets. This whisker plot clearly shows that the $\Lambda$CDM model is well recovered within the 68\% CL.

Similar to the previously discussed parametrizations, CPL-A is also statistically disfavored compared to the $\Lambda$CDM scenario, as shown in Table~\ref{tab:bayesian1}, but improves the $\chi^2$ analysis, as shown in Table~\ref{table-CPLA}.

\subsection{CPL-B}

This is the last parametrization of this series, and in Table~\ref{table-CPLB} we display the constraints that are obtained from various datasets similar to the previous cases. Figs.~\ref{Fig:CMB},~\ref{Fig:CMB+BAO+SN},~\ref{fig:whisker}, and~\ref{Fig:CMB+BAO+SN2} summarize all the findings graphically.

In a similar fashion, we start by looking at the constraints from the CMB alone dataset. Similar to the previous parametrizations, our observations on the DE parameters, namely, $w_{\rm 0, DE}$ and $w_{\rm a, DE}$, remain consistent. This means that a phantom nature of $w_{\rm 0, DE}$ through its mean value is mildly supported by Planck alone, but that possibility is not supported by any of the ground-based CMB experiments considered in this article. However, all three CMB datasets are perfectly in agreement with a cosmological constant. On the other hand, we find that $w_{\rm a, DE}$ attains an upper limit by Planck alone, but for the remaining CMB experiments, $w_{\rm a}$ is well constrained at 68\% CL ($w_{\rm a, DE} = -0.8^{+1.3}_{-1.1}$ for ATW and $w_{\rm a, DE} = -1.0 \pm 1.1$ for STW). Additionally, we notice that due to the existing geometric degeneracy between $H_0$ and $\Omega_{\rm m}$, Planck alone returns a lower value of $\Omega_{\rm m}$ but a higher value of $H_0$, while in the cases with ATW and STW, $H_0$ is lower and, as a consequence, $\Omega_{\rm m}$ is higher. We refer to Fig.~\ref{Fig:CMB} for a graphical understanding of some of the key parameters of this parametrization.

The inclusion of BAO to all three CMB datasets does not offer anything significantly new compared to the constraints observed with the earlier parametrizations. Concerning the DE parameters, we notice that the mean values of $w_{\rm 0, DE}$ are quintessential, and within 68\% CL, $w_{\rm 0, DE} > -1$ strictly for Planck+BAO and ATW+BAO, while within 68\% CL for STW+BAO, a phantom crossing ($w_{\rm 0, DE} < -1$) is allowed. For the remaining DE parameter, $w_{\rm a, DE}$, we notice that $w_{\rm a, DE} \neq 0$, and in particular, $w_{\rm a, DE} < 0$ at more than 68\% CL for Planck+BAO and ATW+BAO. Hence, a dynamical nature of $w_{\rm DE}$ is indicated for both Planck+BAO and ATW+BAO at slightly more than 68\% CL, and in addition, a phantom nature of $w_{\rm DE}$ in the past is also suggested. However, for STW+BAO, $w_{\rm a, DE} = 0$ is allowed within 68\% CL.

Finally, when SNeIa data are added to CMB (Planck/ATW/STW)+BAO, we do not find any significant changes in the constraints compared to the earlier parametrizations. Effectively, a quintessential DE is preferred for all three datasets as indicated by the mean values of $w_{\rm 0, DE}$; however, the cosmological constant $w_{\rm 0, DE} = -1$ is well consistent within 68\% CL. On the other hand, Planck+BAO+SNeIa favors a dynamical DE with $w_{\rm a, DE} \neq 0$ at more than 68\% CL, but this is consistent with its null value within 95\% CL. For the remaining two datasets, namely, ATW+BAO+SNeIa and STW+BAO+SNeIa, $w_{\rm a, DE} = 0$ is consistent within 68\% CL.

\begin{table*}[!t]
\begin{center}
\renewcommand{\arraystretch}{1}
\caption{Summary of $\ln B_{ij}$ values calculated for all the dynamical DE parametrizations with respect to the $\Lambda$CDM model. The negative value of $\ln B_{ij}$ indicates that $\Lambda$CDM is preferred over the model.  }
\label{tab:bayesian1}
\resizebox{0.8\textwidth}{!}{  
\begin{tabular}{cc|cc|cc|cc}
\hline\hline
Dataset & Model & $\ln B_{ij}$ & Evidence for $\Lambda$CDM & $\ln B_{ij}$ & Evidence for Exp & $\ln B_{ij}$ & Evidence for CPL  \\ \hline
Planck & Exponential & $-0.2$ & Inconclusive & --- & --- & $-0.2$ & inconclusive \\
Planck+BAO & Exponential & $-3.7$ & Moderate & --- & ---  & $-0.4$ & inconclusive \\ 
Planck+BAO+SNeIa & Exponential & $-5.0$ & Strong & --- & ---  & $-0.1$ & inconclusive\\ 
ATW & Exponential & $-1.6$ & Weak & --- & ---  & $-0.3$ & inconclusive\\
ATW+BAO & Exponential & $-4.2$ & Moderate & --- & ---  & $-0.1$ & inconclusive\\ 
ATW+BAO+SNeIa & Exponential & $-7.1$ & Strong & --- & --- & $-1.4$ & Weak \\ 
STW & Exponential & $-1.9$ & Weak & --- & --- & $-0.5$ & inconclusive \\
STW+BAO & Exponential & $-4.8$ & Moderate & --- & --- & $-0.1$ & inconclusive \\ 
STW+BAO+SNeIa & Exponential & $-6.1$ & Strong & --- & --- & $-0.4$ & inconclusive \\ \hline\hline

Planck & CPL & $-0.02$ & inconclusive & $0.2$ & inconclusive & --- & ---\\ 
Planck+BAO & CPL & $-3.2$ & Moderate & $0.4$ & inconclusive & --- & ---\\ 
Planck+BAO+SNeIa & CPL & $-4.9$ & Moderate & $0.1$ & inconclusive & --- & ---\\ 
ATW & CPL & $-1.3$ & Weak & $0.3$ & inconclusive  & --- & ---\\ 
ATW+BAO & CPL & $-4.1$ & Moderate & $0.1$ & inconclusive & --- & ---\\ 
ATW+BAO+SNeIa & CPL & $-5.7$ & Strong & $1.4$ & Weak & --- & ---\\ 
STW & CPL & $-1.4$ & Weak & $0.5$ & inconclusive & --- & ---\\ 
STW+BAO & CPL & $-4.7$ & Moderate & $0.1$ & inconclusive & --- & ---\\ 
STW+BAO+SNeIa & CPL & $-5.8$ & Strong & $0.4$ & inconclusive & --- & ---\\ \hline\hline

Planck & CPL-A & $-0.5$ & inconclusive & $-0.3$ & inconclusive & $-0.4$ & inconclusive\\ 
Planck+BAO & CPL-A & $-3.3$ & Moderate & $0.4$ & inconclusive & $-0.07$ & inconclusive  \\ 
Planck+BAO+SNeIa & CPL-A & $-5.0$ & Strong & $0.04$ & inconclusive & $-0.08$ & inconclusive \\ 
ATW & CPL-A & $-1.8$ & Weak & $-0.2$ & inconclusive & $-0.5$ & inconclusive \\ 
ATW+BAO & CPL-A & $-4.0$ & Moderate & $0.2$ & inconclusive & $0.04$ & inconclusive \\ 
ATW+BAO+SNeIa & CPL-A & $-6.0$ & Strong & $1.2$ & Weak & $-0.2$ & inconclusive \\ 
STW & CPL-A & $-1.6$ & Weak & $0.3$ & inconclusive & $-0.3$ & inconclusive \\ 
STW+BAO & CPL-A & $-4.5$ & Moderate & $0.2$ & inconclusive & $0.1$ & inconclusive \\ 
STW+BAO+SNeIa & CPL-A & $-6.0$ & Strong & $0.2$ & inconclusive & $-0.1$ & inconclusive\\ \hline\hline

Planck & CPL-B & $-0.1$ & inconclusive & $0.1$ & inconclusive & $-0.1$ & inconclusive \\ 
Planck+BAO & CPL-B & $-3.3$ & Moderate & $0.3$ & inconclusive & $-0.1$ & inconclusive \\ 
Planck+BAO+SNeIa & CPL-B & $-4.7$ & Moderate & $0.3$ & inconclusive & $0.2$ & inconclusive \\ 
ATW & CPL-B & $-1.8$ & Weak & $-0.2$ & inconclusive & $-0.5$ & inconclusive \\
ATW+BAO & CPL-B & $-4.1$ & Moderate & $0.1$ & inconclusive & $-0.05$ & inconclusive \\ 
ATW+BAO+SNeIa & CPL-B & $-7.1$ & Strong & $0.03$ & inconclusive & $-1.3$ & Weak \\ 
STW & CPL-B & $-1.7$ & Weak & $0.2$ & inconclusive & $-0.3$ & inconclusive \\ 
STW+BAO & CPL-B & $-4.5$ & Moderate & $0.3$ & inconclusive & $0.1$ & inconclusive \\ 
STW+BAO+SNeIa & CPL-B & $-6.1$ & Strong & $-0.009$ & inconclusive & $-0.4$ & inconclusive \\
\hline\hline
\end{tabular}
 }  
\end{center}
\end{table*}

\subsection{Comparison of the parametrizations}
\label{sec-comparison}

Having all the results in a nutshell, in this section we compare the proposed DE parametrizations. We refer to Figs.~\ref{Fig:CMB},~\ref{Fig:CMB+BAO+SN},~\ref{Fig:CMB+BAO+SN2}, and~\ref{fig:whisker} for a quick comparison between the parametrizations, which clearly show that it is difficult to distinguish between them for all the datasets explored here. However, it is interesting to note that while for the Planck and STW datasets (top and bottom panels in Fig.~\ref{Fig:CMB}), all the parametrizations give the same constraints on the DE parameters, for ATW (middle panel in Fig.~\ref{Fig:CMB}), the additional terms in the expansion (namely CPL-A and CPL-B) help in constraining $w_{\rm a, DE}$, which is otherwise unconstrained for the CPL case. As expected, with the addition of low-redshift data (BAO and SNeIa), all the differences disappear and the DE parameters are very well constrained.

Additionally, we have performed the Bayesian evidence analysis~\cite{Heavens:2017hkr,Heavens:2017afc} to understand the goodness of fit of the models with respect to some reference model. In Table~\ref{tab:bayesian1}, we have presented the values of $\ln B_{ij}$ considering $\Lambda$CDM, Exponential, and CPL as reference models. According to these results, we find that $\Lambda$CDM is preferred over all the DE parametrizations, while both the Exponential and the CPL parametrizations are indistinguishable (inconclusive evidence) among themselves and with the extended cases CPL-A and CPL-B. The only exception is for ATW+BAO+SNeIa, where the CPL and CPL-A parametrizations are weakly preferred over the Exponential one, but the CPL parametrization is weakly preferred over the CPL-A case.

\section{Summary and Conclusions}
\label{sec-summary}

The nature of dark energy has remained elusive despite extensive astronomical surveys. While the standard $\Lambda$CDM cosmology has effectively described the late-time accelerating expansion and the distribution of dark matter in the universe, emerging cosmological tensions between early (based on $\Lambda$CDM) and late-time (model-independent) cosmological probes have called for revisions. Therefore, any modification to the $\Lambda$CDM model is welcome as long as observational data does not definitively rule out such modifications.

In this article, we explore a well-established way of modification: the parametrization of the DE equation of state. Parametrizing the DE sector offers numerous possibilities, with one of the simplest approaches being to vary the equation of state of DE, $w_{\rm DE}$, in different ways and then assess the viability of the proposed parametrized models. Along this line, the widely recognized parametrization is the Chevallier-Polarski-Linder (CPL) model, which involves a Taylor expansion of $w_{\rm DE}$ around the present epoch. While CPL has been extensively used and often serves as the primary choice for many astronomical surveys when investigating the properties of DE, in our current study, we opt for a generalized parametrization: the Exponential parametrization. This choice is motivated by its ability to encompass the CPL parametrization within its first-order approximation.
Additionally, the higher-order approximations of this general parametrization, namely, CPL-A and CPL-B, have only two free parameters, which are economically less expensive compared to DE parametrizations with more than two parameters. Overall, we have considered four parametrizations of the DE equation of state: Exponential, CPL, CPL-A, and CPL-B, and constrained them using CMB data from Planck, ACT, SPT, and their combinations with BAO, as well as the Pantheon sample from SNeIa. Tables~\ref{table-exp},~\ref{table-CPL},~\ref{table-CPLA}, and~\ref{table-CPLB}, and Figs.~\ref{Fig:CMB},~\ref{Fig:CMB+BAO+SN},~\ref{fig:whisker}, and~\ref{Fig:CMB+BAO+SN2} summarize the constraints on all these parametrizations.

The main conclusion of this article regarding the key cosmological parameters is as follows:

\begin{itemize}

     \item $\mathbf{(w_{\rm 0, DE})}$ $-$ Planck alone indicates a slight (but not statistically significant) phantom inclination of DE (through the mean value of $w_{\rm 0, DE}$) for all the parametrizations, which goes away for the other two ground-based CMB experiments. However, within 68\% CL, Planck, ATW, and STW are in agreement with a cosmological constant. The inclusion of BAO makes a difference. For all the parametrizations, $w_{\rm 0, DE}$ remains strictly in the quintessence regime ($w_{\rm 0, DE} > -1$) at slightly more than 68\% CL for both Planck+BAO and ATW+BAO, but STW+BAO allows a phantom crossing ($w_{\rm 0, DE} < -1$) at slightly more than 68\% CL. When SNeIa data are included, for all three combined analyses, the mean values of $w_{\rm 0, DE}$ for all the parametrizations get very close to $-1$ but they all remain quintessential, allowing $w_{\rm 0, DE} < -1$ within 68\% CL.

     \item  $\mathbf{(w_{\rm a, DE})}$ $-$ The parameter $w_{\rm a, DE}$ remains unconstrained for Planck alone, while it becomes constrained for ATW (except in the CPL parametrization, where $w_{\rm a, DE}$ remains unconstrained), STW, and other combined datasets in the presence of BAO and SNeIa. Evidence of $w_{\rm a, DE} \neq 0$ is found for several datasets in different parametrizations. We noticed that for the Exponential parametrization, $w_{\rm a, DE} \neq 0$ at slightly more than 68\% CL for ATW, Planck+BAO, ATW+BAO, and Planck+BAO+SNeIa. For the CPL parametrization, the same is true for Planck+BAO and ATW+BAO. For both CPL-A and CPL-B, Planck+BAO, ATW+BAO, and Planck+BAO+SNeIa favor a mild preference of $w_{\rm a, DE} \neq 0$ at slightly more than 68\% CL. Thus, the preference for dynamical DE within the present DE parametrizations is supported by several cosmological probes.

    \item $\mathbf{H_0}$ $-$ The most stringent constraints on $H_0$ are obtained by Planck+BAO+SNeIa, ATW+BAO+SNeIa, and STW+BAO+SNeIa, where a mild reduction in its tension ($H_0 \gtrsim 68.4$ km/s/Mpc) is found, even if still above $3\sigma$.

    \item $\mathbf{S_8}$ $-$ With the exception of Planck alone, which reports $S_8 \sim 0.78$ for all the parametrizations, we do not find anything noteworthy for the remaining datasets.

\end{itemize}

The contrast of multipole CMB probes across various dark energy equation of state parameters, as explored in this article, presents a compelling narrative. Particularly intriguing is the scrutiny on the current state of $w_{\rm a, DE}$ in light of the findings from two ground-based CMB experiments.

In summary, the findings of this study show an interesting perspective: while $\Lambda$CDM serves as the boundary of present dark energy parametrizations, the potential dynamical nature of dark energy remains a viable consideration in interpreting the results. This conclusion gains further support from recent developments, such as those arising from the new DESI data~\cite{DESI:2024mwx}.\\

\section*{Acknowledgments}
We thank William Giar\`e for the help and useful discussions. MN gratefully thanks Saeed Fakhry and Mina
Ghodsi Yengejeh for the useful discussion and constructive comments.
SP acknowledges the financial support from the Department of Science and Technology (DST), Govt. of India under the Scheme   ``Fund for Improvement of S\&T Infrastructure (FIST)'' (File No. SR/FST/MS-I/2019/41). 
EDV acknowledges support from the Royal Society through a Royal Society Dorothy Hodgkin Research Fellowship. 
This article is based upon work from the COST Action CA21136 ``Addressing observational tensions in cosmology with systematics and fundamental physics (CosmoVerse), supported by COST (European Cooperation in Science and Technology).

\bibliography{biblio}

\begin{thebibliography}{87}%
\makeatletter
\providecommand \@ifxundefined [1]{%
 \@ifx{#1\undefined}
}%
\providecommand \@ifnum [1]{%
 \ifnum #1\expandafter \@firstoftwo
 \else \expandafter \@secondoftwo
 \fi
}%
\providecommand \@ifx [1]{%
 \ifx #1\expandafter \@firstoftwo
 \else \expandafter \@secondoftwo
 \fi
}%
\providecommand \natexlab [1]{#1}%
\providecommand \enquote  [1]{``#1''}%
\providecommand \bibnamefont  [1]{#1}%
\providecommand \bibfnamefont [1]{#1}%
\providecommand \citenamefont [1]{#1}%
\providecommand \href@noop [0]{\@secondoftwo}%
\providecommand \href [0]{\begingroup \@sanitize@url \@href}%
\providecommand \@href[1]{\@@startlink{#1}\@@href}%
\providecommand \@@href[1]{\endgroup#1\@@endlink}%
\providecommand \@sanitize@url [0]{\catcode `\\12\catcode `\$12\catcode
  `\&12\catcode `\#12\catcode `\^12\catcode `\_12\catcode `\%12\relax}%
\providecommand \@@startlink[1]{}%
\providecommand \@@endlink[0]{}%
\providecommand \url  [0]{\begingroup\@sanitize@url \@url }%
\providecommand \@url [1]{\endgroup\@href {#1}{\urlprefix }}%
\providecommand \urlprefix  [0]{URL }%
\providecommand \Eprint [0]{\href }%
\providecommand \doibase [0]{http://dx.doi.org/}%
\providecommand \selectlanguage [0]{\@gobble}%
\providecommand \bibinfo  [0]{\@secondoftwo}%
\providecommand \bibfield  [0]{\@secondoftwo}%
\providecommand \translation [1]{[#1]}%
\providecommand \BibitemOpen [0]{}%
\providecommand \bibitemStop [0]{}%
\providecommand \bibitemNoStop [0]{.\EOS\space}%
\providecommand \EOS [0]{\spacefactor3000\relax}%
\providecommand \BibitemShut  [1]{\csname bibitem#1\endcsname}%
\let\auto@bib@innerbib\@empty
\bibitem [{\citenamefont {Riess}\ \emph {et~al.}(1998)\citenamefont {Riess}
  \emph {et~al.}}]{SupernovaSearchTeam:1998fmf}%
  \BibitemOpen
  \bibfield  {author} {\bibinfo {author} {\bibfnamefont {A.~G.}\ \bibnamefont
  {Riess}} \emph {et~al.} (\bibinfo {collaboration} {Supernova Search Team}),\
  }\href {\doibase 10.1086/300499} {\bibfield  {journal} {\bibinfo  {journal}
  {Astron. J.}\ }\textbf {\bibinfo {volume} {116}},\ \bibinfo {pages} {1009}
  (\bibinfo {year} {1998})},\ \Eprint {http://arxiv.org/abs/astro-ph/9805201}
  {arXiv:astro-ph/9805201} \BibitemShut {NoStop}%
\bibitem [{\citenamefont {Perlmutter}\ \emph {et~al.}(1999)\citenamefont
  {Perlmutter} \emph {et~al.}}]{SupernovaCosmologyProject:1998vns}%
  \BibitemOpen
  \bibfield  {author} {\bibinfo {author} {\bibfnamefont {S.}~\bibnamefont
  {Perlmutter}} \emph {et~al.} (\bibinfo {collaboration} {Supernova Cosmology
  Project}),\ }\href {\doibase 10.1086/307221} {\bibfield  {journal} {\bibinfo
  {journal} {Astrophys. J.}\ }\textbf {\bibinfo {volume} {517}},\ \bibinfo
  {pages} {565} (\bibinfo {year} {1999})},\ \Eprint
  {http://arxiv.org/abs/astro-ph/9812133} {arXiv:astro-ph/9812133} \BibitemShut
  {NoStop}%
\bibitem [{\citenamefont {Copeland}\ \emph {et~al.}(2006)\citenamefont
  {Copeland}, \citenamefont {Sami},\ and\ \citenamefont
  {Tsujikawa}}]{Copeland:2006wr}%
  \BibitemOpen
  \bibfield  {author} {\bibinfo {author} {\bibfnamefont {E.~J.}\ \bibnamefont
  {Copeland}}, \bibinfo {author} {\bibfnamefont {M.}~\bibnamefont {Sami}}, \
  and\ \bibinfo {author} {\bibfnamefont {S.}~\bibnamefont {Tsujikawa}},\ }\href
  {\doibase 10.1142/S021827180600942X} {\bibfield  {journal} {\bibinfo
  {journal} {Int. J. Mod. Phys. D}\ }\textbf {\bibinfo {volume} {15}},\
  \bibinfo {pages} {1753} (\bibinfo {year} {2006})},\ \Eprint
  {http://arxiv.org/abs/hep-th/0603057} {arXiv:hep-th/0603057} \BibitemShut
  {NoStop}%
\bibitem [{\citenamefont {Bamba}\ \emph {et~al.}(2012)\citenamefont {Bamba},
  \citenamefont {Capozziello}, \citenamefont {Nojiri},\ and\ \citenamefont
  {Odintsov}}]{Bamba:2012cp}%
  \BibitemOpen
  \bibfield  {author} {\bibinfo {author} {\bibfnamefont {K.}~\bibnamefont
  {Bamba}}, \bibinfo {author} {\bibfnamefont {S.}~\bibnamefont {Capozziello}},
  \bibinfo {author} {\bibfnamefont {S.}~\bibnamefont {Nojiri}}, \ and\ \bibinfo
  {author} {\bibfnamefont {S.~D.}\ \bibnamefont {Odintsov}},\ }\href {\doibase
  10.1007/s10509-012-1181-8} {\bibfield  {journal} {\bibinfo  {journal}
  {Astrophys. Space Sci.}\ }\textbf {\bibinfo {volume} {342}},\ \bibinfo
  {pages} {155} (\bibinfo {year} {2012})},\ \Eprint
  {http://arxiv.org/abs/1205.3421} {arXiv:1205.3421 [gr-qc]} \BibitemShut
  {NoStop}%
\bibitem [{\citenamefont {Nojiri}\ and\ \citenamefont
  {Odintsov}(2006{\natexlab{a}})}]{Nojiri:2006ri}%
  \BibitemOpen
  \bibfield  {author} {\bibinfo {author} {\bibfnamefont {S.}~\bibnamefont
  {Nojiri}}\ and\ \bibinfo {author} {\bibfnamefont {S.~D.}\ \bibnamefont
  {Odintsov}},\ }\href {\doibase 10.1142/S0219887807001928} {\bibfield
  {journal} {\bibinfo  {journal} {eConf}\ }\textbf {\bibinfo {volume}
  {C0602061}},\ \bibinfo {pages} {06} (\bibinfo {year} {2006}{\natexlab{a}})},\
  \Eprint {http://arxiv.org/abs/hep-th/0601213} {arXiv:hep-th/0601213}
  \BibitemShut {NoStop}%
\bibitem [{\citenamefont {Sotiriou}\ and\ \citenamefont
  {Faraoni}(2010)}]{Sotiriou:2008rp}%
  \BibitemOpen
  \bibfield  {author} {\bibinfo {author} {\bibfnamefont {T.~P.}\ \bibnamefont
  {Sotiriou}}\ and\ \bibinfo {author} {\bibfnamefont {V.}~\bibnamefont
  {Faraoni}},\ }\href {\doibase 10.1103/RevModPhys.82.451} {\bibfield
  {journal} {\bibinfo  {journal} {Rev. Mod. Phys.}\ }\textbf {\bibinfo {volume}
  {82}},\ \bibinfo {pages} {451} (\bibinfo {year} {2010})},\ \Eprint
  {http://arxiv.org/abs/0805.1726} {arXiv:0805.1726 [gr-qc]} \BibitemShut
  {NoStop}%
\bibitem [{\citenamefont {De~Felice}\ and\ \citenamefont
  {Tsujikawa}(2010)}]{DeFelice:2010aj}%
  \BibitemOpen
  \bibfield  {author} {\bibinfo {author} {\bibfnamefont {A.}~\bibnamefont
  {De~Felice}}\ and\ \bibinfo {author} {\bibfnamefont {S.}~\bibnamefont
  {Tsujikawa}},\ }\href {\doibase 10.12942/lrr-2010-3} {\bibfield  {journal}
  {\bibinfo  {journal} {Living Rev. Rel.}\ }\textbf {\bibinfo {volume} {13}},\
  \bibinfo {pages} {3} (\bibinfo {year} {2010})},\ \Eprint
  {http://arxiv.org/abs/1002.4928} {arXiv:1002.4928 [gr-qc]} \BibitemShut
  {NoStop}%
\bibitem [{\citenamefont {Clifton}\ \emph {et~al.}(2012)\citenamefont
  {Clifton}, \citenamefont {Ferreira}, \citenamefont {Padilla},\ and\
  \citenamefont {Skordis}}]{Clifton:2011jh}%
  \BibitemOpen
  \bibfield  {author} {\bibinfo {author} {\bibfnamefont {T.}~\bibnamefont
  {Clifton}}, \bibinfo {author} {\bibfnamefont {P.~G.}\ \bibnamefont
  {Ferreira}}, \bibinfo {author} {\bibfnamefont {A.}~\bibnamefont {Padilla}}, \
  and\ \bibinfo {author} {\bibfnamefont {C.}~\bibnamefont {Skordis}},\ }\href
  {\doibase 10.1016/j.physrep.2012.01.001} {\bibfield  {journal} {\bibinfo
  {journal} {Phys. Rept.}\ }\textbf {\bibinfo {volume} {513}},\ \bibinfo
  {pages} {1} (\bibinfo {year} {2012})},\ \Eprint
  {http://arxiv.org/abs/1106.2476} {arXiv:1106.2476 [astro-ph.CO]} \BibitemShut
  {NoStop}%
\bibitem [{\citenamefont {Capozziello}\ and\ \citenamefont
  {De~Laurentis}(2011)}]{Capozziello:2011et}%
  \BibitemOpen
  \bibfield  {author} {\bibinfo {author} {\bibfnamefont {S.}~\bibnamefont
  {Capozziello}}\ and\ \bibinfo {author} {\bibfnamefont {M.}~\bibnamefont
  {De~Laurentis}},\ }\href {\doibase 10.1016/j.physrep.2011.09.003} {\bibfield
  {journal} {\bibinfo  {journal} {Phys. Rept.}\ }\textbf {\bibinfo {volume}
  {509}},\ \bibinfo {pages} {167} (\bibinfo {year} {2011})},\ \Eprint
  {http://arxiv.org/abs/1108.6266} {arXiv:1108.6266 [gr-qc]} \BibitemShut
  {NoStop}%
\bibitem [{\citenamefont {Cai}\ \emph {et~al.}(2016)\citenamefont {Cai},
  \citenamefont {Capozziello}, \citenamefont {De~Laurentis},\ and\
  \citenamefont {Saridakis}}]{Cai:2015emx}%
  \BibitemOpen
  \bibfield  {author} {\bibinfo {author} {\bibfnamefont {Y.-F.}\ \bibnamefont
  {Cai}}, \bibinfo {author} {\bibfnamefont {S.}~\bibnamefont {Capozziello}},
  \bibinfo {author} {\bibfnamefont {M.}~\bibnamefont {De~Laurentis}}, \ and\
  \bibinfo {author} {\bibfnamefont {E.~N.}\ \bibnamefont {Saridakis}},\ }\href
  {\doibase 10.1088/0034-4885/79/10/106901} {\enquote {\bibinfo {title} {{f(T)
  teleparallel gravity and cosmology}},}\ } (\bibinfo {year} {2016}),\ \Eprint
  {http://arxiv.org/abs/1511.07586} {arXiv:1511.07586 [gr-qc]} \BibitemShut
  {NoStop}%
\bibitem [{\citenamefont {Nojiri}\ \emph {et~al.}(2017)\citenamefont {Nojiri},
  \citenamefont {Odintsov},\ and\ \citenamefont {Oikonomou}}]{Nojiri:2017ncd}%
  \BibitemOpen
  \bibfield  {author} {\bibinfo {author} {\bibfnamefont {S.}~\bibnamefont
  {Nojiri}}, \bibinfo {author} {\bibfnamefont {S.}~\bibnamefont {Odintsov}}, \
  and\ \bibinfo {author} {\bibfnamefont {V.}~\bibnamefont {Oikonomou}},\ }\href
  {\doibase 10.1016/j.physrep.2017.06.001} {\bibfield  {journal} {\bibinfo
  {journal} {Phys. Rept.}\ }\textbf {\bibinfo {volume} {692}},\ \bibinfo
  {pages} {1} (\bibinfo {year} {2017})},\ \Eprint
  {http://arxiv.org/abs/1705.11098} {arXiv:1705.11098 [gr-qc]} \BibitemShut
  {NoStop}%
\bibitem [{\citenamefont {Abdalla}\ \emph {et~al.}(2022)\citenamefont {Abdalla}
  \emph {et~al.}}]{Abdalla:2022yfr}%
  \BibitemOpen
  \bibfield  {author} {\bibinfo {author} {\bibfnamefont {E.}~\bibnamefont
  {Abdalla}} \emph {et~al.},\ }\href {\doibase 10.1016/j.jheap.2022.04.002}
  {\bibfield  {journal} {\bibinfo  {journal} {JHEAp}\ }\textbf {\bibinfo
  {volume} {34}},\ \bibinfo {pages} {49} (\bibinfo {year} {2022})},\ \Eprint
  {http://arxiv.org/abs/2203.06142} {arXiv:2203.06142 [astro-ph.CO]}
  \BibitemShut {NoStop}%
\bibitem [{\citenamefont {Efstathiou}(1999)}]{Efstathiou:1999tm}%
  \BibitemOpen
  \bibfield  {author} {\bibinfo {author} {\bibfnamefont {G.}~\bibnamefont
  {Efstathiou}},\ }\href {\doibase 10.1046/j.1365-8711.1999.02997.x} {\bibfield
   {journal} {\bibinfo  {journal} {Mon. Not. Roy. Astron. Soc.}\ }\textbf
  {\bibinfo {volume} {310}},\ \bibinfo {pages} {842} (\bibinfo {year}
  {1999})},\ \Eprint {http://arxiv.org/abs/astro-ph/9904356}
  {arXiv:astro-ph/9904356} \BibitemShut {NoStop}%
\bibitem [{\citenamefont {Chevallier}\ and\ \citenamefont
  {Polarski}(2001)}]{Chevallier:2000qy}%
  \BibitemOpen
  \bibfield  {author} {\bibinfo {author} {\bibfnamefont {M.}~\bibnamefont
  {Chevallier}}\ and\ \bibinfo {author} {\bibfnamefont {D.}~\bibnamefont
  {Polarski}},\ }\href {\doibase 10.1142/S0218271801000822} {\bibfield
  {journal} {\bibinfo  {journal} {Int. J. Mod. Phys.}\ }\textbf {\bibinfo
  {volume} {D10}},\ \bibinfo {pages} {213} (\bibinfo {year} {2001})},\ \Eprint
  {http://arxiv.org/abs/gr-qc/0009008} {arXiv:gr-qc/0009008 [gr-qc]}
  \BibitemShut {NoStop}%
\bibitem [{\citenamefont {Astier}(2001)}]{Astier:2000as}%
  \BibitemOpen
  \bibfield  {author} {\bibinfo {author} {\bibfnamefont {P.}~\bibnamefont
  {Astier}},\ }\href {\doibase 10.1016/S0370-2693(01)00072-7} {\bibfield
  {journal} {\bibinfo  {journal} {Phys. Lett. B}\ }\textbf {\bibinfo {volume}
  {500}},\ \bibinfo {pages} {8} (\bibinfo {year} {2001})},\ \Eprint
  {http://arxiv.org/abs/astro-ph/0008306} {arXiv:astro-ph/0008306} \BibitemShut
  {NoStop}%
\bibitem [{\citenamefont {Weller}\ and\ \citenamefont
  {Albrecht}(2002)}]{Weller:2001gf}%
  \BibitemOpen
  \bibfield  {author} {\bibinfo {author} {\bibfnamefont {J.}~\bibnamefont
  {Weller}}\ and\ \bibinfo {author} {\bibfnamefont {A.}~\bibnamefont
  {Albrecht}},\ }\href {\doibase 10.1103/PhysRevD.65.103512} {\bibfield
  {journal} {\bibinfo  {journal} {Phys. Rev. D}\ }\textbf {\bibinfo {volume}
  {65}},\ \bibinfo {pages} {103512} (\bibinfo {year} {2002})},\ \Eprint
  {http://arxiv.org/abs/astro-ph/0106079} {arXiv:astro-ph/0106079} \BibitemShut
  {NoStop}%
\bibitem [{\citenamefont {Linder}(2003)}]{Linder:2002et}%
  \BibitemOpen
  \bibfield  {author} {\bibinfo {author} {\bibfnamefont {E.~V.}\ \bibnamefont
  {Linder}},\ }\href {\doibase 10.1103/PhysRevLett.90.091301} {\bibfield
  {journal} {\bibinfo  {journal} {Phys. Rev. Lett.}\ }\textbf {\bibinfo
  {volume} {90}},\ \bibinfo {pages} {091301} (\bibinfo {year} {2003})},\
  \Eprint {http://arxiv.org/abs/astro-ph/0208512} {arXiv:astro-ph/0208512
  [astro-ph]} \BibitemShut {NoStop}%
\bibitem [{\citenamefont {Rubano}\ \emph {et~al.}(2003)\citenamefont {Rubano},
  \citenamefont {Scudellaro}, \citenamefont {Piedipalumbo},\ and\ \citenamefont
  {Capozziello}}]{Rubano:2003er}%
  \BibitemOpen
  \bibfield  {author} {\bibinfo {author} {\bibfnamefont {C.}~\bibnamefont
  {Rubano}}, \bibinfo {author} {\bibfnamefont {P.}~\bibnamefont {Scudellaro}},
  \bibinfo {author} {\bibfnamefont {E.}~\bibnamefont {Piedipalumbo}}, \ and\
  \bibinfo {author} {\bibfnamefont {S.}~\bibnamefont {Capozziello}},\ }\href
  {\doibase 10.1103/PhysRevD.68.123501} {\bibfield  {journal} {\bibinfo
  {journal} {Phys. Rev. D}\ }\textbf {\bibinfo {volume} {68}},\ \bibinfo
  {pages} {123501} (\bibinfo {year} {2003})},\ \Eprint
  {http://arxiv.org/abs/astro-ph/0311535} {arXiv:astro-ph/0311535} \BibitemShut
  {NoStop}%
\bibitem [{\citenamefont {Guo}\ \emph {et~al.}(2005)\citenamefont {Guo},
  \citenamefont {Piao}, \citenamefont {Zhang},\ and\ \citenamefont
  {Zhang}}]{Guo:2004fq}%
  \BibitemOpen
  \bibfield  {author} {\bibinfo {author} {\bibfnamefont {Z.-K.}\ \bibnamefont
  {Guo}}, \bibinfo {author} {\bibfnamefont {Y.-S.}\ \bibnamefont {Piao}},
  \bibinfo {author} {\bibfnamefont {X.-M.}\ \bibnamefont {Zhang}}, \ and\
  \bibinfo {author} {\bibfnamefont {Y.-Z.}\ \bibnamefont {Zhang}},\ }\href
  {\doibase 10.1016/j.physletb.2005.01.017} {\bibfield  {journal} {\bibinfo
  {journal} {Phys. Lett. B}\ }\textbf {\bibinfo {volume} {608}},\ \bibinfo
  {pages} {177} (\bibinfo {year} {2005})},\ \Eprint
  {http://arxiv.org/abs/astro-ph/0410654} {arXiv:astro-ph/0410654} \BibitemShut
  {NoStop}%
\bibitem [{\citenamefont {Feng}\ \emph {et~al.}(2006)\citenamefont {Feng},
  \citenamefont {Li}, \citenamefont {Piao},\ and\ \citenamefont
  {Zhang}}]{Feng:2004ff}%
  \BibitemOpen
  \bibfield  {author} {\bibinfo {author} {\bibfnamefont {B.}~\bibnamefont
  {Feng}}, \bibinfo {author} {\bibfnamefont {M.}~\bibnamefont {Li}}, \bibinfo
  {author} {\bibfnamefont {Y.-S.}\ \bibnamefont {Piao}}, \ and\ \bibinfo
  {author} {\bibfnamefont {X.}~\bibnamefont {Zhang}},\ }\href {\doibase
  10.1016/j.physletb.2006.01.066} {\bibfield  {journal} {\bibinfo  {journal}
  {Phys. Lett. B}\ }\textbf {\bibinfo {volume} {634}},\ \bibinfo {pages} {101}
  (\bibinfo {year} {2006})},\ \Eprint {http://arxiv.org/abs/astro-ph/0407432}
  {arXiv:astro-ph/0407432} \BibitemShut {NoStop}%
\bibitem [{\citenamefont {Nesseris}\ and\ \citenamefont
  {Perivolaropoulos}(2005)}]{Nesseris:2005ur}%
  \BibitemOpen
  \bibfield  {author} {\bibinfo {author} {\bibfnamefont {S.}~\bibnamefont
  {Nesseris}}\ and\ \bibinfo {author} {\bibfnamefont {L.}~\bibnamefont
  {Perivolaropoulos}},\ }\href {\doibase 10.1103/PhysRevD.72.123519} {\bibfield
   {journal} {\bibinfo  {journal} {Phys. Rev. D}\ }\textbf {\bibinfo {volume}
  {72}},\ \bibinfo {pages} {123519} (\bibinfo {year} {2005})},\ \Eprint
  {http://arxiv.org/abs/astro-ph/0511040} {arXiv:astro-ph/0511040} \BibitemShut
  {NoStop}%
\bibitem [{\citenamefont {Jassal}\ \emph {et~al.}(2005)\citenamefont {Jassal},
  \citenamefont {Bagla},\ and\ \citenamefont {Padmanabhan}}]{Jassal:2005qc}%
  \BibitemOpen
  \bibfield  {author} {\bibinfo {author} {\bibfnamefont {H.~K.}\ \bibnamefont
  {Jassal}}, \bibinfo {author} {\bibfnamefont {J.~S.}\ \bibnamefont {Bagla}}, \
  and\ \bibinfo {author} {\bibfnamefont {T.}~\bibnamefont {Padmanabhan}},\
  }\href {\doibase 10.1103/PhysRevD.72.103503} {\bibfield  {journal} {\bibinfo
  {journal} {Phys. Rev. D}\ }\textbf {\bibinfo {volume} {72}},\ \bibinfo
  {pages} {103503} (\bibinfo {year} {2005})},\ \Eprint
  {http://arxiv.org/abs/astro-ph/0506748} {arXiv:astro-ph/0506748} \BibitemShut
  {NoStop}%
\bibitem [{\citenamefont {Gong}\ and\ \citenamefont
  {Zhang}(2005)}]{Gong:2005de}%
  \BibitemOpen
  \bibfield  {author} {\bibinfo {author} {\bibfnamefont {Y.-g.}\ \bibnamefont
  {Gong}}\ and\ \bibinfo {author} {\bibfnamefont {Y.-Z.}\ \bibnamefont
  {Zhang}},\ }\href {\doibase 10.1103/PhysRevD.72.043518} {\bibfield  {journal}
  {\bibinfo  {journal} {Phys. Rev. D}\ }\textbf {\bibinfo {volume} {72}},\
  \bibinfo {pages} {043518} (\bibinfo {year} {2005})},\ \Eprint
  {http://arxiv.org/abs/astro-ph/0502262} {arXiv:astro-ph/0502262} \BibitemShut
  {NoStop}%
\bibitem [{\citenamefont {Nojiri}\ and\ \citenamefont
  {Odintsov}(2006{\natexlab{b}})}]{Nojiri:2006ww}%
  \BibitemOpen
  \bibfield  {author} {\bibinfo {author} {\bibfnamefont {S.}~\bibnamefont
  {Nojiri}}\ and\ \bibinfo {author} {\bibfnamefont {S.~D.}\ \bibnamefont
  {Odintsov}},\ }\href {\doibase 10.1016/j.physletb.2006.04.026} {\bibfield
  {journal} {\bibinfo  {journal} {Phys. Lett. B}\ }\textbf {\bibinfo {volume}
  {637}},\ \bibinfo {pages} {139} (\bibinfo {year} {2006}{\natexlab{b}})},\
  \Eprint {http://arxiv.org/abs/hep-th/0603062} {arXiv:hep-th/0603062}
  \BibitemShut {NoStop}%
\bibitem [{\citenamefont {Kurek}\ \emph {et~al.}(2008)\citenamefont {Kurek},
  \citenamefont {Hrycyna},\ and\ \citenamefont {Szydlowski}}]{Kurek:2007bu}%
  \BibitemOpen
  \bibfield  {author} {\bibinfo {author} {\bibfnamefont {A.}~\bibnamefont
  {Kurek}}, \bibinfo {author} {\bibfnamefont {O.}~\bibnamefont {Hrycyna}}, \
  and\ \bibinfo {author} {\bibfnamefont {M.}~\bibnamefont {Szydlowski}},\
  }\href {\doibase 10.1016/j.physletb.2007.10.074} {\bibfield  {journal}
  {\bibinfo  {journal} {Phys. Lett. B}\ }\textbf {\bibinfo {volume} {659}},\
  \bibinfo {pages} {14} (\bibinfo {year} {2008})},\ \Eprint
  {http://arxiv.org/abs/0707.0292} {arXiv:0707.0292 [astro-ph]} \BibitemShut
  {NoStop}%
\bibitem [{\citenamefont {Barboza}\ and\ \citenamefont
  {Alcaniz}(2008)}]{Barboza:2008rh}%
  \BibitemOpen
  \bibfield  {author} {\bibinfo {author} {\bibfnamefont {E.~M.}\ \bibnamefont
  {Barboza}, \bibfnamefont {Jr.}}\ and\ \bibinfo {author} {\bibfnamefont
  {J.~S.}\ \bibnamefont {Alcaniz}},\ }\href {\doibase
  10.1016/j.physletb.2008.08.012} {\bibfield  {journal} {\bibinfo  {journal}
  {Phys. Lett. B}\ }\textbf {\bibinfo {volume} {666}},\ \bibinfo {pages} {415}
  (\bibinfo {year} {2008})},\ \Eprint {http://arxiv.org/abs/0805.1713}
  {arXiv:0805.1713 [astro-ph]} \BibitemShut {NoStop}%
\bibitem [{\citenamefont {Kurek}\ \emph {et~al.}(2010)\citenamefont {Kurek},
  \citenamefont {Hrycyna},\ and\ \citenamefont {Szydlowski}}]{Kurek:2008qt}%
  \BibitemOpen
  \bibfield  {author} {\bibinfo {author} {\bibfnamefont {A.}~\bibnamefont
  {Kurek}}, \bibinfo {author} {\bibfnamefont {O.}~\bibnamefont {Hrycyna}}, \
  and\ \bibinfo {author} {\bibfnamefont {M.}~\bibnamefont {Szydlowski}},\
  }\href {\doibase 10.1016/j.physletb.2010.05.061} {\bibfield  {journal}
  {\bibinfo  {journal} {Phys. Lett. B}\ }\textbf {\bibinfo {volume} {690}},\
  \bibinfo {pages} {337} (\bibinfo {year} {2010})},\ \Eprint
  {http://arxiv.org/abs/0805.4005} {arXiv:0805.4005 [astro-ph]} \BibitemShut
  {NoStop}%
\bibitem [{\citenamefont {Barboza}\ \emph {et~al.}(2009)\citenamefont
  {Barboza}, \citenamefont {Alcaniz}, \citenamefont {Zhu},\ and\ \citenamefont
  {Silva}}]{Barboza:2009ks}%
  \BibitemOpen
  \bibfield  {author} {\bibinfo {author} {\bibfnamefont {E.~M.}\ \bibnamefont
  {Barboza}}, \bibinfo {author} {\bibfnamefont {J.~S.}\ \bibnamefont
  {Alcaniz}}, \bibinfo {author} {\bibfnamefont {Z.~H.}\ \bibnamefont {Zhu}}, \
  and\ \bibinfo {author} {\bibfnamefont {R.}~\bibnamefont {Silva}},\ }\href
  {\doibase 10.1103/PhysRevD.80.043521} {\bibfield  {journal} {\bibinfo
  {journal} {Phys. Rev. D}\ }\textbf {\bibinfo {volume} {80}},\ \bibinfo
  {pages} {043521} (\bibinfo {year} {2009})},\ \Eprint
  {http://arxiv.org/abs/0905.4052} {arXiv:0905.4052 [astro-ph.CO]} \BibitemShut
  {NoStop}%
\bibitem [{\citenamefont {Li}\ and\ \citenamefont {Zhang}(2011)}]{Li:2011dr}%
  \BibitemOpen
  \bibfield  {author} {\bibinfo {author} {\bibfnamefont {H.}~\bibnamefont
  {Li}}\ and\ \bibinfo {author} {\bibfnamefont {X.}~\bibnamefont {Zhang}},\
  }\href {\doibase 10.1016/j.physletb.2011.07.069} {\bibfield  {journal}
  {\bibinfo  {journal} {Phys. Lett. B}\ }\textbf {\bibinfo {volume} {703}},\
  \bibinfo {pages} {119} (\bibinfo {year} {2011})},\ \Eprint
  {http://arxiv.org/abs/1106.5658} {arXiv:1106.5658 [astro-ph.CO]} \BibitemShut
  {NoStop}%
\bibitem [{\citenamefont {Li}\ and\ \citenamefont {Zhang}(2012)}]{Li:2012vn}%
  \BibitemOpen
  \bibfield  {author} {\bibinfo {author} {\bibfnamefont {H.}~\bibnamefont
  {Li}}\ and\ \bibinfo {author} {\bibfnamefont {X.}~\bibnamefont {Zhang}},\
  }\href {\doibase 10.1016/j.physletb.2012.06.030} {\bibfield  {journal}
  {\bibinfo  {journal} {Phys. Lett. B}\ }\textbf {\bibinfo {volume} {713}},\
  \bibinfo {pages} {160} (\bibinfo {year} {2012})},\ \Eprint
  {http://arxiv.org/abs/1202.4071} {arXiv:1202.4071 [astro-ph.CO]} \BibitemShut
  {NoStop}%
\bibitem [{\citenamefont {Pace}\ \emph {et~al.}(2012)\citenamefont {Pace},
  \citenamefont {Fedeli}, \citenamefont {Moscardini},\ and\ \citenamefont
  {Bartelmann}}]{Pace:2011kb}%
  \BibitemOpen
  \bibfield  {author} {\bibinfo {author} {\bibfnamefont {F.}~\bibnamefont
  {Pace}}, \bibinfo {author} {\bibfnamefont {C.}~\bibnamefont {Fedeli}},
  \bibinfo {author} {\bibfnamefont {L.}~\bibnamefont {Moscardini}}, \ and\
  \bibinfo {author} {\bibfnamefont {M.}~\bibnamefont {Bartelmann}},\ }\href
  {\doibase 10.1111/j.1365-2966.2012.20692.x} {\bibfield  {journal} {\bibinfo
  {journal} {Mon. Not. Roy. Astron. Soc.}\ }\textbf {\bibinfo {volume} {422}},\
  \bibinfo {pages} {1186} (\bibinfo {year} {2012})},\ \Eprint
  {http://arxiv.org/abs/1111.1556} {arXiv:1111.1556 [astro-ph.CO]} \BibitemShut
  {NoStop}%
\bibitem [{\citenamefont {Feng}\ and\ \citenamefont {Lu}(2011)}]{Feng:2011zzo}%
  \BibitemOpen
  \bibfield  {author} {\bibinfo {author} {\bibfnamefont {L.}~\bibnamefont
  {Feng}}\ and\ \bibinfo {author} {\bibfnamefont {T.}~\bibnamefont {Lu}},\
  }\href {\doibase 10.1088/1475-7516/2011/11/034} {\bibfield  {journal}
  {\bibinfo  {journal} {JCAP}\ }\textbf {\bibinfo {volume} {11}},\ \bibinfo
  {pages} {034} (\bibinfo {year} {2011})},\ \Eprint
  {http://arxiv.org/abs/1203.1784} {arXiv:1203.1784 [astro-ph.CO]} \BibitemShut
  {NoStop}%
\bibitem [{\citenamefont {Feng}\ \emph {et~al.}(2012)\citenamefont {Feng},
  \citenamefont {Shen}, \citenamefont {Li},\ and\ \citenamefont
  {Li}}]{Feng:2012gf}%
  \BibitemOpen
  \bibfield  {author} {\bibinfo {author} {\bibfnamefont {C.-J.}\ \bibnamefont
  {Feng}}, \bibinfo {author} {\bibfnamefont {X.-Y.}\ \bibnamefont {Shen}},
  \bibinfo {author} {\bibfnamefont {P.}~\bibnamefont {Li}}, \ and\ \bibinfo
  {author} {\bibfnamefont {X.-Z.}\ \bibnamefont {Li}},\ }\href {\doibase
  10.1088/1475-7516/2012/09/023} {\bibfield  {journal} {\bibinfo  {journal}
  {JCAP}\ }\textbf {\bibinfo {volume} {09}},\ \bibinfo {pages} {023} (\bibinfo
  {year} {2012})},\ \Eprint {http://arxiv.org/abs/1206.0063} {arXiv:1206.0063
  [astro-ph.CO]} \BibitemShut {NoStop}%
\bibitem [{\citenamefont {Akarsu}\ \emph {et~al.}(2015)\citenamefont {Akarsu},
  \citenamefont {Dereli},\ and\ \citenamefont {Vazquez}}]{Akarsu:2015yea}%
  \BibitemOpen
  \bibfield  {author} {\bibinfo {author} {\bibfnamefont {O.}~\bibnamefont
  {Akarsu}}, \bibinfo {author} {\bibfnamefont {T.}~\bibnamefont {Dereli}}, \
  and\ \bibinfo {author} {\bibfnamefont {J.~A.}\ \bibnamefont {Vazquez}},\
  }\href {\doibase 10.1088/1475-7516/2015/06/049} {\bibfield  {journal}
  {\bibinfo  {journal} {JCAP}\ }\textbf {\bibinfo {volume} {06}},\ \bibinfo
  {pages} {049} (\bibinfo {year} {2015})},\ \Eprint
  {http://arxiv.org/abs/1501.07598} {arXiv:1501.07598 [astro-ph.CO]}
  \BibitemShut {NoStop}%
\bibitem [{\citenamefont {Ch\'avez}\ \emph {et~al.}(2016)\citenamefont
  {Ch\'avez}, \citenamefont {Plionis}, \citenamefont {Basilakos}, \citenamefont
  {Terlevich}, \citenamefont {Terlevich}, \citenamefont {Melnick},
  \citenamefont {Bresolin},\ and\ \citenamefont
  {Gonz\'alez-Mor\'an}}]{Chavez:2016epc}%
  \BibitemOpen
  \bibfield  {author} {\bibinfo {author} {\bibfnamefont {R.}~\bibnamefont
  {Ch\'avez}}, \bibinfo {author} {\bibfnamefont {M.}~\bibnamefont {Plionis}},
  \bibinfo {author} {\bibfnamefont {S.}~\bibnamefont {Basilakos}}, \bibinfo
  {author} {\bibfnamefont {R.}~\bibnamefont {Terlevich}}, \bibinfo {author}
  {\bibfnamefont {E.}~\bibnamefont {Terlevich}}, \bibinfo {author}
  {\bibfnamefont {J.}~\bibnamefont {Melnick}}, \bibinfo {author} {\bibfnamefont
  {F.}~\bibnamefont {Bresolin}}, \ and\ \bibinfo {author} {\bibfnamefont
  {A.~L.}\ \bibnamefont {Gonz\'alez-Mor\'an}},\ }\href {\doibase
  10.1093/mnras/stw1813} {\bibfield  {journal} {\bibinfo  {journal} {Mon. Not.
  Roy. Astron. Soc.}\ }\textbf {\bibinfo {volume} {462}},\ \bibinfo {pages}
  {2431} (\bibinfo {year} {2016})},\ \Eprint {http://arxiv.org/abs/1607.06458}
  {arXiv:1607.06458 [astro-ph.CO]} \BibitemShut {NoStop}%
\bibitem [{\citenamefont {Pantazis}\ \emph {et~al.}(2016)\citenamefont
  {Pantazis}, \citenamefont {Nesseris},\ and\ \citenamefont
  {Perivolaropoulos}}]{Pantazis:2016nky}%
  \BibitemOpen
  \bibfield  {author} {\bibinfo {author} {\bibfnamefont {G.}~\bibnamefont
  {Pantazis}}, \bibinfo {author} {\bibfnamefont {S.}~\bibnamefont {Nesseris}},
  \ and\ \bibinfo {author} {\bibfnamefont {L.}~\bibnamefont
  {Perivolaropoulos}},\ }\href {\doibase 10.1103/PhysRevD.93.103503} {\bibfield
   {journal} {\bibinfo  {journal} {Phys. Rev. D}\ }\textbf {\bibinfo {volume}
  {93}},\ \bibinfo {pages} {103503} (\bibinfo {year} {2016})},\ \Eprint
  {http://arxiv.org/abs/1603.02164} {arXiv:1603.02164 [astro-ph.CO]}
  \BibitemShut {NoStop}%
\bibitem [{\citenamefont {Yang}\ \emph {et~al.}(2017)\citenamefont {Yang},
  \citenamefont {Nunes}, \citenamefont {Pan},\ and\ \citenamefont
  {Mota}}]{Yang:2017amu}%
  \BibitemOpen
  \bibfield  {author} {\bibinfo {author} {\bibfnamefont {W.}~\bibnamefont
  {Yang}}, \bibinfo {author} {\bibfnamefont {R.~C.}\ \bibnamefont {Nunes}},
  \bibinfo {author} {\bibfnamefont {S.}~\bibnamefont {Pan}}, \ and\ \bibinfo
  {author} {\bibfnamefont {D.~F.}\ \bibnamefont {Mota}},\ }\href {\doibase
  10.1103/PhysRevD.95.103522} {\bibfield  {journal} {\bibinfo  {journal} {Phys.
  Rev. D}\ }\textbf {\bibinfo {volume} {95}},\ \bibinfo {pages} {103522}
  (\bibinfo {year} {2017})},\ \Eprint {http://arxiv.org/abs/1703.02556}
  {arXiv:1703.02556 [astro-ph.CO]} \BibitemShut {NoStop}%
\bibitem [{\citenamefont {Yang}\ \emph {et~al.}(2018)\citenamefont {Yang},
  \citenamefont {Pan},\ and\ \citenamefont {Paliathanasis}}]{Yang:2017alx}%
  \BibitemOpen
  \bibfield  {author} {\bibinfo {author} {\bibfnamefont {W.}~\bibnamefont
  {Yang}}, \bibinfo {author} {\bibfnamefont {S.}~\bibnamefont {Pan}}, \ and\
  \bibinfo {author} {\bibfnamefont {A.}~\bibnamefont {Paliathanasis}},\ }\href
  {\doibase 10.1093/mnras/sty019} {\bibfield  {journal} {\bibinfo  {journal}
  {Mon. Not. Roy. Astron. Soc.}\ }\textbf {\bibinfo {volume} {475}},\ \bibinfo
  {pages} {2605} (\bibinfo {year} {2018})},\ \Eprint
  {http://arxiv.org/abs/1708.01717} {arXiv:1708.01717 [gr-qc]} \BibitemShut
  {NoStop}%
\bibitem [{\citenamefont {Pan}\ \emph {et~al.}(2018)\citenamefont {Pan},
  \citenamefont {Saridakis},\ and\ \citenamefont {Yang}}]{Pan:2017zoh}%
  \BibitemOpen
  \bibfield  {author} {\bibinfo {author} {\bibfnamefont {S.}~\bibnamefont
  {Pan}}, \bibinfo {author} {\bibfnamefont {E.~N.}\ \bibnamefont {Saridakis}},
  \ and\ \bibinfo {author} {\bibfnamefont {W.}~\bibnamefont {Yang}},\ }\href
  {\doibase 10.1103/PhysRevD.98.063510} {\bibfield  {journal} {\bibinfo
  {journal} {Phys. Rev. D}\ }\textbf {\bibinfo {volume} {98}},\ \bibinfo
  {pages} {063510} (\bibinfo {year} {2018})},\ \Eprint
  {http://arxiv.org/abs/1712.05746} {arXiv:1712.05746 [astro-ph.CO]}
  \BibitemShut {NoStop}%
\bibitem [{\citenamefont {Rezaei}\ \emph {et~al.}(2017)\citenamefont {Rezaei},
  \citenamefont {Malekjani}, \citenamefont {Basilakos}, \citenamefont
  {Mehrabi},\ and\ \citenamefont {Mota}}]{Rezaei:2017yyj}%
  \BibitemOpen
  \bibfield  {author} {\bibinfo {author} {\bibfnamefont {M.}~\bibnamefont
  {Rezaei}}, \bibinfo {author} {\bibfnamefont {M.}~\bibnamefont {Malekjani}},
  \bibinfo {author} {\bibfnamefont {S.}~\bibnamefont {Basilakos}}, \bibinfo
  {author} {\bibfnamefont {A.}~\bibnamefont {Mehrabi}}, \ and\ \bibinfo
  {author} {\bibfnamefont {D.~F.}\ \bibnamefont {Mota}},\ }\href {\doibase
  10.3847/1538-4357/aa7898} {\bibfield  {journal} {\bibinfo  {journal}
  {Astrophys. J.}\ }\textbf {\bibinfo {volume} {843}},\ \bibinfo {pages} {65}
  (\bibinfo {year} {2017})},\ \Eprint {http://arxiv.org/abs/1706.02537}
  {arXiv:1706.02537 [astro-ph.CO]} \BibitemShut {NoStop}%
\bibitem [{\citenamefont {Mehrabi}\ and\ \citenamefont
  {Basilakos}(2018)}]{Mehrabi:2018oke}%
  \BibitemOpen
  \bibfield  {author} {\bibinfo {author} {\bibfnamefont {A.}~\bibnamefont
  {Mehrabi}}\ and\ \bibinfo {author} {\bibfnamefont {S.}~\bibnamefont
  {Basilakos}},\ }\href {\doibase 10.1140/epjc/s10052-018-6368-x} {\bibfield
  {journal} {\bibinfo  {journal} {Eur. Phys. J. C}\ }\textbf {\bibinfo {volume}
  {78}},\ \bibinfo {pages} {889} (\bibinfo {year} {2018})},\ \Eprint
  {http://arxiv.org/abs/1804.10794} {arXiv:1804.10794 [astro-ph.CO]}
  \BibitemShut {NoStop}%
\bibitem [{\citenamefont {Yang}\ \emph
  {et~al.}(2019{\natexlab{a}})\citenamefont {Yang}, \citenamefont {Pan},
  \citenamefont {Di~Valentino}, \citenamefont {Saridakis},\ and\ \citenamefont
  {Chakraborty}}]{Yang:2018qmz}%
  \BibitemOpen
  \bibfield  {author} {\bibinfo {author} {\bibfnamefont {W.}~\bibnamefont
  {Yang}}, \bibinfo {author} {\bibfnamefont {S.}~\bibnamefont {Pan}}, \bibinfo
  {author} {\bibfnamefont {E.}~\bibnamefont {Di~Valentino}}, \bibinfo {author}
  {\bibfnamefont {E.~N.}\ \bibnamefont {Saridakis}}, \ and\ \bibinfo {author}
  {\bibfnamefont {S.}~\bibnamefont {Chakraborty}},\ }\href {\doibase
  10.1103/PhysRevD.99.043543} {\bibfield  {journal} {\bibinfo  {journal} {Phys.
  Rev.}\ }\textbf {\bibinfo {volume} {D99}},\ \bibinfo {pages} {043543}
  (\bibinfo {year} {2019}{\natexlab{a}})},\ \Eprint
  {http://arxiv.org/abs/1810.05141} {arXiv:1810.05141 [astro-ph.CO]}
  \BibitemShut {NoStop}%
\bibitem [{\citenamefont {Yang}\ \emph
  {et~al.}(2019{\natexlab{b}})\citenamefont {Yang}, \citenamefont {Pan},
  \citenamefont {Di~Valentino},\ and\ \citenamefont
  {Saridakis}}]{Yang:2018prh}%
  \BibitemOpen
  \bibfield  {author} {\bibinfo {author} {\bibfnamefont {W.}~\bibnamefont
  {Yang}}, \bibinfo {author} {\bibfnamefont {S.}~\bibnamefont {Pan}}, \bibinfo
  {author} {\bibfnamefont {E.}~\bibnamefont {Di~Valentino}}, \ and\ \bibinfo
  {author} {\bibfnamefont {E.~N.}\ \bibnamefont {Saridakis}},\ }\href {\doibase
  10.3390/universe5110219} {\bibfield  {journal} {\bibinfo  {journal}
  {Universe}\ }\textbf {\bibinfo {volume} {5}},\ \bibinfo {pages} {219}
  (\bibinfo {year} {2019}{\natexlab{b}})},\ \Eprint
  {http://arxiv.org/abs/1811.06932} {arXiv:1811.06932 [astro-ph.CO]}
  \BibitemShut {NoStop}%
\bibitem [{\citenamefont {Panotopoulos}\ and\ \citenamefont
  {Rinc\'on}(2018)}]{Panotopoulos:2018sso}%
  \BibitemOpen
  \bibfield  {author} {\bibinfo {author} {\bibfnamefont {G.}~\bibnamefont
  {Panotopoulos}}\ and\ \bibinfo {author} {\bibfnamefont {A.}~\bibnamefont
  {Rinc\'on}},\ }\href {\doibase 10.1103/PhysRevD.97.103509} {\bibfield
  {journal} {\bibinfo  {journal} {Phys. Rev. D}\ }\textbf {\bibinfo {volume}
  {97}},\ \bibinfo {pages} {103509} (\bibinfo {year} {2018})},\ \Eprint
  {http://arxiv.org/abs/1804.11208} {arXiv:1804.11208 [astro-ph.CO]}
  \BibitemShut {NoStop}%
\bibitem [{\citenamefont {Pan}\ \emph {et~al.}(2020{\natexlab{a}})\citenamefont
  {Pan}, \citenamefont {Yang}, \citenamefont {Di~Valentino}, \citenamefont
  {Shafieloo},\ and\ \citenamefont {Chakraborty}}]{Pan:2019hac}%
  \BibitemOpen
  \bibfield  {author} {\bibinfo {author} {\bibfnamefont {S.}~\bibnamefont
  {Pan}}, \bibinfo {author} {\bibfnamefont {W.}~\bibnamefont {Yang}}, \bibinfo
  {author} {\bibfnamefont {E.}~\bibnamefont {Di~Valentino}}, \bibinfo {author}
  {\bibfnamefont {A.}~\bibnamefont {Shafieloo}}, \ and\ \bibinfo {author}
  {\bibfnamefont {S.}~\bibnamefont {Chakraborty}},\ }\href {\doibase
  10.1088/1475-7516/2020/06/062} {\bibfield  {journal} {\bibinfo  {journal}
  {JCAP}\ }\textbf {\bibinfo {volume} {06}},\ \bibinfo {pages} {062} (\bibinfo
  {year} {2020}{\natexlab{a}})},\ \Eprint {http://arxiv.org/abs/1907.12551}
  {arXiv:1907.12551 [astro-ph.CO]} \BibitemShut {NoStop}%
\bibitem [{\citenamefont {Chudaykin}\ \emph {et~al.}(2021)\citenamefont
  {Chudaykin}, \citenamefont {Dolgikh},\ and\ \citenamefont
  {Ivanov}}]{Chudaykin:2020ghx}%
  \BibitemOpen
  \bibfield  {author} {\bibinfo {author} {\bibfnamefont {A.}~\bibnamefont
  {Chudaykin}}, \bibinfo {author} {\bibfnamefont {K.}~\bibnamefont {Dolgikh}},
  \ and\ \bibinfo {author} {\bibfnamefont {M.~M.}\ \bibnamefont {Ivanov}},\
  }\href {\doibase 10.1103/PhysRevD.103.023507} {\bibfield  {journal} {\bibinfo
   {journal} {Phys. Rev. D}\ }\textbf {\bibinfo {volume} {103}},\ \bibinfo
  {pages} {023507} (\bibinfo {year} {2021})},\ \Eprint
  {http://arxiv.org/abs/2009.10106} {arXiv:2009.10106 [astro-ph.CO]}
  \BibitemShut {NoStop}%
\bibitem [{\citenamefont {Yang}\ \emph
  {et~al.}(2021{\natexlab{a}})\citenamefont {Yang}, \citenamefont
  {Di~Valentino}, \citenamefont {Pan}, \citenamefont {Wu},\ and\ \citenamefont
  {Lu}}]{Yang:2021flj}%
  \BibitemOpen
  \bibfield  {author} {\bibinfo {author} {\bibfnamefont {W.}~\bibnamefont
  {Yang}}, \bibinfo {author} {\bibfnamefont {E.}~\bibnamefont {Di~Valentino}},
  \bibinfo {author} {\bibfnamefont {S.}~\bibnamefont {Pan}}, \bibinfo {author}
  {\bibfnamefont {Y.}~\bibnamefont {Wu}}, \ and\ \bibinfo {author}
  {\bibfnamefont {J.}~\bibnamefont {Lu}},\ }\href {\doibase
  10.1093/mnras/staa3914} {\bibfield  {journal} {\bibinfo  {journal} {Mon. Not.
  Roy. Astron. Soc.}\ }\textbf {\bibinfo {volume} {501}},\ \bibinfo {pages}
  {5845} (\bibinfo {year} {2021}{\natexlab{a}})},\ \Eprint
  {http://arxiv.org/abs/2101.02168} {arXiv:2101.02168 [astro-ph.CO]}
  \BibitemShut {NoStop}%
\bibitem [{\citenamefont {Yang}\ \emph
  {et~al.}(2021{\natexlab{b}})\citenamefont {Yang}, \citenamefont
  {Di~Valentino}, \citenamefont {Pan}, \citenamefont {Shafieloo},\ and\
  \citenamefont {Li}}]{Yang:2021eud}%
  \BibitemOpen
  \bibfield  {author} {\bibinfo {author} {\bibfnamefont {W.}~\bibnamefont
  {Yang}}, \bibinfo {author} {\bibfnamefont {E.}~\bibnamefont {Di~Valentino}},
  \bibinfo {author} {\bibfnamefont {S.}~\bibnamefont {Pan}}, \bibinfo {author}
  {\bibfnamefont {A.}~\bibnamefont {Shafieloo}}, \ and\ \bibinfo {author}
  {\bibfnamefont {X.}~\bibnamefont {Li}},\ }\href@noop {} {\bibfield  {journal}
  {\bibinfo  {journal} {2103.03815}\ } (\bibinfo {year}
  {2021}{\natexlab{b}})}\BibitemShut {NoStop}%
\bibitem [{\citenamefont {Yang}\ \emph
  {et~al.}(2021{\natexlab{c}})\citenamefont {Yang}, \citenamefont
  {Di~Valentino}, \citenamefont {Pan},\ and\ \citenamefont
  {Mena}}]{Yang:2020ope}%
  \BibitemOpen
  \bibfield  {author} {\bibinfo {author} {\bibfnamefont {W.}~\bibnamefont
  {Yang}}, \bibinfo {author} {\bibfnamefont {E.}~\bibnamefont {Di~Valentino}},
  \bibinfo {author} {\bibfnamefont {S.}~\bibnamefont {Pan}}, \ and\ \bibinfo
  {author} {\bibfnamefont {O.}~\bibnamefont {Mena}},\ }\href {\doibase
  10.1016/j.dark.2020.100762} {\bibfield  {journal} {\bibinfo  {journal} {Phys.
  Dark Univ.}\ }\textbf {\bibinfo {volume} {31}},\ \bibinfo {pages} {100762}
  (\bibinfo {year} {2021}{\natexlab{c}})},\ \Eprint
  {http://arxiv.org/abs/2007.02927} {arXiv:2007.02927 [astro-ph.CO]}
  \BibitemShut {NoStop}%
\bibitem [{\citenamefont {Benaoum}\ \emph {et~al.}(2020)\citenamefont
  {Benaoum}, \citenamefont {Yang}, \citenamefont {Pan},\ and\ \citenamefont
  {Di~Valentino}}]{Benaoum:2020qsi}%
  \BibitemOpen
  \bibfield  {author} {\bibinfo {author} {\bibfnamefont {H.~B.}\ \bibnamefont
  {Benaoum}}, \bibinfo {author} {\bibfnamefont {W.}~\bibnamefont {Yang}},
  \bibinfo {author} {\bibfnamefont {S.}~\bibnamefont {Pan}}, \ and\ \bibinfo
  {author} {\bibfnamefont {E.}~\bibnamefont {Di~Valentino}},\ }\href@noop {} {\
   (\bibinfo {year} {2020})},\ \Eprint {http://arxiv.org/abs/2008.09098}
  {arXiv:2008.09098 [gr-qc]} \BibitemShut {NoStop}%
\bibitem [{\citenamefont {Wang}(2022)}]{Wang:2022xdw}%
  \BibitemOpen
  \bibfield  {author} {\bibinfo {author} {\bibfnamefont {D.}~\bibnamefont
  {Wang}},\ }\href {\doibase 10.1103/PhysRevD.106.063515} {\bibfield  {journal}
  {\bibinfo  {journal} {Phys. Rev. D}\ }\textbf {\bibinfo {volume} {106}},\
  \bibinfo {pages} {063515} (\bibinfo {year} {2022})},\ \Eprint
  {http://arxiv.org/abs/2207.07164} {arXiv:2207.07164 [astro-ph.CO]}
  \BibitemShut {NoStop}%
\bibitem [{\citenamefont {Yang}\ \emph {et~al.}(2023)\citenamefont {Yang},
  \citenamefont {Giar\`e}, \citenamefont {Pan}, \citenamefont {Di~Valentino},
  \citenamefont {Melchiorri},\ and\ \citenamefont {Silk}}]{Yang:2022kho}%
  \BibitemOpen
  \bibfield  {author} {\bibinfo {author} {\bibfnamefont {W.}~\bibnamefont
  {Yang}}, \bibinfo {author} {\bibfnamefont {W.}~\bibnamefont {Giar\`e}},
  \bibinfo {author} {\bibfnamefont {S.}~\bibnamefont {Pan}}, \bibinfo {author}
  {\bibfnamefont {E.}~\bibnamefont {Di~Valentino}}, \bibinfo {author}
  {\bibfnamefont {A.}~\bibnamefont {Melchiorri}}, \ and\ \bibinfo {author}
  {\bibfnamefont {J.}~\bibnamefont {Silk}},\ }\href {\doibase
  10.1103/PhysRevD.107.063509} {\bibfield  {journal} {\bibinfo  {journal}
  {Phys. Rev. D}\ }\textbf {\bibinfo {volume} {107}},\ \bibinfo {pages}
  {063509} (\bibinfo {year} {2023})},\ \Eprint
  {http://arxiv.org/abs/2210.09865} {arXiv:2210.09865 [astro-ph.CO]}
  \BibitemShut {NoStop}%
\bibitem [{\citenamefont {Rezaei}\ \emph {et~al.}(2024)\citenamefont {Rezaei},
  \citenamefont {Pan}, \citenamefont {Yang},\ and\ \citenamefont
  {Mota}}]{Rezaei:2023xkj}%
  \BibitemOpen
  \bibfield  {author} {\bibinfo {author} {\bibfnamefont {M.}~\bibnamefont
  {Rezaei}}, \bibinfo {author} {\bibfnamefont {S.}~\bibnamefont {Pan}},
  \bibinfo {author} {\bibfnamefont {W.}~\bibnamefont {Yang}}, \ and\ \bibinfo
  {author} {\bibfnamefont {D.~F.}\ \bibnamefont {Mota}},\ }\href {\doibase
  10.1088/1475-7516/2024/01/052} {\bibfield  {journal} {\bibinfo  {journal}
  {JCAP}\ }\textbf {\bibinfo {volume} {01}},\ \bibinfo {pages} {052} (\bibinfo
  {year} {2024})},\ \Eprint {http://arxiv.org/abs/2305.18544} {arXiv:2305.18544
  [astro-ph.CO]} \BibitemShut {NoStop}%
\bibitem [{\citenamefont {Escamilla}\ \emph
  {et~al.}(2024{\natexlab{a}})\citenamefont {Escamilla}, \citenamefont
  {Giar\`e}, \citenamefont {Di~Valentino}, \citenamefont {Nunes},\ and\
  \citenamefont {Vagnozzi}}]{Escamilla:2023oce}%
  \BibitemOpen
  \bibfield  {author} {\bibinfo {author} {\bibfnamefont {L.~A.}\ \bibnamefont
  {Escamilla}}, \bibinfo {author} {\bibfnamefont {W.}~\bibnamefont {Giar\`e}},
  \bibinfo {author} {\bibfnamefont {E.}~\bibnamefont {Di~Valentino}}, \bibinfo
  {author} {\bibfnamefont {R.~C.}\ \bibnamefont {Nunes}}, \ and\ \bibinfo
  {author} {\bibfnamefont {S.}~\bibnamefont {Vagnozzi}},\ }\href {\doibase
  10.1088/1475-7516/2024/05/091} {\bibfield  {journal} {\bibinfo  {journal}
  {JCAP}\ }\textbf {\bibinfo {volume} {05}},\ \bibinfo {pages} {091} (\bibinfo
  {year} {2024}{\natexlab{a}})},\ \Eprint {http://arxiv.org/abs/2307.14802}
  {arXiv:2307.14802 [astro-ph.CO]} \BibitemShut {NoStop}%
\bibitem [{\citenamefont {Adil}\ \emph {et~al.}(2024)\citenamefont {Adil},
  \citenamefont {Akarsu}, \citenamefont {Di~Valentino}, \citenamefont {Nunes},
  \citenamefont {\"Oz\"ulker}, \citenamefont {Sen},\ and\ \citenamefont
  {Specogna}}]{Adil:2023exv}%
  \BibitemOpen
  \bibfield  {author} {\bibinfo {author} {\bibfnamefont {S.~A.}\ \bibnamefont
  {Adil}}, \bibinfo {author} {\bibfnamefont {O.}~\bibnamefont {Akarsu}},
  \bibinfo {author} {\bibfnamefont {E.}~\bibnamefont {Di~Valentino}}, \bibinfo
  {author} {\bibfnamefont {R.~C.}\ \bibnamefont {Nunes}}, \bibinfo {author}
  {\bibfnamefont {E.}~\bibnamefont {\"Oz\"ulker}}, \bibinfo {author}
  {\bibfnamefont {A.~A.}\ \bibnamefont {Sen}}, \ and\ \bibinfo {author}
  {\bibfnamefont {E.}~\bibnamefont {Specogna}},\ }\href {\doibase
  10.1103/PhysRevD.109.023527} {\bibfield  {journal} {\bibinfo  {journal}
  {Phys. Rev. D}\ }\textbf {\bibinfo {volume} {109}},\ \bibinfo {pages}
  {023527} (\bibinfo {year} {2024})},\ \Eprint
  {http://arxiv.org/abs/2306.08046} {arXiv:2306.08046 [astro-ph.CO]}
  \BibitemShut {NoStop}%
\bibitem [{\citenamefont {Rezaei}(2024)}]{Rezaei:2024vtg}%
  \BibitemOpen
  \bibfield  {author} {\bibinfo {author} {\bibfnamefont {M.}~\bibnamefont
  {Rezaei}},\ }\href {\doibase 10.3847/1538-4357/ad3963} {\bibfield  {journal}
  {\bibinfo  {journal} {Astrophys. J.}\ }\textbf {\bibinfo {volume} {967}},\
  \bibinfo {pages} {2} (\bibinfo {year} {2024})},\ \Eprint
  {http://arxiv.org/abs/2403.18968} {arXiv:2403.18968 [astro-ph.CO]}
  \BibitemShut {NoStop}%
\bibitem [{\citenamefont {Escamilla}\ \emph
  {et~al.}(2024{\natexlab{b}})\citenamefont {Escamilla}, \citenamefont {Pan},
  \citenamefont {Di~Valentino}, \citenamefont {Paliathanasis}, \citenamefont
  {V\'azquez},\ and\ \citenamefont {Yang}}]{Escamilla:2024olw}%
  \BibitemOpen
  \bibfield  {author} {\bibinfo {author} {\bibfnamefont {L.~A.}\ \bibnamefont
  {Escamilla}}, \bibinfo {author} {\bibfnamefont {S.}~\bibnamefont {Pan}},
  \bibinfo {author} {\bibfnamefont {E.}~\bibnamefont {Di~Valentino}}, \bibinfo
  {author} {\bibfnamefont {A.}~\bibnamefont {Paliathanasis}}, \bibinfo {author}
  {\bibfnamefont {J.~A.}\ \bibnamefont {V\'azquez}}, \ and\ \bibinfo {author}
  {\bibfnamefont {W.}~\bibnamefont {Yang}},\ }\href@noop {} {\bibfield
  {journal} {\bibinfo  {journal} {2404.00181}\ } (\bibinfo {year}
  {2024}{\natexlab{b}})}\BibitemShut {NoStop}%
\bibitem [{\citenamefont {Moshafi}\ \emph {et~al.}(2024)\citenamefont
  {Moshafi}, \citenamefont {Talebian}, \citenamefont {Yusofi},\ and\
  \citenamefont {Di~Valentino}}]{Moshafi:2024guo}%
  \BibitemOpen
  \bibfield  {author} {\bibinfo {author} {\bibfnamefont {H.}~\bibnamefont
  {Moshafi}}, \bibinfo {author} {\bibfnamefont {A.}~\bibnamefont {Talebian}},
  \bibinfo {author} {\bibfnamefont {E.}~\bibnamefont {Yusofi}}, \ and\ \bibinfo
  {author} {\bibfnamefont {E.}~\bibnamefont {Di~Valentino}},\ }\href {\doibase
  10.1016/j.dark.2024.101524} {\bibfield  {journal} {\bibinfo  {journal} {Phys.
  Dark Univ.}\ }\textbf {\bibinfo {volume} {45}},\ \bibinfo {pages} {101524}
  (\bibinfo {year} {2024})},\ \Eprint {http://arxiv.org/abs/2403.02000}
  {arXiv:2403.02000 [astro-ph.CO]} \BibitemShut {NoStop}%
\bibitem [{\citenamefont {Reyhani}\ \emph {et~al.}(2024)\citenamefont
  {Reyhani}, \citenamefont {Najafi}, \citenamefont {Firouzjaee},\ and\
  \citenamefont {Di~Valentino}}]{Reyhani:2024cnr}%
  \BibitemOpen
  \bibfield  {author} {\bibinfo {author} {\bibfnamefont {M.}~\bibnamefont
  {Reyhani}}, \bibinfo {author} {\bibfnamefont {M.}~\bibnamefont {Najafi}},
  \bibinfo {author} {\bibfnamefont {J.~T.}\ \bibnamefont {Firouzjaee}}, \ and\
  \bibinfo {author} {\bibfnamefont {E.}~\bibnamefont {Di~Valentino}},\ }\href
  {\doibase 10.1016/j.dark.2024.101477} {\bibfield  {journal} {\bibinfo
  {journal} {Phys. Dark Univ.}\ }\textbf {\bibinfo {volume} {44}},\ \bibinfo
  {pages} {101477} (\bibinfo {year} {2024})},\ \Eprint
  {http://arxiv.org/abs/2403.15202} {arXiv:2403.15202 [astro-ph.CO]}
  \BibitemShut {NoStop}%
\bibitem [{\citenamefont {Dimakis}\ \emph {et~al.}(2016)\citenamefont
  {Dimakis}, \citenamefont {Karagiorgos}, \citenamefont {Zampeli},
  \citenamefont {Paliathanasis}, \citenamefont {Christodoulakis},\ and\
  \citenamefont {Terzis}}]{Dimakis:2016mip}%
  \BibitemOpen
  \bibfield  {author} {\bibinfo {author} {\bibfnamefont {N.}~\bibnamefont
  {Dimakis}}, \bibinfo {author} {\bibfnamefont {A.}~\bibnamefont
  {Karagiorgos}}, \bibinfo {author} {\bibfnamefont {A.}~\bibnamefont
  {Zampeli}}, \bibinfo {author} {\bibfnamefont {A.}~\bibnamefont
  {Paliathanasis}}, \bibinfo {author} {\bibfnamefont {T.}~\bibnamefont
  {Christodoulakis}}, \ and\ \bibinfo {author} {\bibfnamefont {P.~A.}\
  \bibnamefont {Terzis}},\ }\href {\doibase 10.1103/PhysRevD.93.123518}
  {\bibfield  {journal} {\bibinfo  {journal} {Phys. Rev. D}\ }\textbf {\bibinfo
  {volume} {93}},\ \bibinfo {pages} {123518} (\bibinfo {year} {2016})},\
  \Eprint {http://arxiv.org/abs/1604.05168} {arXiv:1604.05168 [gr-qc]}
  \BibitemShut {NoStop}%
\bibitem [{\citenamefont {Pan}\ \emph {et~al.}(2020{\natexlab{b}})\citenamefont
  {Pan}, \citenamefont {Yang},\ and\ \citenamefont
  {Paliathanasis}}]{Pan:2019brc}%
  \BibitemOpen
  \bibfield  {author} {\bibinfo {author} {\bibfnamefont {S.}~\bibnamefont
  {Pan}}, \bibinfo {author} {\bibfnamefont {W.}~\bibnamefont {Yang}}, \ and\
  \bibinfo {author} {\bibfnamefont {A.}~\bibnamefont {Paliathanasis}},\ }\href
  {\doibase 10.1140/epjc/s10052-020-7832-y} {\bibfield  {journal} {\bibinfo
  {journal} {Eur. Phys. J. C}\ }\textbf {\bibinfo {volume} {80}},\ \bibinfo
  {pages} {274} (\bibinfo {year} {2020}{\natexlab{b}})},\ \Eprint
  {http://arxiv.org/abs/1902.07108} {arXiv:1902.07108 [astro-ph.CO]}
  \BibitemShut {NoStop}%
\bibitem [{\citenamefont {Mukhanov}\ \emph {et~al.}(1992)\citenamefont
  {Mukhanov}, \citenamefont {Feldman},\ and\ \citenamefont
  {Brandenberger}}]{Mukhanov:1990me}%
  \BibitemOpen
  \bibfield  {author} {\bibinfo {author} {\bibfnamefont {V.~F.}\ \bibnamefont
  {Mukhanov}}, \bibinfo {author} {\bibfnamefont {H.~A.}\ \bibnamefont
  {Feldman}}, \ and\ \bibinfo {author} {\bibfnamefont {R.~H.}\ \bibnamefont
  {Brandenberger}},\ }\href {\doibase 10.1016/0370-1573(92)90044-Z} {\bibfield
  {journal} {\bibinfo  {journal} {Phys. Rept.}\ }\textbf {\bibinfo {volume}
  {215}},\ \bibinfo {pages} {203} (\bibinfo {year} {1992})}\BibitemShut
  {NoStop}%
\bibitem [{\citenamefont {Ma}\ and\ \citenamefont
  {Bertschinger}(1995)}]{Ma:1995ey}%
  \BibitemOpen
  \bibfield  {author} {\bibinfo {author} {\bibfnamefont {C.-P.}\ \bibnamefont
  {Ma}}\ and\ \bibinfo {author} {\bibfnamefont {E.}~\bibnamefont
  {Bertschinger}},\ }\href {\doibase 10.1086/176550} {\bibfield  {journal}
  {\bibinfo  {journal} {Astrophys. J.}\ }\textbf {\bibinfo {volume} {455}},\
  \bibinfo {pages} {7} (\bibinfo {year} {1995})},\ \Eprint
  {http://arxiv.org/abs/astro-ph/9506072} {arXiv:astro-ph/9506072} \BibitemShut
  {NoStop}%
\bibitem [{\citenamefont {Malik}\ and\ \citenamefont
  {Wands}(2009)}]{Malik:2008im}%
  \BibitemOpen
  \bibfield  {author} {\bibinfo {author} {\bibfnamefont {K.~A.}\ \bibnamefont
  {Malik}}\ and\ \bibinfo {author} {\bibfnamefont {D.}~\bibnamefont {Wands}},\
  }\href {\doibase 10.1016/j.physrep.2009.03.001} {\bibfield  {journal}
  {\bibinfo  {journal} {Phys. Rept.}\ }\textbf {\bibinfo {volume} {475}},\
  \bibinfo {pages} {1} (\bibinfo {year} {2009})},\ \Eprint
  {http://arxiv.org/abs/0809.4944} {arXiv:0809.4944 [astro-ph]} \BibitemShut
  {NoStop}%
\bibitem [{\citenamefont {Valiviita}\ \emph {et~al.}(2008)\citenamefont
  {Valiviita}, \citenamefont {Majerotto},\ and\ \citenamefont
  {Maartens}}]{Valiviita:2008iv}%
  \BibitemOpen
  \bibfield  {author} {\bibinfo {author} {\bibfnamefont {J.}~\bibnamefont
  {Valiviita}}, \bibinfo {author} {\bibfnamefont {E.}~\bibnamefont
  {Majerotto}}, \ and\ \bibinfo {author} {\bibfnamefont {R.}~\bibnamefont
  {Maartens}},\ }\href {\doibase 10.1088/1475-7516/2008/07/020} {\bibfield
  {journal} {\bibinfo  {journal} {JCAP}\ }\textbf {\bibinfo {volume} {07}},\
  \bibinfo {pages} {020} (\bibinfo {year} {2008})},\ \Eprint
  {http://arxiv.org/abs/0804.0232} {arXiv:0804.0232 [astro-ph]} \BibitemShut
  {NoStop}%
\bibitem [{\citenamefont {Aghanim}\ \emph
  {et~al.}(2020{\natexlab{a}})\citenamefont {Aghanim} \emph
  {et~al.}}]{Aghanim:2019ame}%
  \BibitemOpen
  \bibfield  {author} {\bibinfo {author} {\bibfnamefont {N.}~\bibnamefont
  {Aghanim}} \emph {et~al.} (\bibinfo {collaboration} {Planck}),\ }\href
  {\doibase 10.1051/0004-6361/201936386} {\bibfield  {journal} {\bibinfo
  {journal} {Astron. Astrophys.}\ }\textbf {\bibinfo {volume} {641}},\ \bibinfo
  {pages} {A5} (\bibinfo {year} {2020}{\natexlab{a}})},\ \Eprint
  {http://arxiv.org/abs/1907.12875} {arXiv:1907.12875 [astro-ph.CO]}
  \BibitemShut {NoStop}%
\bibitem [{\citenamefont {Aghanim}\ \emph
  {et~al.}(2020{\natexlab{b}})\citenamefont {Aghanim} \emph
  {et~al.}}]{Aghanim:2018oex}%
  \BibitemOpen
  \bibfield  {author} {\bibinfo {author} {\bibfnamefont {N.}~\bibnamefont
  {Aghanim}} \emph {et~al.} (\bibinfo {collaboration} {Planck}),\ }\href
  {\doibase 10.1051/0004-6361/201833886} {\bibfield  {journal} {\bibinfo
  {journal} {Astron. Astrophys.}\ }\textbf {\bibinfo {volume} {641}},\ \bibinfo
  {pages} {A8} (\bibinfo {year} {2020}{\natexlab{b}})},\ \Eprint
  {http://arxiv.org/abs/1807.06210} {arXiv:1807.06210 [astro-ph.CO]}
  \BibitemShut {NoStop}%
\bibitem [{\citenamefont {Aiola}\ \emph {et~al.}(2020)\citenamefont {Aiola}
  \emph {et~al.}}]{ACT:2020gnv}%
  \BibitemOpen
  \bibfield  {author} {\bibinfo {author} {\bibfnamefont {S.}~\bibnamefont
  {Aiola}} \emph {et~al.} (\bibinfo {collaboration} {ACT}),\ }\href {\doibase
  10.1088/1475-7516/2020/12/047} {\bibfield  {journal} {\bibinfo  {journal}
  {JCAP}\ }\textbf {\bibinfo {volume} {12}},\ \bibinfo {pages} {047} (\bibinfo
  {year} {2020})},\ \Eprint {http://arxiv.org/abs/2007.07288} {arXiv:2007.07288
  [astro-ph.CO]} \BibitemShut {NoStop}%
\bibitem [{\citenamefont {Choi}\ \emph {et~al.}(2020)\citenamefont {Choi} \emph
  {et~al.}}]{ACT:2020frw}%
  \BibitemOpen
  \bibfield  {author} {\bibinfo {author} {\bibfnamefont {S.~K.}\ \bibnamefont
  {Choi}} \emph {et~al.} (\bibinfo {collaboration} {ACT}),\ }\href {\doibase
  10.1088/1475-7516/2020/12/045} {\bibfield  {journal} {\bibinfo  {journal}
  {JCAP}\ }\textbf {\bibinfo {volume} {12}},\ \bibinfo {pages} {045} (\bibinfo
  {year} {2020})},\ \Eprint {http://arxiv.org/abs/2007.07289} {arXiv:2007.07289
  [astro-ph.CO]} \BibitemShut {NoStop}%
\bibitem [{\citenamefont {Balkenhol}\ \emph {et~al.}(2023)\citenamefont
  {Balkenhol} \emph {et~al.}}]{SPT-3G:2022hvq}%
  \BibitemOpen
  \bibfield  {author} {\bibinfo {author} {\bibfnamefont {L.}~\bibnamefont
  {Balkenhol}} \emph {et~al.} (\bibinfo {collaboration} {SPT-3G}),\ }\href
  {\doibase 10.1103/PhysRevD.108.023510} {\bibfield  {journal} {\bibinfo
  {journal} {Phys. Rev. D}\ }\textbf {\bibinfo {volume} {108}},\ \bibinfo
  {pages} {023510} (\bibinfo {year} {2023})},\ \Eprint
  {http://arxiv.org/abs/2212.05642} {arXiv:2212.05642 [astro-ph.CO]}
  \BibitemShut {NoStop}%
\bibitem [{\citenamefont {Bennett}\ \emph {et~al.}(2013)\citenamefont {Bennett}
  \emph {et~al.}}]{WMAP:2012fli}%
  \BibitemOpen
  \bibfield  {author} {\bibinfo {author} {\bibfnamefont {C.~L.}\ \bibnamefont
  {Bennett}} \emph {et~al.} (\bibinfo {collaboration} {WMAP}),\ }\href
  {\doibase 10.1088/0067-0049/208/2/20} {\bibfield  {journal} {\bibinfo
  {journal} {Astrophys. J. Suppl.}\ }\textbf {\bibinfo {volume} {208}},\
  \bibinfo {pages} {20} (\bibinfo {year} {2013})},\ \Eprint
  {http://arxiv.org/abs/1212.5225} {arXiv:1212.5225 [astro-ph.CO]} \BibitemShut
  {NoStop}%
\bibitem [{\citenamefont {Ade}\ \emph {et~al.}(2014)\citenamefont {Ade} \emph
  {et~al.}}]{Planck:2013win}%
  \BibitemOpen
  \bibfield  {author} {\bibinfo {author} {\bibfnamefont {P.~A.~R.}\
  \bibnamefont {Ade}} \emph {et~al.} (\bibinfo {collaboration} {Planck}),\
  }\href {\doibase 10.1051/0004-6361/201321573} {\bibfield  {journal} {\bibinfo
   {journal} {Astron. Astrophys.}\ }\textbf {\bibinfo {volume} {571}},\
  \bibinfo {pages} {A15} (\bibinfo {year} {2014})},\ \Eprint
  {http://arxiv.org/abs/1303.5075} {arXiv:1303.5075 [astro-ph.CO]} \BibitemShut
  {NoStop}%
\bibitem [{\citenamefont {Alam}\ \emph {et~al.}(2021)\citenamefont {Alam} \emph
  {et~al.}}]{Alam:2020sor}%
  \BibitemOpen
  \bibfield  {author} {\bibinfo {author} {\bibfnamefont {S.}~\bibnamefont
  {Alam}} \emph {et~al.} (\bibinfo {collaboration} {eBOSS}),\ }\href {\doibase
  10.1103/PhysRevD.103.083533} {\bibfield  {journal} {\bibinfo  {journal}
  {Phys. Rev. D}\ }\textbf {\bibinfo {volume} {103}},\ \bibinfo {pages}
  {083533} (\bibinfo {year} {2021})},\ \Eprint
  {http://arxiv.org/abs/2007.08991} {arXiv:2007.08991 [astro-ph.CO]}
  \BibitemShut {NoStop}%
\bibitem [{\citenamefont {Scolnic}\ \emph {et~al.}(2018)\citenamefont {Scolnic}
  \emph {et~al.}}]{Scolnic:2017caz}%
  \BibitemOpen
  \bibfield  {author} {\bibinfo {author} {\bibfnamefont {D.~M.}\ \bibnamefont
  {Scolnic}} \emph {et~al.},\ }\href {\doibase 10.3847/1538-4357/aab9bb}
  {\bibfield  {journal} {\bibinfo  {journal} {Astrophys. J.}\ }\textbf
  {\bibinfo {volume} {859}},\ \bibinfo {pages} {101} (\bibinfo {year}
  {2018})},\ \Eprint {http://arxiv.org/abs/1710.00845} {arXiv:1710.00845
  [astro-ph.CO]} \BibitemShut {NoStop}%
\bibitem [{\citenamefont {Lewis}\ \emph {et~al.}(2000)\citenamefont {Lewis},
  \citenamefont {Challinor},\ and\ \citenamefont {Lasenby}}]{Lewis:1999bs}%
  \BibitemOpen
  \bibfield  {author} {\bibinfo {author} {\bibfnamefont {A.}~\bibnamefont
  {Lewis}}, \bibinfo {author} {\bibfnamefont {A.}~\bibnamefont {Challinor}}, \
  and\ \bibinfo {author} {\bibfnamefont {A.}~\bibnamefont {Lasenby}},\ }\href
  {\doibase 10.1086/309179} {\bibfield  {journal} {\bibinfo  {journal}
  {Astrophys. J.}\ }\textbf {\bibinfo {volume} {538}},\ \bibinfo {pages} {473}
  (\bibinfo {year} {2000})},\ \Eprint {http://arxiv.org/abs/astro-ph/9911177}
  {arXiv:astro-ph/9911177 [astro-ph]} \BibitemShut {NoStop}%
\bibitem [{\citenamefont {Howlett}\ \emph {et~al.}(2012)\citenamefont
  {Howlett}, \citenamefont {Lewis}, \citenamefont {Hall},\ and\ \citenamefont
  {Challinor}}]{Howlett:2012mh}%
  \BibitemOpen
  \bibfield  {author} {\bibinfo {author} {\bibfnamefont {C.}~\bibnamefont
  {Howlett}}, \bibinfo {author} {\bibfnamefont {A.}~\bibnamefont {Lewis}},
  \bibinfo {author} {\bibfnamefont {A.}~\bibnamefont {Hall}}, \ and\ \bibinfo
  {author} {\bibfnamefont {A.}~\bibnamefont {Challinor}},\ }\href {\doibase
  10.1088/1475-7516/2012/04/027} {\bibfield  {journal} {\bibinfo  {journal}
  {JCAP}\ }\textbf {\bibinfo {volume} {1204}},\ \bibinfo {pages} {027}
  (\bibinfo {year} {2012})},\ \Eprint {http://arxiv.org/abs/1201.3654}
  {arXiv:1201.3654 [astro-ph.CO]} \BibitemShut {NoStop}%
\bibitem [{\citenamefont {Lewis}\ and\ \citenamefont
  {Bridle}(2002)}]{Lewis:2002ah}%
  \BibitemOpen
  \bibfield  {author} {\bibinfo {author} {\bibfnamefont {A.}~\bibnamefont
  {Lewis}}\ and\ \bibinfo {author} {\bibfnamefont {S.}~\bibnamefont {Bridle}},\
  }\href {\doibase 10.1103/PhysRevD.66.103511} {\bibfield  {journal} {\bibinfo
  {journal} {Phys. Rev.}\ }\textbf {\bibinfo {volume} {D66}},\ \bibinfo {pages}
  {103511} (\bibinfo {year} {2002})},\ \Eprint
  {http://arxiv.org/abs/astro-ph/0205436} {arXiv:astro-ph/0205436 [astro-ph]}
  \BibitemShut {NoStop}%
\bibitem [{\citenamefont {Lewis}(2013)}]{Lewis:2013hha}%
  \BibitemOpen
  \bibfield  {author} {\bibinfo {author} {\bibfnamefont {A.}~\bibnamefont
  {Lewis}},\ }\href {\doibase 10.1103/PhysRevD.87.103529} {\bibfield  {journal}
  {\bibinfo  {journal} {Phys. Rev.}\ }\textbf {\bibinfo {volume} {D87}},\
  \bibinfo {pages} {103529} (\bibinfo {year} {2013})},\ \Eprint
  {http://arxiv.org/abs/1304.4473} {arXiv:1304.4473 [astro-ph.CO]} \BibitemShut
  {NoStop}%
\bibitem [{\citenamefont {{Neal}}(2005)}]{Neal:2005}%
  \BibitemOpen
  \bibfield  {author} {\bibinfo {author} {\bibfnamefont {R.~M.}\ \bibnamefont
  {{Neal}}},\ }\href {https://arxiv.org/abs/math/0502099} {\bibfield  {journal}
  {\bibinfo  {journal} {ArXiv Mathematics e-prints}\ } (\bibinfo {year}
  {2005})},\ \Eprint {http://arxiv.org/abs/math/0502099} {math/0502099}
  \BibitemShut {NoStop}%
\bibitem [{\citenamefont {Gelman}\ and\ \citenamefont
  {Rubin}(1992)}]{Gelman:1992zz}%
  \BibitemOpen
  \bibfield  {author} {\bibinfo {author} {\bibfnamefont {A.}~\bibnamefont
  {Gelman}}\ and\ \bibinfo {author} {\bibfnamefont {D.~B.}\ \bibnamefont
  {Rubin}},\ }\href {\doibase 10.1214/ss/1177011136} {\bibfield  {journal}
  {\bibinfo  {journal} {Statist. Sci.}\ }\textbf {\bibinfo {volume} {7}},\
  \bibinfo {pages} {457} (\bibinfo {year} {1992})}\BibitemShut {NoStop}%
\bibitem [{\citenamefont {Heavens}\ \emph
  {et~al.}(2017{\natexlab{a}})\citenamefont {Heavens}, \citenamefont {Fantaye},
  \citenamefont {Sellentin}, \citenamefont {Eggers}, \citenamefont {Hosenie},
  \citenamefont {Kroon},\ and\ \citenamefont {Mootoovaloo}}]{Heavens:2017hkr}%
  \BibitemOpen
  \bibfield  {author} {\bibinfo {author} {\bibfnamefont {A.}~\bibnamefont
  {Heavens}}, \bibinfo {author} {\bibfnamefont {Y.}~\bibnamefont {Fantaye}},
  \bibinfo {author} {\bibfnamefont {E.}~\bibnamefont {Sellentin}}, \bibinfo
  {author} {\bibfnamefont {H.}~\bibnamefont {Eggers}}, \bibinfo {author}
  {\bibfnamefont {Z.}~\bibnamefont {Hosenie}}, \bibinfo {author} {\bibfnamefont
  {S.}~\bibnamefont {Kroon}}, \ and\ \bibinfo {author} {\bibfnamefont
  {A.}~\bibnamefont {Mootoovaloo}},\ }\href {\doibase
  10.1103/PhysRevLett.119.101301} {\bibfield  {journal} {\bibinfo  {journal}
  {Phys. Rev. Lett.}\ }\textbf {\bibinfo {volume} {119}},\ \bibinfo {pages}
  {101301} (\bibinfo {year} {2017}{\natexlab{a}})},\ \Eprint
  {http://arxiv.org/abs/1704.03467} {arXiv:1704.03467 [astro-ph.CO]}
  \BibitemShut {NoStop}%
\bibitem [{\citenamefont {Heavens}\ \emph
  {et~al.}(2017{\natexlab{b}})\citenamefont {Heavens}, \citenamefont {Fantaye},
  \citenamefont {Mootoovaloo}, \citenamefont {Eggers}, \citenamefont {Hosenie},
  \citenamefont {Kroon},\ and\ \citenamefont {Sellentin}}]{Heavens:2017afc}%
  \BibitemOpen
  \bibfield  {author} {\bibinfo {author} {\bibfnamefont {A.}~\bibnamefont
  {Heavens}}, \bibinfo {author} {\bibfnamefont {Y.}~\bibnamefont {Fantaye}},
  \bibinfo {author} {\bibfnamefont {A.}~\bibnamefont {Mootoovaloo}}, \bibinfo
  {author} {\bibfnamefont {H.}~\bibnamefont {Eggers}}, \bibinfo {author}
  {\bibfnamefont {Z.}~\bibnamefont {Hosenie}}, \bibinfo {author} {\bibfnamefont
  {S.}~\bibnamefont {Kroon}}, \ and\ \bibinfo {author} {\bibfnamefont
  {E.}~\bibnamefont {Sellentin}},\ }\href@noop {} {\  (\bibinfo {year}
  {2017}{\natexlab{b}})},\ \Eprint {http://arxiv.org/abs/1704.03472}
  {arXiv:1704.03472 [stat.CO]} \BibitemShut {NoStop}%
\bibitem [{\citenamefont {Kass}\ and\ \citenamefont
  {Raftery}(1995)}]{Kass:1995loi}%
  \BibitemOpen
  \bibfield  {author} {\bibinfo {author} {\bibfnamefont {R.~E.}\ \bibnamefont
  {Kass}}\ and\ \bibinfo {author} {\bibfnamefont {A.~E.}\ \bibnamefont
  {Raftery}},\ }\href {\doibase 10.1080/01621459.1995.10476572} {\bibfield
  {journal} {\bibinfo  {journal} {J. Am. Statist. Assoc.}\ }\textbf {\bibinfo
  {volume} {90}},\ \bibinfo {pages} {773} (\bibinfo {year} {1995})}\BibitemShut
  {NoStop}%
\bibitem [{\citenamefont {Trotta}(2008)}]{Trotta:2008qt}%
  \BibitemOpen
  \bibfield  {author} {\bibinfo {author} {\bibfnamefont {R.}~\bibnamefont
  {Trotta}},\ }\href {\doibase 10.1080/00107510802066753} {\bibfield  {journal}
  {\bibinfo  {journal} {Contemp. Phys.}\ }\textbf {\bibinfo {volume} {49}},\
  \bibinfo {pages} {71} (\bibinfo {year} {2008})},\ \Eprint
  {http://arxiv.org/abs/0803.4089} {arXiv:0803.4089 [astro-ph]} \BibitemShut
  {NoStop}%
\bibitem [{\citenamefont {Riess}\ \emph {et~al.}(2021)\citenamefont {Riess}
  \emph {et~al.}}]{Riess:2021jrx}%
  \BibitemOpen
  \bibfield  {author} {\bibinfo {author} {\bibfnamefont {A.~G.}\ \bibnamefont
  {Riess}} \emph {et~al.},\ }\href@noop {} {\  (\bibinfo {year} {2021})},\
  \Eprint {http://arxiv.org/abs/2112.04510} {arXiv:2112.04510 [astro-ph.CO]}
  \BibitemShut {NoStop}%
\bibitem [{\citenamefont {Aghanim}\ \emph
  {et~al.}(2020{\natexlab{c}})\citenamefont {Aghanim} \emph
  {et~al.}}]{Planck:2018vyg}%
  \BibitemOpen
  \bibfield  {author} {\bibinfo {author} {\bibfnamefont {N.}~\bibnamefont
  {Aghanim}} \emph {et~al.} (\bibinfo {collaboration} {Planck}),\ }\href
  {\doibase 10.1051/0004-6361/201833910} {\bibfield  {journal} {\bibinfo
  {journal} {Astron. Astrophys.}\ }\textbf {\bibinfo {volume} {641}},\ \bibinfo
  {pages} {A6} (\bibinfo {year} {2020}{\natexlab{c}})},\ \bibinfo {note}
  {[Erratum: Astron.Astrophys. 652, C4 (2021)]},\ \Eprint
  {http://arxiv.org/abs/1807.06209} {arXiv:1807.06209 [astro-ph.CO]}
  \BibitemShut {NoStop}%
\bibitem [{\citenamefont {Adame}\ \emph {et~al.}(2024)\citenamefont {Adame}
  \emph {et~al.}}]{DESI:2024mwx}%
  \BibitemOpen
  \bibfield  {author} {\bibinfo {author} {\bibfnamefont {A.~G.}\ \bibnamefont
  {Adame}} \emph {et~al.} (\bibinfo {collaboration} {DESI}),\ }\href@noop {} {\
   (\bibinfo {year} {2024})},\ \Eprint {http://arxiv.org/abs/2404.03002}
  {arXiv:2404.03002 [astro-ph.CO]} \BibitemShut {NoStop}%
\end{thebibliography}%
\end{document}